\documentclass[12pt,cite,epsfig]{article}
\usepackage{times,epsfig}
\newcommand{\nc}{\newcommand}
\nc{\bb}{\bibitem}
\nc{\be}{\begin{equation}}
\nc{\ee}{\end{equation}}
\nc{\pa}{\partial}
\nc{\parsym} {\stackrel{\leftrightarrow}{\pa}}
\nc{\ra}{\rightarrow}
\nc{\la}{\leftarrow}
\nc{\etp}{{\eta^\prime}}
\nc{\omg}{\omega}
\nc{\ggam}{\gamma \gamma}
\nc{\gam}{\gamma }
\nc{\bea}{\begin{eqnarray}}
\nc{\eea}{\end{eqnarray}}
\nc{\beas}{\begin{eqnarray*}}
\nc{\eeas}{\end{eqnarray*}}
\nc{\non}{\nonumber}

\def\hhha{\rule[-3.mm]{0.mm}{7.mm}}
\def\hhhb{\rule[-3.mm]{0.mm}{9.mm}}

\def\hhhu{\rule[-3.mm]{0.mm}{12.mm}}
\def\hhhv{\rule[-3.mm]{0.mm}{9.mm}}
\def\hhhq{\rule[-3.mm]{0.mm}{9.mm}}

\begin{document}
\begin{titlepage}
\vbox{~~~ \\
                                   \null \hfill LPNHE 2009--05\\

\title{A Global Treatment Of VMD Physics Up To The $\phi$~:\\
I.~~ $e^+e^-$ Annihilations, Anomalies And Vector Meson
Partial Widths
   }
\author{
M.~Benayoun, P.~David, L.~ DelBuono, O. Leitner \\
\small{ LPNHE Paris VI/VII, IN2P3/CNRS, F-75252 Paris, France }\\
}
\date{\today}
\maketitle
\begin{abstract}
The HLS Model, equipped with a mechanism providing the breaking of U(3)/SU(3)
symmetry and an isospin symmetry breaking leading naturally to
vector meson mixing, has been recently shown to successfully account for
$e^+ e^- \ra \pi^+\pi^-$ cross section and for the dipion spectrum in $\tau$ decay.
The present study shows that the 
full anomalous sector of the HLS model can be considered and is validated by the
experimental data. Indeed, this extended model provides
a successful simultaneous  fit to  the $e^+ e^- \ra \pi^+\pi^-$ data together
with the available data on $e^+ e^- \ra \pi^0\gamma$, $e^+ e^- \ra \eta\gamma$ and 
$e^+ e^- \ra \pi^0 \pi^+\pi^-$ cross sections. It is shown that the fit
of these data sets  also predicts an accurate description of the
$\eta/\eta^\prime \ra \pi^+ \pi^- \gamma$ decays  fully consistent
with the reported  information on their branching fractions and spectra.  
Finally, one also derives from our global fits products of widths of the form 
$\Gamma (V \ra f_1)\Gamma(V \ra e^+ e^-) $ 
and ratios of the form  $\Gamma (V \ra f_1)/\Gamma (V \ra f_2)$ describing
decays of vector mesons to several non--leptonic final states.
 
\end{abstract}
}
\end{titlepage}

\section{Introduction}
\indent \indent
A Lagrangian model based on the Hidden Local Symmetry Model \cite{HLSOrigin,HLSRef} 
has been used \cite{taupaper} in order to account simultaneously for 
the $e^+e^- \ra \pi^+ \pi^-$ and $\tau^\pm \ra \pi^\pm \pi^0 \nu_\tau$ data.
This model has been  supplied with a SU(3)/U(3) symmetry breaking mechanism \cite{Heath1,rad},
essentially relying on the BKY \cite{BKY} mechanism. In order to apply this model
simultaneously  to annihilation data and to the $\pi \pi$ spectrum from
$\tau$ decays, an isospin breaking mechanism has been defined, which is also a specific
implementation of the vector meson mixing. This is needed  at least in order to 
account for the $\omg \ra \pi^+ \pi^-$ decay which is an important signal in
$e^+e^- \ra \pi^+ \pi^-$ data. Moreover, a global fit to $e^+e^-$ and $\tau$ data needs
a reasonably general isospin symmetry breaking mechanism, active in $e^+e^-$ annihilations
and which  can be switched off for the $\tau$ decay. 

The peculiar aspect of the mixing  model defined in 
\cite{taupaper} is the use of several decay modes which serve
to constrain the parameters introduced by these breaking schemes. 
Then, the anomalous decay 
modes of the form{\footnote{ Throughout this paper, we denote by $P$ and $V$, resp. the basic
pseudoscalar  and vector meson {\it nonets}. The  explicit form of their matrix representation can be found 
in several papers \cite{HLSOrigin,HLSRef,Heath1,rad} and is reminded in \cite{taupaper}. The electromagnetic 
field is denoted by $A$. All definitions and notations in the present paper closely follow
those in  \cite{taupaper}.}  $VP\gamma$ and the two photon decay widths of the 
$\pi^0,~\eta,~\eta^\prime$ mesons  play an important role.

As such, the model is already overconstrained and has been successfully used \cite{taupaper}.
However, it cannot apply to $e^+e^-$ annihilation
channels other than $\pi^+ \pi^-$, $K^+ K^-$ and $K^0 \overline{K}^0$,  and processes like
$e^+e^- \ra \pi^0 \gamma$,  $e^+e^- \ra \eta \gamma$,  $e^+e^- \ra \eta^\prime \gamma$
or $e^+e^- \ra \pi^0 \pi^+ \pi^-$, remain beyond its scope.
Indeed, even if  $V \ra P \gamma$ decay widths  are an important part of these processes, they
do not exhaust the  physics content of the annihilation processes.

Therefore, having a model able to describe all these processes (and  possibly 
others, as for instance  $e^+e^- \ra \pi^0 \omega$) is certainly valuable. This is already important 
from a physics point of view  but, more technically, this may allow
to check the relative consistency of various data sets within the same framework and, therefore,
the systematics affecting the data sets.

An extension of the HLS model is possible by introducing the anomalous Lagrangian
pieces identified long ago \cite{FKTUY,HLSRef} and relying also on the above  mentioned breaking 
schemes reminded or defined in \cite{taupaper}. This could allow interesting improvements
in estimating the hadronic contribution to the muon anomalous magnetic moment $a_\mu$.
Indeed, as  basically  several  crucial physics parameters are common to all low energy annihilations,
considering all annihilation processes altogether may turn out to increase the statistics and, then,
improve uncertainties.   In this way, 
one may expect a better determination of $e^+e^- \ra \pi^+ \pi^-$ using all other
annihilation processes. Conversely, the huge
experimental effort invested in order to get accurate $e^+e^- \ra \pi^+ \pi^-$ data
may help in better estimating contributions to $a_\mu$ from other processes,
$e^+e^- \ra \pi^0 \pi^+ \pi^-$  for instance. However, the systematics present in all these
processes may limit the expected improvements.  
 
The purpose of the present study  is  to construct such a model and test its ability to account
for the above mentioned processes beside $e^+e^- \ra \pi^+ \pi^-$. This study will be split up 
into two parts for ease of reading. In the following, we will built the model and apply it to
several annihilation processes and decay width information. An accompanying   paper is devoted to 
analyzing the dipion
spectra in $\tau$ decay and some non--perturbative contributions to $a_\mu$.

We leave aside from the present study the $e^+e^- \ra K^+ K^-/K^0 \overline{K}^0$
annihilation channels
which are known to raise a problem thoroughly examined in \cite{BGPter} and may call
for a solution beyond the scope of effective Lagrangians \cite{Voloshin}. Indeed,
the ratio of the partial widths of $\phi$ decays to a charged or neutral kaon pair
looks inconsistent with expectations. This question raises an issue which deserves
a specific study at the level of cross sections  which we have only started.

The paper is organized as follows.  Section \ref{ExtendedModel}   reminds the
ingredients  to be used in order to construct  the Lagrangian of our Extended Model~:
these can be found in  \cite{FKTUY,HLSRef}. Without further constraints, this
leads to introducing three more free parameters.
In Section \ref{WZWBrk}, we require these anomalous Lagrangians to recover identically
amplitudes at the chiral limit as given by the standard Wess--Zumino--Witten (WZW) 
Lagrangians \cite{WZ,Witten}~; this leads to constraining some of the parameters  
above and, then, to reduce the additional parameter freedom of the model. In Sections
\ref{Reminder1} and \ref{Reminder2}, we briefly remind the isospin breaking model
of \cite{taupaper}, mostly in order to simplify notations. Section \ref{stat}
is devoted to our treatment of correlated systematic errors which introduce 
uncertainties on the global scale of spectra~; we also emphasize
the way to check the scale information provided with the various data sets.
A method  to account for possible data discrepancies is also given. In Section \ref{pipi}, the 
Extended Model is shown to keep intact the description
of the $e^+e^- \ra \pi^+ \pi^-$ data examined in \cite{taupaper}, together
with the unavoidable set of 17 radiative and leptonic decays. 
 
 In Section \ref{Pgamma}  the real use of the  Extended Model begins
 with fitting the $e^+e^- \ra \pi^0 \gamma$ and $e^+e^- \ra \eta \gamma$ data
 simultaneously with our usual set of $e^+e^- \ra \pi^+ \pi^-$ data
 \cite{taupaper}. While introducing these data sets, one has to remove 
  all $V \pi^0 \gamma$  and $V \eta \gamma$ decay  pieces of information 
  from the set of decay partial widths used together with annihilations,
  in order to avoid redundancies. At this step, the new Lagrangian pieces 
  are shown to correctly account for the data.  In Section   \ref{pipiKLOE},
  one performs a  fit of the $e^+e^- \ra \pi^+ \pi^-$ data  collected
  by the KLOE Collaboration using the Initial State
  Radiation Method (ISR). This is seemingly  the first published fit to
  the KLOE data \cite{KLOE,KLOE_ISR1}. The next step, including
  the $e^+e^- \ra \pi^0 \pi^+ \pi^-$  annihilation data in our fit procedure,
  is the subject of Section 
  \ref{3pionNSK}, where a discussion on  the various data samples is
  done. In Section \ref{numRes} numerical values for $\omg$ and
  $\phi$ meson parameters are given and the exact strength of the 
  $APPP$ and $VPPP$ couplings is emphasized. 
  
  The aim of  Section \ref{BoxAnomalies}
  is to show
  that the Extended Model, with parameters fixed by fitting to the cross 
  section listed above, allows to accurately predict the properties
  (spectrum lineshape, partial width) of the decay channels
  $\eta/\eta^\prime \ra \pi^+ \pi^- \gamma$, both dominated by the box anomalies
  of QCD. 
   In Section \ref{partWidths}, we focus on vector meson partial widths and discuss
the question of getting their absolute values from $e^+ e^-$ data only,
where models are sensitive -- for each vector meson -- to only the product of its leptonic decay 
width and of its partial width to the final state of the $e^+ e^-$ annihilation.
Finally, a few  concluding remarks are collected in the Conclusion 
to the present part. Some reminders or results are given in Appendices, in order
to keep the main text as clear as possible.
  
\section{The Extended HLS Lagrangian Model}
\label{ExtendedModel}

\indent \indent
The most general Lagrangian of the HLS model \cite{HLSOrigin,HLSRef,FKTUY}
involves anomalous and non--anomalous sectors. It can be written~:
\be
{\cal L}= {\cal L}_{HLS}  + {\cal L}_{YM} + {\cal L}_{WZW}+ {\cal L}_{FKTUY}
\label{eq1}
\ee

The piece denoted ${\cal L}_{HLS}$ can be written ${\cal L}_{A} + a {\cal L}_{V}$,
where $a$ is the specific HLS parameter, expected equal to 2  if vector dominance is 
fully satisfied. This Lagrangian piece allows  a direct 
coupling of photons to pseudoscalar mesons to survive with a magnitude proportional to $a-2$. 
This piece, which accounts for the non--anomalous sector, has been given in expanded form in 
\cite{Heath1}. A brief survey of the HLS model can be found in Appendix
A of \cite{taupaper} and is enough for the purpose of the present paper.
A recent and comprehensive review can be found in \cite{HLSRef}, for instance.

The parts of the HLS Lagrangian ${\cal L}_{HLS}$ specific of the
$e^+e^- $ physics (${\cal L}_{VMD}$) and  $\tau$ decays  (${\cal L}_{\tau}$)
have also been given in \cite{taupaper}~; they are reproduced with almost no comment
in the  Appendix (see Section \ref{HLSLagrangian}) in order to avoid too frequent 
references to an external paper. 

${\cal L}_{YM}$  is the Yang--Mills  piece which accounts for the vector 
mesons and exhibits the non--abelian  group structure of the vector fields (see Eq. (D.1)
in  \cite{taupaper}). 

\vspace{0.7cm}
\indent \indent
${\cal L}_{WZW}$ and ${\cal L}_{FKTUY}$ are the anomalous Wess--Zumino--Witten \cite{WZ,Witten}
and FKTUY  \cite{FKTUY,HLSRef}   Lagrangian pieces which account for parity violating decays.
Very briefly, the anomalous Lagrangian can be formally written~:
\be
{\cal L}_{anomalous}={\cal L}_{WZW} + \sum_{i=1}^{4}c_i~{\cal L}_i
\label{eq2}
\ee
where the $c_i$'s are arbitrary constants weighting the additional FKTUY anomalous Lagrangians.
Limiting oneself to the photon and vector meson couplings, ${\cal L}_{anomalous}$ can be cast in the 
form\cite{HLSRef}~:
\be
{\cal L}_{anomalous}={\cal L}_{VVP}+{\cal L}_{AVP}+{\cal L}_{AAP}+{\cal L}_{VPPP}+{\cal L}_{APPP}
\label{eq3}
\ee
where $A$ denotes the electromagnetic field. The Lagrangian pieces occuring in Eq. (\ref{eq3})
are\footnote{For clarity, the new constant parameters are denoted exactly as they are
defined in  \cite{HLSRef}.} \cite{HLSRef}~:
\be
\left \{
\begin{array}{lll}
{\cal L}_{VVP}=& \displaystyle - \frac{N_c g^2}{4 \pi^2 f_\pi} ~c_3
 \epsilon^{\mu \nu \alpha \beta}{\rm Tr}[ \partial_\mu V_\nu \partial_\alpha V_\beta P] \\[0.5cm]
 {\cal L}_{AVP}=& \displaystyle - \frac{N_c ge }{8 \pi^2 f_\pi} ~(c_3- c_4)
 \epsilon^{\mu \nu \alpha \beta}\partial_\mu A_\nu
 {\rm Tr}[ \{ \partial_\alpha V_\beta , Q \} P] \\[0.5cm]
 {\cal L}_{AAP}=& \displaystyle - \frac{N_c e^2 }{4 \pi^2 f_\pi} ~(1- c_4)
 \epsilon^{\mu \nu \alpha \beta}\partial_\mu A_\nu \partial_\alpha A_\beta{\rm Tr}[Q^2 P]\\[0.5cm]
 {\cal L}_{VPPP}=& \displaystyle - i \frac{N_c g }{4 \pi^2 f_\pi^3} (c_1-c_2-c_3)
   \epsilon^{\mu \nu \alpha \beta}{\rm Tr}[V_\mu \partial_\nu P \partial_\alpha P \partial_\beta P]\\[0.5cm]
  {\cal L}_{APPP}=& \displaystyle - i \frac{N_c e}{3 \pi^2 f_\pi^3} [1- \frac{3}{4}(c_1-c_2+c_4)]  
  \epsilon^{\mu \nu \alpha \beta} A_\mu{\rm Tr}[ Q \partial_\nu P \partial_\alpha P \partial_\beta P]
\end{array}
\right .
\label{eq4}
\ee
where we have extracted the universal vector meson coupling constant  $g$ 
from the definition of the vector field matrix (see \cite{taupaper}
for our definitions), $Q$ denotes the quark charge matrix and $\{ \partial_\alpha V_\beta , Q \} $  is
the anticommutator of its arguments. $N_c$ is the number of quark colors,  3 in our case.
The original WZW terms here can be easily identified in Eqs (\ref{eq4}) and one
also observes that these Lagrangians depend on only 3 (FKTUY) parameter combinations~: $c_1-c_2$, $c_3$ and $c_4$. 
We may refer to the first three Lagrangians in Eqs. (\ref{eq4}) as Triangle Anomaly and the last two 
as Box Anomaly Lagrangians.
\vspace{0.7cm}

\indent \indent
If full Vector Meson Dominance were fulfilled, one would have~:
\be
c_3=c_4=1~~~~~{\rm and}~~~~~ c_1-c_2=\frac{1}{3}~~~~~.
\label{eq5}
\ee
Indeed, one can easily check that these conditions turn out to cancel the 3 Lagrangian
pieces in Eqs. (\ref{eq4}) involving the photon field ($A$). In this case, photons
connect  to hadrons only through the $V-A$ transitions allowed by the non--anomalous
HLS Lagrangian (see Appendix \ref{HLSLagrangian}).

However, relying on a limited subset of data, it has been argued in \cite{FKTUY,HLSRef} that 
some experimental data may instead favor~:
\be
c_3=c_4=c_1-c_2=1
\label{eq6}
\ee

Previous studies especially devoted to  $VP\gamma$ and $P\gamma\gamma$ partial widths \cite{rad} 
performed assuming Eqs. (\ref{eq6}) were succesfull enough to  conclude that these conditions were
well accepted by the data. Our work
\cite{box} on a simultaneous account of $\eta/\eta^\prime \ra \gamma \gamma$ and
$\eta/\eta^\prime \ra \pi^+ \pi^- \gamma$ led to the same conclusion. This gave a strong support
to the relevance of the box anomaly phenomenon in pseudoscalar meson decays.
 
 Therefore, the present status is that full VMD  is well accepted by all
 existing data only in the Triangle Anomaly sector, while the
 $\eta/\eta^\prime \ra \pi^+ \pi^- \gamma$ decay modes prevent extending
 full VMD to the Box Anomaly sector.

However, the statement $ c_3=c_4=c_1-c_2=1$  cannot be considered as firmly established
without being motivated by a detailed study of all relevant data. If any,
such a conclusion  should only follow from a study involving the 
annihilation processes $e^+ e^- \ra \pi^0 \gamma,~\eta \gamma,~\eta^\prime \gamma,~\pi^+\pi^-\pi^0$ 
in their full 
intricacy, beside the 
properties of the decay modes $\eta/\eta^\prime \ra \gamma \gamma$ and 
$\eta/\eta^\prime \ra \pi^+ \pi^- \gamma$. As  anomalous transitions also involve non--anomalous pieces,
the data set to examine should certainly include the $e^+ e^- \ra  \pi^+ \pi^-$
annihilation process in order to define precisely the other model parameters. 

Whether such a program is
realistic and can be practically worked out is the subject of the present paper. It implies to go much beyond
what has been done recently in \cite{taupaper} and formerly in  studies devoted to radiative decays, our \cite{rad,box}  
for instance.

\section{U(3)/SU(3) Breaking And The WZW Conditions}
\label{WZWBrk}
\indent \indent We will not rediscuss here the U(3)/SU(3) breaking conditions in the non--anomalous sector
which have been extensively discussed in our \cite{Heath1,taupaper} following the idea of Bando, Kugo and
Yamawaki \cite{BKY}, referred to as BKY mechanism. A detailed summary of the   U(3)/SU(3) 
(BKY) breaking scheme is given in Appendix C  of \cite{taupaper} where the corresponding 
breaking for the Yang--Mills piece has been also considered (see Appendix D herein). Let us only remind 
that the pseudoscalar bare field matrix $P~(=P_8+ P_0)$ undergoes a renormalization to $P^\prime$~:
\be
\left \{
\begin{array}{lll}
P^\prime \equiv  P_8^\prime+ x  P_0^\prime =X_A^{1/2}(P_8+ P_0)X_A^{1/2}\\[0.5cm]
X_A={\rm Diag}(1,1,z_A)
\end{array}
\right .
\label{eq7}
\ee
where $x \ne 1$  carries the (U(3)) nonet symmetry breaking{\footnote{See \cite{WZWChPT} for the relation
between this parameter and determinant terms \cite{tHooft} explicitly breaking the U$_A$(1) symmetry.}}. 
The other breaking parameter has a well defined expression in terms of the kaon and pion decay constants~: 
$z_A=[f_K/f_\pi]^2$.
Correspondingly, the renormalized vector field matrix $V^\prime$ is given in terms of the bare one $V$
\cite{taupaper} by~:
\be
\left \{
\begin{array}{lll}
V^\prime_\mu =X_T^{1/2}V_\mu X_T^{1/2}\\[0.5cm]
X_T={\rm Diag}(1,1,z_T)
\end{array}
\right .
\label{eq8}
\ee

 Considering a HLS Lagrangian as general as allowed from its derivation \cite{FKTUY,HLSRef} leads to
 introduce {\it a priori} three more basic parameters ($c_1-c_2$, $c_3$ and $c_4$) beside the
 standard $g$, $a$ (and $e$)  occuring already in the non--anomalous Lagrangian. 
 
 However, one may impose a natural constraint to this Extended HLS Model by requesting that it
provides amplitudes identical to those derived from the WZW Lagrangian alone 
  at the chiral point. 
 These amplitudes may only be modified by U(3)/SU(3) breaking effects affecting 
 the pseudoscalar sector \cite{rad,taupaper}.
   Instead, additional SU(3) breaking effects which may affect the trace terms in Eqs. 
 (\ref{eq4}), as well as the dependence upon the $c_i$'s, should drop out at the chiral
 point.
 
 We perform the SU(3) breaking of each Lagrangian in Eqs. (\ref{eq4}) using a prescription 
 inspired from Bramon, Grau and Pancheri \cite{BGP,BGPbis}. This motivates the following
 breaking scheme~:
 
 \be
\left \{
\begin{array}{lll}
{\cal L}_{VVP}~:& \displaystyle {\rm Tr}[ \partial_\mu V_\nu \partial_\alpha V_\beta P] 
\Longrightarrow {\rm Tr}[ \partial_\mu V_\nu X_W \partial_\alpha V_\beta P] \\[0.5cm]
 {\cal L}_{AVP}~:& \displaystyle \partial_\mu A_\nu {\rm Tr}[ \{ \partial_\alpha V_\beta , Q \} P] 
 \Longrightarrow \partial_\mu A_\nu {\rm Tr}[ \{ \partial_\alpha V_\beta , Q X_U\} P]   \\[0.5cm]
 {\cal L}_{AAP}~:& \displaystyle \partial_\mu A_\nu \partial_\alpha A_\beta{\rm Tr}[Q^2 P]
  \Longrightarrow \partial_\mu A_\nu \partial_\alpha A_\beta{\rm Tr}[X_R Q^2 P]
 \\[0.5cm]
 {\cal L}_{VPPP}~:& \displaystyle {\rm Tr}[V_\mu \partial_\nu P \partial_\alpha P \partial_\beta P]
   \Longrightarrow {\rm Tr}[X_K V_\mu \partial_\nu P \partial_\alpha P \partial_\beta P]\\[0.5cm]
  {\cal L}_{APPP}~:& \displaystyle A_\mu {\rm Tr}[ Q \partial_\nu P \partial_\alpha P \partial_\beta P]
    \Longrightarrow {\rm Tr}[ X_L Q \partial_\nu P \partial_\alpha P \partial_\beta P]
\end{array}
\right .
\label{eq9}
\ee

This might introduce as many as 5 more (breaking) parameters ($z_W,~z_U,~z_R,~z_K,~z_L$) as the  five 
matrices introduced here are supposed to
carry the same form than $X_A$ or $X_T$  reminded above.  We now show that, 
actually, it is not the case.

\subsection{The Triangle Anomaly Sector}
\indent \indent 
 Using the  Lagrangians in Eqs. (\ref{eq4}) broken as shown in Eqs. (\ref{eq9}) just above 
and the $\gamma - V$ transition amplitudes as given by the non--anomalous
 Lagrangian, one can derive the amplitudes~:
 \be
\begin{array}{llll}
T_X=T(X \ra \gamma \gamma)~\epsilon^{\mu \nu \alpha \beta}q^1_\mu q^2_\nu 
\epsilon^1_\alpha \epsilon^2_\beta ~~~,~~ & (X =\pi^0, ~\eta,~\eta^\prime)
\end{array}
\label{eq10}
\ee
One gets first~:
\be
T(\pi^0 \ra \gamma \gamma)= \displaystyle
-i \frac{\alpha}{\pi f_\pi} \left[1+ 2 (c_3 - c_4)
\right]  
\label{eq11}
\ee
which implies $c_3 = c_4$ in order to recover the usual WZW term.
Therefore, the ${\cal L}_{AVP}$ Lagrangian vanishes. Using this condition, 
one similarly  derives~:
\be
\left \{
\begin{array}{lll}
\displaystyle 
T(\eta_0 \ra \gamma \gamma)=
-x\frac{i\alpha}{\pi f_\pi}\sqrt{\frac{2}{3}} 
\left[
\frac{ 
(1-c_3)(z_W z_T^2 - z_R)
}{3 z_A}
+\frac{ 
z_W z_T^2+5 z_A}{3 z_A}
\right]\\[0.5cm]
T(\eta_8 \ra \gamma \gamma)=\displaystyle 
-\frac{i\alpha}{\pi f_\pi}\sqrt{\frac{1}{3}} 
\left[
\frac{ 
-2 (1-c_3)( z_W z_T^2 - z_R)
}{3 z_A}
+\frac{ 
5 z_A -2 z_W z_T^2}{3 z_A}
\right]
\end{array}
\right .
\label{eq12}
\ee
for the singlet and octet parts of the $\eta$ and $\eta^\prime$ mesons.
In order to recover the usual WZW expressions (see for instance \cite{rad,box} or
more recently \cite{taupaper}), one clearly needs to request $z_W z_T^2=1$,
which was first phenomenologically found as a numerical constraint arising
from fits \cite{rad}. If, additionally, one requests $z_R=1$ (which may look
quite natural), then the standard WZW amplitudes for $\eta_{0/8} \ra \gamma \gamma$
are recovered without requiring any further constraint on $c_3$.  We are thus led to choose
as constraints on the parameters~:
\be
z_W z_T^2= z_R=1 ~~~,~~~  c_3=c_4  
\label{eq13}
\ee
 
 \subsection{The Box Anomaly Sector}
\indent \indent  As  measurements related with box anomalies only involve  couplings of the form
$X \pi^+ \pi^- \gamma$ with $X=\pi^0,~\eta,~\eta^\prime$, we only focus on this sector.
Let us list the Lagrangian pieces relevant for this purpose.

\begin{itemize}
\item  A part only of the VVP Lagrangian plays a role in these couplings~:
\be
\hspace{-2.5cm}
\left\{
\begin{array}{lll}
 {\cal L}_{VVP}= &  \displaystyle C \epsilon^{\mu \nu \alpha \beta} 
 \left\{
 \left [\partial_\mu \rho^I_\nu \partial_\alpha \rho^I_\beta +
 \partial_\mu \omg^I_\nu \partial_\alpha \omg^I_\beta \right ]
\left [\frac{\eta_8}{2\sqrt{3}} +\frac{x \eta_0}{\sqrt{6}} \right ]
+ \frac{z_W z_T^2}{z_A}
\partial_\mu \phi^I_\nu \partial_\alpha \phi^I_\beta 
\left [-\frac{\eta_8}{\sqrt{3}} +\frac{x \eta_0}{\sqrt{6}} \right ]
\right \}\\[0.5cm]
  C=&\displaystyle -\frac{N_c g^2 c_3}{8 \pi^2 f_\pi}
\end{array}
\right .
\label{eq14}
\ee
in terms of ideal fields.

Actually, the $\gamma VP$ couplings relevant 
for our purpose  can be derived from the following
effective piece constructed from Eq. (\ref{eq14}) above on the one hand, and from the 
$\gamma - V$ transitions of the non--anomalous Lagrangian \cite{taupaper} on the other hand
(see Appendix \ref{HLSLagrangian} below)~:
\be
\hspace{-2.5cm}
\left\{
\begin{array}{lll}
{\cal L}_{AVP}^\prime= \displaystyle C^\prime \epsilon^{\mu \nu \alpha \beta}
F_{\mu \nu} \partial_\alpha A_\beta~~~~~~~~,~~~
 C^\prime= \displaystyle -\frac{N_c g~e c_3}{12 \pi^2 f_\pi}
 \\[0.5cm]
\displaystyle  F_{\mu \nu}=
\left[  \frac{1}{2} \pi^0 + \frac{\sqrt{3}}{2} \eta_8 + x \sqrt{\frac{3}{2}} \eta_0
 \right]~\partial_\mu \rho^I_\nu +
\left[  \frac{3}{2} \pi^0 + \frac{1}{2\sqrt{3}} \eta_8 +  \frac{x}{\sqrt{6}} \eta_0
 \right]~\partial_\mu \omg^I_\nu +
 \left[\frac{1}{z_A} \sqrt{\frac{2}{3}}  \eta_8 - \frac{x}{z_A\sqrt{3}}\eta_0
  \right]~\partial_\mu \phi^I_\nu 
 \\[0.5cm]

\end{array}
\right .
\label{eq15}
\ee
For the purpose of constructing the $X \pi^+ \pi^- \gamma$ coupling at the chiral point,
this piece fully replaces Eq. (\ref{eq14}). It will be used together with the transition
$\rho^I \ra \pi^+ \pi^- $ given by the non--anomalous Lagrangian\footnote{Anticipating
on what is reminded in the next Section and relying on \cite{taupaper}, this is justified
by the fact that ideal vector fields and physical vector fields coincide at $s=0$. Then,
at $s=0$, the couplings of the physical $\omg$ and $\phi$ mesons identically vanish, $s$ being
the square of the momentum carried by the vector meson.}.

\item  The VPPP piece of relevance is~:
\be
\hspace{-2.5cm}
\left\{
\begin{array}{lll}
{\cal L}_{VPPP}= &  \displaystyle - iD \epsilon^{\mu \nu \alpha \beta} 
 \left\{
 \rho^I_\mu \left[
 \frac{\sqrt{3}}{12} \partial_\nu \eta_8 +\frac{x\sqrt{6}}{12} \partial_\nu \eta_0
 \right] + \frac{3}{4}  \omg^I_\mu  \partial_\nu \pi^0 \right \}
\partial_\alpha \pi^-\partial_\beta \pi^+ \\[0.5cm]
 D=&\displaystyle -\frac{N_c g}{4 \pi^2 f_\pi^3} (c_1-c_2-c_3)
\end{array}
\right .
\label{eq16}
\ee

\item  The relevant APPP piece writes~:
\be
\hspace{-2.5cm}
\left\{
\begin{array}{lll}
{\cal L}_{APPP}= &  \displaystyle -iE \epsilon^{\mu \nu \alpha \beta} A_\mu
\left[ \frac{1}{4} \partial_\nu \pi^0 + \frac{\sqrt{3}}{12} \partial_\nu \eta_8
+ \frac{x\sqrt{6}}{12} \partial_\nu \eta_0\right] 
\partial_\alpha \pi^-\partial_\beta \pi^+ \\[0.5cm]
 E=&\displaystyle -\frac{N_c e}{3 \pi^2 f_\pi^3} 
 \left[ 1 -\frac{3}{4}(c_1-c_2+c_4) \right]
\end{array}
\right .
\label{eq17}
\ee
\end{itemize}
\hspace{1.cm}

\indent \indent
With these pieces at hand, one can compute the amplitudes at  the chiral point.
One first gets~:
\be
  T (\pi^0 \ra \pi^+ \pi^- \gamma)= \displaystyle \frac{ieN_c}{12 \pi^2 f_\pi^3} \left [
  1+ \frac{3}{4} (c_3 - c_4)
 \right]
\label{eq18}
\ee
which coincides with the amplitude expected from the WZW Lagrangian alone if $c_3=c_4$.
Assuming this condition, one can easily derive at the chiral point~: 
\be
\left\{
\begin{array}{lll}
T(\eta_8 \ra \pi^+ \pi^- \gamma) = 
\displaystyle \frac{i e N_c}{12 \pi^2 f_\pi^3} \frac{1}{\sqrt{3}}
~~~~~,~~& T(\eta_0 \ra \pi^+ \pi^- \gamma) = 
\displaystyle \frac{i e N_c}{12 \pi^2 f_\pi^3} \sqrt{\frac{2}{3}} ~x
\end{array}
\right .
\label{eq19}
\ee
which proves that the WZW usual (U(3)/SU(3) broken)  amplitudes \cite{box} at the chiral point
are recovered.
One may notice that the breaking matrices $X_K$ and $X_L$ play no role in the $X \pi^+ \pi^- \gamma$ 
sector.

Therefore, assuming the conditions summarized in Eq. (\ref{eq13})  leaves us with 2 more 
physics parameters unconstrained, compared with the previous version 
of our model~: $c_3$ and $c_1-c_2$. Using the information previously defined, 
the Extended Lagrangian we propose 
is~: 
\be
\left\{
\begin{array}{lll}
{\cal L}= {\cal L}_A(z_A)+ a{\cal L}_V(z_V)+ {\cal L}_{YM}(z_T) + \\[0.5cm]
~~~~+{\cal L}_{VVP}(z_W)+ {\cal L}_{AAP}(z_R=1)+ 
{\cal L}_{VPPP}(z_K=1)+ {\cal L}_{APPP}(z_L=1)\\[0.5cm]
c_4=c_3\\[0.5cm]
z_Wz_T^2=1
\end{array}
\right .
\label{eq20} 
\ee
where each SU(3) breaking parameter is exhibited with its value (when fixed). Out of these, the breaking
parameters to be determined from data are $z_A,~z_V,~z_T,x$, keeping in mind that $z_A=[f_K/f_\pi]^2$ should always be
satisfied\footnote{It could even be directly fixed to its experimental value $[f_K/f_\pi]^2=1.495 \pm0.031$
without degrading the fits.}.

 Such a price for the extension of the HLS model we propose looks acceptable,
taking into account that we plan to describe all available anomalous processes
beside $e^+e^- \ra \pi^+ \pi^-$ and $\tau^\pm \ra \pi^\pm \pi^0 \nu_\tau$ already accounted for
within this model. This covers the 
 $e^+e^- \ra  \pi^0 \gamma, ~\eta \gamma$,~$\pi^+ \pi^- \pi^0$ 
cross sections  and the box anomalous processes $\eta/\eta^\prime \ra \pi^+ \pi^- \gamma$.
A priori   $e^+e^- \ra  \eta^\prime \gamma $ data will fall into this extended scope when they will become
available\footnote{A process  like $e^+e^- \ra  \omg \pi^0$ is of this kind, however, one has
first to carefully study the effects of scalars before any use of the corresponding data.}.

\section{Isospin Breaking And The ($\rho~,\omg,~\phi$) Mixing}
\label{Reminder1}
\indent \indent
We adopt here the ($\rho~,\omg,~\phi$) mixing developed in \cite{taupaper}. We only remind here a few
properties for completeness and to simplify the notations. Let us define ~:
\be
\left \{
\begin{array}{lll}
\epsilon_1(s)=g_{\rho K K}^2 (\Pi_+(s)-\Pi_0(s))\\[0.5cm]
\epsilon_2(s)=g_{\rho K K}^2 (\Pi_+(s)+\Pi_0(s))\\[0.5cm]
\Pi_{\pi \pi}(s) = g_{\rho \pi \pi}^2 \Pi^\prime(s)
\end{array}
\label{eq21}
\right .
\ee
where $\Pi^\prime(s)$ denotes the $\pi^+ \pi^-$ amputated\footnote{{\it i.e.} with unit 
coupling constants.} loop function, $\Pi_+(s)$ and $\Pi_0(s)$  the charged
and neutral amputated kaon loops. The coupling constants occuring in Eqs. (\ref{eq21}) fulfill 
$g_{\rho \pi \pi}=2 z_A g_{\rho K K}=a g/2$, in terms of the basic parameters of the model
and of the SU(3) symmetry breaking  parameter $z_A \equiv [f_K/f_\pi]^2$.
These loop functions are analytic functions each real on the real
$s$--axis below the corresponding threshold. At the limit of equal charged and neutral kaon masses,
$\epsilon_1(s)$ vanishes~; on the other hand, both $\epsilon_1(s)$ and $\epsilon_2(s)$ have small 
magnitudes \cite{taupaper}  in the whole $s$--region we are interested 
in (from the 2--pion threshold to the $\phi$ mass). Instead, $\Pi_{\pi \pi}(s)$ is basically
the $\rho$ self mass and is thus known to be significant in this $s$--region. The functions in Eqs.
(\ref{eq21}) are given by dispersion relations and contain each a polynomial in $s$ 
chosen \cite{taupaper} of second degree and vanishing at the origin.
These functions mainly serve to define three complex quantities\footnote{
The functions $\alpha(s)$, $\beta(s)$, $\gamma(s)$  defined here may be used all along this
paper without exhibiting their functional ($s$) dependence. This notation makes easier
reading the formulae given in Section 8 of \cite{taupaper} which  are used in the present study.
In order to avoid confusion, the fine structure constant is always denoted $\alpha_{em}$.}~:
\be
\left (
\begin{array}{lll}
\alpha(s)\\[0.5cm]
\beta(s)\\[0.5cm]
\gamma(s)
\end{array}
\right ) = 
\left (
\begin{array}{cll}
\displaystyle \frac{\epsilon_1(s)}{\Pi_{\pi \pi}^\rho(s)-\epsilon_2(s)} \\[0.5cm]
\displaystyle   \frac{\mu \epsilon_1(s) }{(1-z_V) m^2 + \Pi_{\pi \pi}(s) -\mu^2 \epsilon_2(s)}\\[0.5cm]
\displaystyle  \frac{\mu \epsilon_2(s) }{(1-z_V) m^2 + (1-\mu^2) \epsilon_2(s)}\\[0.5cm]
\end{array}
\right ) 
\label{eq22}
\ee
where $z_V$ is a parameter involved in the SU(3) symmetry breaking of the ${\cal L}_V$
part of the non--anomalous HLS Lagrangian \cite{Heath1,rad,taupaper}. We have defined
$\mu=z_V \sqrt{2}$, and $m^2=a g^2f_\pi^2$  is the Higgs--Kibble  $\rho$ meson mass squared 
generated by the spontaneous
breakdown of the HLS model. The complex 
quantities $\alpha(s)$, $\beta(s)$, $\gamma(s)$ are, in some sense, "angles" which describe
the ($\rho~,\omg,~\phi$) mixing. At lowest (first) order in $\epsilon_1(s)$ and $\epsilon_2(s)$, the
mixing scheme is given by~:
\be
\left (
\begin{array}{lll}
\rho^I\\[0.5cm]
\omg^I\\[0.5cm]
\phi^I
\end{array}
\right ) = 
\left (
\begin{array}{cll}
\displaystyle  \rho^0 -\alpha \omg+\beta \phi\\[0.5cm]
\displaystyle  \omg +\alpha \rho^0+\gamma \phi\\[0.5cm]
\displaystyle \phi -\beta  \rho^0 -\gamma \omg
\end{array}
\right ) 
\label{eq23}
\ee

These relations exhibit the connexion between the "ideal" vector fields (carrying definite isospin
properties), which are entries of the vector field matrix, and the "physical" fields 
(each a mixture of isospin 0 and 1 fields) which enter the physics processes. One may consider that
$\alpha(s)$ is the $\rho-\omg$ (complex) mixing "angle" and that $\beta(s)$ and $\gamma(s)$ are, resp. the 
$\rho-\phi$ and $\omg-\phi$  mixing "angles". In our model, 
this mixing is $s$--dependent~; for instance, the $\omg-\phi$ mixing angle $\gamma(s)$ has not the same 
value at the $\omg$ mass and at the $\phi$ mass. Moreover, among these angles, $\gamma(s)$ is the single one 
to be practically real  up to the $\phi$ mass region. Finally, as a consequence of  
 $\alpha(0)=\beta(0)=\gamma(0)=0$, physical ($s$--dependent) and ideal fields coincide at $s=0$.

\section{The  $\rho$ Propagators And The $\gamma -V$ Transition Amplitudes}
\label{Reminder2}
\indent \indent The inverse propagators  for the charged and neutral $\rho$ can be 
written~:
\be
\left \{
\begin{array}{lll}
D_ {\rho^0}(s)=s-m^2 -\Pi_{\rho \rho}(s)\\[0.5cm]
D_{\rho^\pm}(s)=s-m^2 -\delta m^2-\Pi_{\rho \rho}^\prime(s)
\end{array}
\right .
\label{eq24}
\ee
allowing for a possible $\rho^\pm-\rho^0$ mass difference. Neglecting 
the effects of mass differences between charged and neutral pseudoscalar 
mesons  in the propagators, the
same self--mass occurs, approximated by \cite{taupaper} ~:
\be
\Pi_{\rho \rho}(s)=\Pi_{\rho \rho}^\prime(s)=\Pi_{\pi \pi}(s) + \epsilon_2(s) \\[0.5cm]
\label{eq25}
\ee

The notations defined in the previous Section allow one to express the 
$\gamma -V$ transition amplitudes in a  more readable way than in \cite{taupaper} (see Section 8
there).  The $\gamma -V$ transition amplitudes  from photon to {\it physical} vector fields 
can be written as $e F_{V\gamma}(s) $ with~:
\be
\displaystyle F_{V\gamma}(s)= f_{V\gamma} - \Pi_{V\gamma}(s)~~~~~~~~~~~,
\label{eq26}
\ee
the constant term can be read off the $(V \cdot A)$ terms in the non-anomalous HLS Lagrangian
and the loop correction depends on the pion and kaon loops. The constant terms become~:
\be
\left \{
\begin{array}{lll}
f_{\rho \gamma} =&  \displaystyle  a g f_\pi^2 ~\left[1 + \frac{1}{3}\alpha + \frac{\mu}{3} \beta \right] \\[0.5cm]
f_{\omg \gamma} =&  \displaystyle  a g f_\pi^2 ~\left[ \frac{1}{3}  -\alpha + \frac{\mu}{3} \gamma \right] \\[0.5cm]
f_{\phi \gamma} =&  \displaystyle  a g f_\pi^2 ~\left[ -\frac{\mu}{3} + \beta + \frac{1}{3} \gamma \right] 
\end{array}
\right .
\label{eq27}
\ee
and the $s$--dependent loop terms are \cite{taupaper}~:
\be
\left \{
\begin{array}{lll}
\Pi_{\rho \gamma} =&  \displaystyle (1-\frac{a}{2}) \frac{\Pi_{\pi\pi}^\gamma(s)}{g_{\rho \pi \pi}}
+ (z_A-\frac{a}{2} -b) \frac{\epsilon_1(s)+\epsilon_2(s)}{g_{\rho \pi \pi}} 
+b \frac{\epsilon_2(s)-\epsilon_1(s)}{g_{\rho \pi \pi}}\\[0.5cm]
\Pi_{\omg \gamma} =&  \displaystyle -(1-\frac{a}{2}) \alpha(s)  \frac{\Pi_{\pi\pi}^\gamma(s)}{g_{\rho \pi \pi}}
+ (z_A-\frac{a}{2} -b) \frac{\epsilon_1(s)+\epsilon_2(s)}{g_{\rho \pi \pi}} 
-b \frac{\epsilon_2(s)-\epsilon_1(s)}{g_{\rho \pi \pi}}\\[0.5cm]
\Pi_{\phi \gamma} =&  \displaystyle (1-\frac{a}{2}) \beta(s)  \frac{\Pi_{\pi\pi}^\gamma(s)}{g_{\rho \pi \pi}}
- (z_A-\frac{a}{2} -b) \mu \frac{\epsilon_1(s)+\epsilon_2(s)}{g_{\rho \pi \pi}} 
+b \mu \frac{\epsilon_2(s)-\epsilon_1(s)}{g_{\rho \pi \pi}}
\end{array}
\right .
\label{eq28}  
\ee
with $\mu=z_V \sqrt{2}$ and $b=a(z_V-1)/6$. $\Pi_{\pi \pi}(s)$ (see Eq. (\ref{eq25})) and 
$\Pi_{\pi\pi}^\gamma(s)$ may
carry different subtraction polynomials \cite{taupaper}. In the fit procedure described
below, as in \cite{taupaper}, their (second degree) subtraction polynomials are chosen 
independently and fit from data. Numerically,  one finds no 
significant correlation among these 2 polynomials.

\section{Combining Statistical and Systematic Uncertainties}
\label{stat}
\indent \indent For any of the data sets we use, there are reported statistical
and systematic errors. One way to proceed   is to add them
in quadrature and define correspondingly a $\chi^2$ to be minimized. If the
errors are large enough, there is no real need to go beyond this simple
treatment.

However, in samples where statistics is large, systematics should be handled a little
bit more carefully. Systematic errors can be split up into two different kinds~:
uncorrelated and correlated uncertainties. It 
is quite traditional to combine uncorrelated systematic errors and statistical
errors in quadrature and we follow this rule as, moreover, several data sets are 
provided with
this combination already performed. When statistical errors are mentioned, 
this combination should be understood, unless explicitly stated.

It remains to handle the correlated systematic errors. As a first statement,
one may interpret these as reflecting a global scale uncertainty which affects the data
set considered. This was already  done in our previous analysis \cite{taupaper}. 
Let us rephrase it with slightly more details, in order to
explain clearly the method which underlies the present study.

\subsection{Scale Uncertainties, A Reminder}
\indent \indent
The way statistical errors and  scale uncertainties combine can be treated
rigorously\footnote{We gratefully ackowledge P. Astier, 
LPNHE Paris 6/7, for several dicussions on this subject.}. 
Let us assume one has a data set $ m:\{ m_i,~i=1,\cdots n \}$ and, correspondingly, 
a model ${\rm M}~:\{ M_i,~i=1,\cdots n \} $. 
Let us assume
given its (symmetric) statistical error covariance matrix $V$ which needs not be 
diagonal. Let us finally assume that a normalization 
scale uncertainty $\lambda$ affects the data~; $\lambda$ 
is supposed to have  as most probable value
0 and  standard deviation $\sigma$.

In this case, the conventional $\chi^2$ is~:
\be
\chi^2= \displaystyle 
\left [ m- M -A \lambda\right ]^T V^{-1} \left [ m-  M -A \lambda\right ]
+\frac{\lambda^2}{\sigma^2}
\label{seq1}
\ee
where $A$ is traditionally the vector  of the model values $M$
 and the other notations
are obvious.  For definiteness, in our study, we preferred using $A=-m$ in order
to avoid having a covariance matrix depending on fit parameters\footnote{
We checked with several of the fit configurations described below  the  
difference between the two possible choices $A=M$ and $A=-m$. We did
not observe differences beyond the 0.3 $\sigma$ level
for the fit parameter values.
}.

In this approach, $\lambda$ is nothing but an additional constrained fit parameter.
One can solve this equation at minimum $\chi^2$ for $\lambda$ using 
$d\chi^2/d\lambda=0$ and substitute the expression for $\lambda$ in Eq. (\ref{seq1}).
This leads to~:
\be
\chi^2= \displaystyle 
\left [ m- M \right ]^T W^{-1}(\sigma^2) \left [ m- M \right ]
\label{seq2}
\ee
where~:
\be
 \displaystyle 
W^{-1}(\sigma^2) = \left [ V + \sigma^2 A A^T \right ]^{-1}=
V^{-1} -\frac{\sigma^2} {1+ \sigma^2 (A^T V^{-1} A)} (V^{-1}A) (V^{-1}A)^T
\label{seq3}
\ee
which corresponds  to  Eqs (46) and (47) in \cite{taupaper}.
If the expected value of $\lambda$ were
some $\lambda_0 \ne 0$, one has to replace in the second term of Eq.(\ref{seq1}) $\lambda^2$ 
 by $(\lambda-\lambda_0)^2 $ and $ m-  M$ would become $m-  M-A \lambda_0$ in Eq.(\ref{seq2}).
For practical use, if a scale uncertainty has been identified,  data are generally corrected
for this and then  $\lambda_0=0$ is justified. 

Eq. (\ref{seq2}) illustrates that a correlated scale error is algebraically related with the 
model~:
\be
\displaystyle \lambda =
\frac{A^T  V^{-1} \left[m-M \right]} 
{A^T V^{-1} A +\displaystyle \frac{1}{\sigma^2}}
\label{seq4}
\ee

If the model depends linearly on parameters to be determined,
the substitution has certainly to be performed in order to avoid the error
covariance matrix of the (fit) parameters having a zero eigenvalue. If the dependence 
is non--linear,
avoiding solving for $\lambda$ only increases errors and produces (spurious)
correlations.

Dealing with  one (or several) data sample(s), the value
of $\lambda$ following from minimizing Eq. (\ref{seq1})
can be confronted with the expected correlated systematic 
uncertainty. In practice, if the mean correlated systematic
error has its correct value, one expects  $|\lambda^{fit}|/\sigma$
small enough. In this case, having checked that $\lambda=0$ is
reasonable as expected, one can impose from start $\lambda\equiv 0$, {\it i.e.}
minimize Eq. (\ref{seq2})~; one should then check that the fit probability and 
the parameter (hidden inside $M$) central values are nearly unchanged while
the magnitude of their errors decreases.
 
In \cite{taupaper}, we performed slightly differently~: the scale uncertainty was 
considered as a random distribution $\delta \lambda$ of zero mean and of standard
deviation $\sigma$ which affects the measurements. One can check that this approach
leads to the same conclusion which is summarized by Eq. (\ref{seq2}). The present approach
only clarifies that the  final fit  should be performed with $\lambda=0$.

\subsection{Checking And Dealing With A Missing Variance}
\indent \indent
However, as well known, identifying and estimating systematic uncertainties 
can be a delicate matter. Some source of systematics could have been missed
or underestimated. When dealing with only one data set, this could well be quite 
transparent, 
as its effects could be absorbed by the other parameter values.
However, while merging different data sets with differents systematics,
the pattern can be quite different and could result in poor fit 
probabilities. Therefore, it is useful to be in position of identifying
a possible missing variance affecting some scale.

Let us assume the correct systematic variance of $\lambda$ be $\sigma^2+\eta^2$
instead of $\sigma^2$. In this case the correct formula is  
Eq. (\ref{seq1}) with $\sigma^2 \rightarrow \sigma^2+\eta^2$. The question
is now how to detect that a piece represented by $\eta^2$ could have  been missed.

One can first check that~:
 \be
\chi^2= \displaystyle 
\left [ m- M -A (\lambda_1+\lambda_2) 
\right ]^T V^{-1} \left [ m-  M -A (\lambda_1+\lambda_2) \right ]
+\frac{\lambda_1^2}{\sigma^2}+\frac{\lambda_2^2}{\eta^2}
\label{seq5}
\ee
allows to recover the right result, provided one treats $\lambda_1$ and $\lambda_2$
as {\it independent} variables of zero mean and of respective variance  
$\sigma^2$ and $\eta^2$.
One can, moreover, check that~:
\be
\chi^2= \displaystyle 
\left [ m- M -A \lambda_2 
\right ]^T W^{-1}(\sigma^2) \left [ m-  M -A \lambda_2 \right ]
+\frac{\lambda_2^2}{\eta^2}
\label{seq6}
\ee
by solving, just as before, for only the identified systematics represented by the couple
($\lambda_1$, $\sigma^2$).
This equation gives us a handle to account for a missing
(part) of the variance which could be revealed by using  
a large ensemble of data sets, each with its own systematics.

A way to check for a  possible missing variance, is to compare
a fitted $\lambda_2$ to the "identified" variance $\sigma^2$. 
A tentative assumption for $\eta^2$ could be to state $\eta^2 \simeq \sigma^2$.
Then, within a numerical fit procedure,
one can check the magnitude of $\lambda_2^{fit}/\sigma$.
If $|\lambda_2^{fit}|/\sigma $ is small enough ($\leq 1 \div 2$)
a missing variance is certainly negligible compared with the identified
one. Otherwise, the fit value of the scale indicates how much the data should be rescaled
in order to match  {\it all other data sets} and the model under test.

Of course, one thus makes an implicit statement~: if the data sets which exhibit a
significant missing variance represent a small minority, one may consider 
that this validates both
a correction for  missing variance  and the model. If, instead, the data sets 
exhibiting a missing variance represent a majority, the model is certainly
invalidated. Intermediate situations, if any, would be uncomfortable~;
this might indicate some unaccounted for physics.
Anyway, the set of data samples we consider does not face us with the latter
 configuration.

Finally, the $\chi^2$ we shall deal with is a sum of  partial $\chi^2$
of the kind shown by Eq. (\ref{seq6}). A first run of the fit procedure helps to
identify which of the $\lambda_2^\alpha$ ($\alpha$ being the data set index)
can be safely dropped out. 
 
Concerning those which have still to be considered,
one can keep the fit parameter as introduced. One should however keep in mind
that this certainly enlarges the variance.  

As a final remark, one should mention a rigorous way to lessen the variance. 
If some $\lambda_2^\alpha $  is considered significant, one should account for it,
as the  corresponding data  set has certainly
not been corrected for this source of uncertainties. 
However, a numerical minimization procedure like
{\sc minuit} provides an estimate of $\lambda_2$ -- we name it $\lambda_2^{fit}$ --
and an estimate of its uncertainty -- we name it $\sigma^2_{miss.}$ --  which can
be accurately known running a code like {\sc minos}. Then, the corrected (partial)
$\chi^2$  to be minimized can be rewritten~:
\be
\chi^2= \displaystyle 
\
\left [ m- M -\lambda_2^{fit} A \right ]^T W^{-1}(\sigma^2 + \sigma^2_{miss}) 
\left [ m- M -\lambda_2^{fit} A \right ]
\label{seq7}
\ee

$\lambda_2^{fit}$ can be estimated from
Eq. (\ref{seq4}) by changing $\sigma^2$ to $\eta^2=\sigma^2_{miss} $.
As one removes one free parameter, one almost certainly lessens the variance.
This last expression depends on the other parameters under fit. A quite acceptable
solution is instead to use the numerical minimizer output\footnote{In our case, the non--linear 
character of the model avoids having a singular parameter error covariance matrix
while keeping the scale among the parameters to be fit.} for both $\lambda_2^{fit}$ and 
$\sigma^2_{miss.}$, which makes the convergence easier. 

\section{The First Step In Modeling}
\label{pipi}
\indent \indent In order to compare the pion form factor in $e^+e^-$ annihilations 
and in $\tau$ decays, one has to account for isospin breaking effects which differ
in both processes. The purpose of  \cite{taupaper} was to show that the mixing
between the $\rho^0$, $\omg$ and $\phi$ mesons is responsible for most of the reported 
difference. We succeeded in determining the mixing model sketched above (mostly the 
"angles" $\alpha(s), ~\beta(s),~\gamma(s)$) using the information provided by  the 
anomalous decays 
of type $VP\gamma$ (and  $P \gamma \gamma$) and the information carried by
the isospin violating decays\footnote{Actually, the $\omg \ra \pi^+ \pi^-$
mode has not to be included as it is already part of the $e^+ e¯ \ra
\pi^+\pi^-$ spectrum.}
 $\omg/\phi \ra \pi \pi$. 
  In order to constrain
more efficiently the parameter set, the partial widths $V \ra e^+e^-$ ($V=\omg,~\phi$)  were
also included. These pieces of information, as well as the $\omg/\phi$ mass and width,
were fixed at their accepted values \cite{RPP2008}.
 We plan to examine the behavior of the Extended Model presented 
above while introducing more and more information to account for. For this purpose,
we first focus on $e^+e^-$ data, and as a first step on the $e^+e^- \ra \pi^+ \pi^-$
data mostly collected at Novosibirsk and already examined in \cite{taupaper}~:
\begin{itemize}
 \item the former data sets collected by the OLYA and CMD collaborations \cite{Barkov}, 
and by DM1 at ACO \cite{DM1}~; these were (and are still) together referred to as  
"old timelike data",
 \item the  data sets more recently collected by the CMD-2 
 \cite{CMD2-1995,CMD2-1995corr,CMD2-1998-1,CMD2-1998-2} and SND \cite{SND-1998}
 Collaborations, referred to globally as "new timelike data"
 \item all the partial widths for the decay processes of type $VP\gamma$,
  $P \gamma \gamma$,  $\omg/\phi \ra e^+e^-$ and $\phi \ra \pi^+ \pi^-$ 
 at their updated recommended values  \cite{RPP2008}. These represent  17 
 pieces of information.
 \end{itemize} 

We gave up including the decay width $\eta^\prime \ra \rho \gamma$ (actually 
$\eta^\prime \ra \pi^+ \pi^- \gamma$) as, in addition to the dominant triangle
anomaly contribution, there is some (small) contamination by the box anomaly 
discussed already above. The decay modes
 $\eta/ \eta^\prime \ra \pi^+ \pi^- \gamma$  are revisited 
 at the end of this study.
 
 The systematic errors on the $e^+ e^- \ra \pi^+ \pi^-$ cross sections just quoted 
 are treated exactly as explained in Section 11.2 of \cite{taupaper}, or as  reminded 
 above in Section \ref{stat}, {\it i.e.}
 by summing in quadrature the statistical errors and the uncorrelated part
 of the systematic errors on the one hand, and, on the other hand, by accounting for the correlated  
 systematic 
 error through a global scale to be fit, at least for cross--check, as also explained in Section 
 \ref{stat}.  The corresponding standard deviations to be introduced in the $\chi^2$ are
$0.4\%$ for the "new timelike data" and $1.0\%$ for the "old timelike data" \cite{simonPriv}. 
 The fitted scale values are expected negligible \cite{taupaper}.
 
 As our data samples contain all data used in order to get the recommended
 mass and width for the  $\omg$ and $\phi$  mesons \cite{RPP2008}, we start here
 by leaving free the $\omg$  mass and width~; this could have easily been avoided
 at this step of our study by using \cite{taupaper} the world average values.   
 
 In addition, the present work uses  the  code delivered by F. Jegerlehner in order to 
 calculate $\alpha_{em}(s)$  which can be downloaded from \cite{Fred0}~; this code,  
 partly documented in \cite{IFSVP}, has been constructed in order 
to improve the estimates of the muon $g-2$ and of $\alpha_{em}(s)$  \cite{Fred1,Fred2,Fred3}. From
there we get the photon (hadronic+leptonic) vacuum polarization (VP) building block, which
is used  by multiplying our model  form factors, generically denoted $F(s)$,
 by the corresponding factor~:
\be
F(s) \Longrightarrow 
\displaystyle 
\frac{1}{1-\Delta \alpha_{em}(s)} F(s) 
\label{eq29} 
\ee

\begin{table}[!htb]
\hspace{-1.cm}
\begin{tabular}{|| c  | c  | c || c  | c | c ||}
\hline
\hline
\hhhu Data Set  & Without VP &  \multicolumn{4}{|c|}{With Vacuum polarisation (VP)} \\
\hhhu $\sharp$ (data $+$ conditions) & NSK & NSK  &  ${\cal +} (\pi^0/\eta)\gamma$  &   {\cal ++} KLOE
& {\cal +++}  $\pi^0 \pi^+ \pi^-$\\
\hline
\hline
Decays  \hhhu       & $7.78/\bf{(17)} $  & $ 7.77/\bf{(17)} $  &  $14.31/\bf{(9)} $ 
& $14.60/\bf{(9)}  $  & $14.70/\bf{(9)} $ \\
\hline
New  \hhhu &  & & &  & \\
Timelike (127+1) & $114.83 $   & $114.09 $ &  $114.12 $&  $127.15 $ &  $127.75 $\\
\hline
Old  \hhhu &  & & & & \\
Timelike (82+1) & $54.80 $    & $53.86 $ &  $50.84 $ &  $49.32 $  &  $49.32 $ \\
\hline
KLOE (60+5) \hhhu    & $- $ & $-  $ &  $-$ & $108.39$  &  $108.21$ \\
\hline
$\pi^0 \gamma$ (86)\hhhu & $- $  &  $- $ & $61.38$  &  $62.07$ &  $65.66 $ \\
\hline
$\eta \gamma$ (182)\hhhu  & $- $&    $- $ & $128.55$ &  $129.73$ &  $135.20$   \\
\hline
\hline
$\chi^2/\rm{dof}$ \hhhu & 177.41/210  & 175.72/210 &  369.50/468 & 491.27/528 & 637.90/653\\
Probability  & 95.0 \%  &95.9 \%  &  99.9\% & 87.2 \% & 65.6 \%\\
\hline
\hline
\end{tabular}
\caption{
\label{T1} The first data column displays the partial $\chi^2$ information while working with the largest set
of partial widths and all available $e^+ e^- \ra \pi^+\pi^-$ data sets (except those of
KLOE) and not taking into account the photon VP. The second data column displays the 
corresponding information while, instead, introducing the photon VP
effects. In the third data column, the data concerning $e^+ e^- \ra \pi^0 \gamma $
and $e^+ e^- \ra \eta \gamma $ are considered together with those on $e^+ e^- \ra \pi^+\pi^-$.
In the following data column, the KLOE data are included  and, correspondingly, the last data column 
gives the fit information
while adding also the  $e^+ e^- \ra \pi^0 \pi^+ \pi^-$ data (see text).  Boldface numbers 
 in the first data line display the number of (independent) partial widths which are included
 in the full  fitted data sets.
}
 
\end{table}

The package alphaQED.uu \cite{Fred0} also provides an estimate of the 
uncertainty on $\Delta \alpha_{em}(s)$.

In the extended model of this paper, the pion form factor has the same expression as in
\cite{taupaper}~:
\be
\displaystyle
F_\pi^e(s) = \left [ (1-\frac{a}{2}) - F_{\rho \gamma}^e(s) g_{\rho \pi \pi} \frac{1}{D_\rho(s)}
- F_{\omega \gamma}^e(s) g_{\omega \pi \pi} \frac{1}{D_\omega(s)}
- F_{\phi \gamma}^e(s) g_{\phi \pi \pi} \frac{1}{D_\phi(s)}
\right]
\label{eq30}
\ee
where $D_\rho(s)$ is defined by the first Eq. (\ref{eq24}) and by Eq. (\ref{eq25}), while
the other propagators are standard fixed width Breit--Wigner formulae. With our new
notation set, the coupling constants simply write~:
\be
\begin{array}{llll}
\displaystyle g_{\rho \pi \pi} = \frac{a g}{2} ~~~,~~&
\displaystyle g_{\omega \pi \pi} = -\frac{a g}{2} \alpha(s)~~~,~~&
\displaystyle g_{\phi \pi \pi} = \frac{a g}{2} \beta(s)
\displaystyle   
\end{array}
\label{eq31}
\ee

On the other hand, the $VP\gamma$ coupling constants entering the corresponding partial widths 
in the extended model are essentially\footnote{Negligible correction terms are 
outlined in Section \ref{partWidths}. } given by those  in Eqs 
(E.1--E.4) of \cite{taupaper} 
{\it multiplied each by the new fit parameter}  $c_3$.  Instead, the couplings constants 
of type  $P \gamma\gamma$  given by Eqs (E.5) in \cite{taupaper} are left unchanged as well
as the leptonic decay widths of the vector mesons.

\vspace{0.7cm}

\indent \indent
For definiteness, we have first performed the fit without including the photon VP. 
The most interesting results are reported in the first data column of Table \ref{T1}. 
The results obtained when introducing the photon $VP$  are given in 
the second data column
in the same Table. One may already conclude that both descriptions provide a quite good
account of the data sets considered. The first data column in Table \ref{T2} displays
the fit value of the parameters having the most intuitive meaning while fitting with the
photon VP. The values found for the fit scale factors exhibit a nice correspondence
with the expectations reported in the experimental papers. One should also note that the
fit value of the newly introduced fit parameter $c_3$ is statistically consistent with 1.
This is in good correspondence with the fits presented in \cite{taupaper}.

We do not discuss any further the intrinsic fit quality and fit parameter values 
as this case becomes interesting only compared to what happens when using additional data sets.
We also do not show plots illustrating this fit quality~: they are visually indistinguishable
from Figure 2 in \cite{taupaper}. One may, however, remark that  the probabilities are more
favorable now~; this should be attributed marginally  to using a different photon hadronic vacuum
polarisation  \cite{Fred0} and, especially, to having withdrawn the spacelike data 
\cite{NA7,fermilab2} from the fit procedure.

One may also note that the 17 accepted decay partial widths  \cite{RPP2008},
which fully determine our symmetry breaking (SU(3)/U(3)/SU(2)) parameters, are  all well accepted
by the fit.  At this stage, as in our previous study  \cite{taupaper},
only the $\rho^0 \ra e^+ e^-$ partial width significantly differs from its PDG value.

One should also note that introducing further cross section data sets has to be accompanied
by the removal of all accepted partial widths \cite{RPP2008}  derived from -- or highly influenced by --
these additional data sets. It is the reason why the number of fit partial widths 
decreases from 17 to 9 as soon as the data on the $e^+ e^- \ra (\pi^0/\eta) \gamma$ annihilations are
considered in addition to $e^+ e^- \ra \pi^+\pi^-$.

\section{Including The $e^+e^- \ra (\pi^0/\eta) \gamma$ Cross Section Data}
\label{Pgamma}
\subsection{Amplitudes and Cross Sections}
\indent \indent Using the Lagrangians given in Appendices \ref{HLSLagrangian}
and \ref{HLSPgamma}, one can derive the transition amplitudes $\gamma^* \ra P \gamma$.
The matrix elements are~:
\be
\left\{
\begin{array}{llll}
T(\gamma^*  \ra \pi^0 \gamma)&= i Y \left [ g~c_3 K_\pi(s) - (1-c_3) L_\pi \right]
\epsilon^{\mu \nu \alpha \beta}
k_\mu \varepsilon_\nu(k) p_\alpha \varepsilon_\beta(p)\\[0.5cm]
T(\gamma^*  \ra \eta_0 \gamma)&= i Y \left [g~ c_3 K_0(s) - (1-c_3) L_0 \right]
\epsilon^{\mu \nu \alpha \beta}
k_\mu \varepsilon_\nu(k) p_\alpha \varepsilon_\beta(p)\\[0.5cm]
T(\gamma^*  \ra \eta_8 \gamma)&= iY  \left [g~ c_3 K_8(s) - (1-c_3) L_8 \right]
\epsilon^{\mu \nu \alpha \beta}
k_\mu \varepsilon_\nu(k) p_\alpha \varepsilon_\beta(p)
\end{array}
\right.
\label{eq32}
\ee
where $Y=-\alpha_{em} N_c/\pi f_\pi$, $k$ is the incoming photon momentum ($k^2=s$),
$p$ the outgoing photon momentum ($p^2=0$) and $N_c=3$. 

We have defined~:
\be
K_P(s) =  \displaystyle\sum_{V_i= \rho^0, \omg,\phi}
\frac{H_{V_i}^{P^j}(s)  F_{V_i \gamma}(s)}{D_{V_i}(s)}~~~~,~~ P=\pi,~\eta_0,~\eta_8
\label{eq33}
\ee
in terms of the $\gamma -V_i$ transition amplitudes $ F_{V_i \gamma}$ (see Eq. (\ref{eq26}))
and of the inverse propagators $D_{V_i}(s)$. $D_{\rho^0}(s)$  is
given by  Eq. (\ref{eq24}), while the other inverse propagators are chosen  of the form 
$s-m_V^2+i m_V \Gamma_V$ for the narrow $\omg$ and $\phi$ mesons. The functions
$H_{V_i}^{P^j}$ which carry the dependence upon the isospin breaking angles $\alpha(s)$,
$\beta(s)$ and  $\gamma(s)$ are given  by Eqs. (\ref{eqB5}) and refer
to physical vector fields. We have also defined  the constants~:
\be
\begin{array}{llll}
L_\pi = \displaystyle \frac{1}{6} ~~,&
L_8 =  \displaystyle 
\frac{1}{6\sqrt{3}} ~ \frac{(5 z_A-2)}{3 z_A} ~~,&
\displaystyle L_0 =  \displaystyle
\frac{x}{3\sqrt{6}} ~ \frac{(5z_A +1)}{3 z_A}
\end{array}
\label{eq34}
\ee
which are terms deriving from the ${\cal L}_{AAP}$ Lagrangian (see Eq. (\ref{eqB6})). 
Defining the three following functions~:
\be
\left\{
\begin{array}{lllll}
R_\pi(s)= Y \left[ g~c_3~ K_\pi(s) - (1-c_3) L_\pi \right ]\\[0.5cm]
\left ( 
\begin{array}{lll}
R_{\eta}(s) \\[0.5cm]
R_{\eta^\prime} (s)
\end{array}
\right )
= ~Y~
\left ( 
\begin{array}{lll}
\cos{\theta_P}  & -\sin{\theta_P}  \\[0.5cm]
\sin{\theta_P}& \cos{\theta_P} 
\end{array}
\right )
~~~
\left ( 
\begin{array}{lll}
g~c_3 K_{8}(s) - (1-c_3) L_8 \\[0.5cm]
g~c_3 K_{0}(s) - (1-c_3) L_0 
\end{array}
\right )
\end{array}
\right.
\label{eq35}
\ee
the $e^+ e^- \ra P \gamma$ cross sections are  ($P=\pi^0,~\eta ,~\eta^\prime$)~:
\be
\sigma(s) = \displaystyle  \frac{\alpha_{em}}{24} \left[\frac{s-m^2_P}{s} \right]^3 R_P^2(s)=
 \frac{3\alpha_{em}^3}{8 \pi^2 f_\pi^2}\left[\frac{s-m^2_P}{s} \right]^3 
 \left | (g~c_3 K_P(s)-(1-c_3) L_P) \right|^2
\label{eq36}
\ee
where $m_P$ is the mass of the pseudoscalar meson produced in the annihilation process.
$K_P$ and $ L_P$ for $P=\eta,\eta^\prime$ are trivially defined from
$K_{0/8}$ and $ L_{0/8}$ using Eqs. (\ref{eq35}).
$\theta_P$, the pseudoscalar mixing angle, is algebraically related 
with the breaking parameters $x$ and $z_A$ reminded above (see Eq. (E7) in
\cite{taupaper}).
Unfortunately, there is presently no available
 cross section data on the $e^+ e^- \ra \eta^\prime \gamma$ channel.
One can only use the $\phi \ra \eta^\prime \gamma$
branching fraction \cite{RPP2008} as constraint~; however, one sees that the cross sections
contain  also definite constant terms (if fits confirm that $c_3 \ne 1$
is significant) and that these constant terms differ for $\eta$ and $\eta^\prime$. 
Therefore, the recommended branching fraction  \cite{RPP2008,KLOEetap}
should be considered with some care, until a consistent analysis of the
corresponding cross section becomes possible.

\subsection{The Data Set Submitted To Fit}
\label{PgammaSub}
\indent \indent There are several data sets on the annihilation processes
$e^+e^- \ra \pi^0 \gamma$ and $e^+e^- \ra \eta \gamma$ available since 1999, 
all collected on VEPP-2M accelerator at Novosibirsk. In our analysis, we
use all the data points up to $\sqrt{s} = 1.05$ GeV.

CMD-2 has recently published a data set on the final states $(\pi^0/\eta) \gamma$
(with $\pi^0/\eta \ra \gamma \gamma$) from 600 to 1380 MeV \cite{CMD2Pg2005}
 with 6 \% systematic error. Previously, the same collaboration has published
 data \cite{CMD2Pg2005} on the $\eta \gamma$ final state covering the same energy 
 range and going through the mode $\eta \ra 3 \pi^0 $~; these data sets have 6.1\%  and 4.1\%
 systematic errors, resp. below and above 950 MeV. We also use their former data
 set \cite{CMD2Pg1999}  on  the $\eta \gamma$ final state, with $\eta \ra \pi^+ \pi^- \pi^0$,
having a systematic error of 4.8\%.

On the other hand, the SND Collaboration has also recently published \cite{sndPg2007}
two different data sets
for the  $\eta \gamma$ final state with $\simeq$ 4.8 \% systematic errors, one
with   $\eta \ra 3 \pi^0$ from 600 to 1360 MeV, the other with $\eta \ra \pi^+ \pi^- \pi^0$
covering an energy range from 755 to 1055 MeV. A sample covering the energy range from
600 to 970 MeV for the $\pi^0 \ra \gamma \gamma$ decay mode was also published 
\cite{sndPg2003}. Other data sets of 14 energy points between 985 and  1039 MeV 
were also published \cite{sndPg2000}
with both final states $(\pi^0/\eta) \gamma $ (and
 $(\pi^0/\eta) \ra 2 \gamma$)  and systematic errors of $2.5$\%.

Altogether, these two Collaborations provide 86 measurement points for
the $e^+e^- \ra \pi^0 \gamma$  cross section and 182 for $e^+e^- \ra \eta \gamma$
for $\sqrt{s}\leq 1.05$ GeV.
 These data are highly valuable in order to build up and thoroughly check our Extended Model
 in the anomalous sector.
 
 In this second step, we consider altogether the three annihilation processes
  $e^+e^- \ra \pi^+ \pi^-$, $e^+e^- \ra \pi^0 \gamma$ and $e^+e^- \ra \eta \gamma$.
  In order to constrain more our isospin symmetry breaking mechanism, we still include in the data 
  set to  be fit a part of the partial widths used in the previous Section and already used in
  \cite{taupaper}.  Only 9 pieces of information are now independent of the present data~: 
  $\rho^\pm \ra \pi^\pm \gamma$, $K^0 \ra K^0 \gamma$, $K^\pm \ra K^\pm \gamma$,
  $\eta^\prime \ra \omg \gamma$, $\phi \ra  \eta^\prime \gamma$,
  $\eta/\eta^\prime \ra \gamma \gamma$ and finally $\phi \ra \pi^+ \pi^-$
  (in modulus and phase), which clearly carry information not statistically
  related with the cross sections considered.
  
  Instead, as the recommended values \cite{RPP2008} for $\omg/\phi \ra e^+e^-$ 
  are information highly influenced by the set of processes considered (and by the
  $e^+e^- \ra \pi^+ \pi^- \pi^0$ data considered later on), it is legitimate
  to let them free. This statement is {\it a fortiori}  valid for the  
  partial decay widths  $\rho^0/\omg /\phi \ra (\pi^0/\eta) \gamma$.

 A last remark~:  the $\omg$ and
 $\phi$ masses and total widths are extracted from the data sets we consider. Therefore, it
 is certainly legitimate to let them vary. At the very end of our procedure, when the data
 on the $e^+e^- \ra \pi^+ \pi^- \pi^0$ annihilation process will have been considered, we will be in 
 position to propose motivated averaged values for this information. Comparing 
 with the results derived using  the S-factor technics of the PDG \cite{RPP2008} would
 become interesting.
 
 Concerning our dealing with systematic errors for the newly introduced cross sections,
 we did not find numerically any need to split up correlated and uncorrelated
 systematic errors, which could have allowed  increasing the fit parameter freedom. We therefore
 have simply added in quadrature systematic and statistical errors in order to compute
 the $\chi^2$ to be minimized.
 
 Finally, the $e^+e^- \ra (\pi^0/\eta) \gamma$ cross sections submitted to fit have to be corrected
 for photon VP effects. This is done, as in the previous  Section,  by using the code provided
 by F. Jegerlehner \cite{Fred0}. 
  
 \subsection{Analysis of the Fit Results}
 \indent \indent 
  The fit has been performed and the fit quality information is reported in the third data column
  of Table \ref{T1} and some fit parameter information is given in the second data column of Table \ref{T2}.

  In Table  \ref{T1}, one sees that the fit quality reached for the $e^+e^- \ra \pi^0 \gamma$
  ($\chi^2/{\rm points} \simeq 61/86$) as well as for $e^+e^- \ra \eta \gamma$ 
  ($\chi^2/{\rm points} \simeq 129/182$) is very good.
  
  Fig. \ref{Fig:pi0g} displays together the fit and data for the annihilation process 
  $e^+e^- \ra \pi^0 \gamma$. The fit description is clearly quite good for these three  data sets.
  Moreover, one should note that the fit values for the $\omg$ and $\phi$ peak locations are
  well centered compared to the data. The situation exhibited by  Fig. \ref{Fig:etag}
 for the annihilation process  $e^+e^- \ra \eta \gamma$  is quite comparable.
 In the $\omg$ peak region, the large fluctuations 
  in the experimental data prevent to be conclusive about the detailed lineshape
  returned by the fit, however, the $\phi$ mass region is nicely accounted for.

In Table \ref{T1}, one should also note a significant increase of the $\chi^2$
contribution of the decay modes~: it practically doubles its value while 
the number of data is reduced by a half. A closer look at the results shows that the $\chi^2$ 
contributions  provided 
by 7 out of the 9 modes sums up to only 2.5, while $\chi^2(\phi \ra \eta^\prime \gamma)=5.1 $ and
$\chi^2(\eta \ra  \gamma \gamma)=6.9 $ , a $2.3 \sigma$ and a $2.6 \sigma$  effect  respectively.
One may not worry too much about the $2.3 \sigma$ difference from the recommented value for 
Br($\phi \ra \eta^\prime \gamma$) for reasons already sketched~; however, a 
$2.6 \sigma$ for Br($\eta \ra  \gamma \gamma$) could call for some comments.

When adding more and more spectra to be fit, the weight of isolated independent partial decay widths 
becomes generally less and less constraining. If needed, one may  
increase the weight of the decay mode of concern inside the full $\chi^2$. For instance, if
instead of adding $\chi^2_{\eta \ra \gamma \gamma}$  to the total $\chi^2$ one adds
$ 4 \times \chi^2_{\eta \ra \gamma \gamma}$ or $ 8 \times \chi^2_{\eta \ra \gamma \gamma}$
the distance to the recommended value becomes resp. $1.10 \sigma$ or $0.72 \sigma$ without
a significant change  to the "decay mode" contribution to the total $\chi^2$ (it increases
by 0.3 unit compared to the datum in Table \ref{T1}).
 As this may look artificial, a more "natural" way would be
to fix $z_A \equiv [f_K/f_\pi]^2=1.495 \pm0.031$ at its  (experimental) central value and use our model equations 
\cite{rad,WZWChPT,taupaper}~:
 \be
 \left \{
 \begin{array}{ll}
G_{\eta \gamma \gamma} = & 
-\displaystyle \frac{\alpha_{em}}{\pi \sqrt{3} f_{\pi}}
\left [ \frac{5-2Z}{3}\cos{\theta_P}-\sqrt{2} 
\frac{5+Z}{3}x \sin{\theta_P} \right ]~~,~~\left( Z=\frac{1}{z_A}\right)\\[0.5cm] 
\tan{\theta_P} = &\displaystyle \sqrt{2} \frac{Z-1}{2Z+1} x
\end{array}
\right .
\label{eq37}
\ee
to connect with the $\Gamma(P \ra \gamma \gamma)$ datum \cite{RPP2008}~: 
 \be
\Gamma(P \ra \gamma \gamma)=  \displaystyle 
\frac{m_P^3}{64 \pi} |G_{P\gamma \gamma}|^2~~~,~~(P=~\pi^0,~\eta,~\eta')~~.
\label{eq38}
\ee
One could also derive the $1~\sigma$ upper and lower bounds for $x$ and $z_A$ 
consistent with the $f_K/f_\pi$ and $\Gamma(P \ra \gamma \gamma)$ data and
force the fit to stay within these limits\footnote{Unless otherwise stated, the
normal running conditions for our fits do not impose bounds to any parameter.}.
Therefore, the problem encountered with $\Gamma(P \ra \gamma \gamma)$ can easily
be accomodated without any trouble.

On the other hand, it is useful to compare the fit parameter values derived in the present case
with  their analogs in the leftmost data column of Table \ref{T2}. One first notes that the scale 
factors affecting
the $e^+e^- \ra \pi^+ \pi^-$  data sets are left unchanged. The second remark concerns the physics
parameters~; most of them vary well within the quoted $1 \sigma$ uncertainties. The uncertainty
for $a$ is  improved  and its value is still significantly different from 2.  The central value for $x$
becomes closer to previous fit results using decay width data only \cite{rad,box}.

One should, however, note that $c_3$ becomes inconsistent with 1 by $\simeq 4.5 \sigma$. 
This is clearly
influenced by the anomalous branching fractions but also by the full cross section lineshapes
of the anomalous annihilation processes we just considered. Forcing $c_3=1$ gives a fit quality
$\chi^2(\phi \ra \eta^\prime \gamma)=383.6/469$, almost as expected\footnote{The $\chi^2$
difference with $\chi^2=369.5$ reported in Table \ref{T1} is found at 14.1 while  one expects
 19.4~; this shows that the minimum is indeed close to parabolic.}.    

It is worth noticing the $\omg$ and $\phi$ parameter values returned by the fit. At this
step -- before using  $\pi^+ \pi^- \pi^0$ data -- we get
$m_\omg =782.45 \pm 0.05$ MeV and $\Gamma_\omg=8.63 \pm 0.08$ MeV on the one hand,
 $m_\phi=1019.25 \pm 0.02$ MeV and $\Gamma_\phi=4.19 \pm 0.05$ MeV on the other hand.
 As will be seen later on, these values are modified while including 3--pion data.

\begin{table}[!htb]
\hspace{-1.5cm}
\begin{tabular}{|| c  | c  | c | c  | c  ||}
\hline
\hline
\hhhb Parameter  & $e^+e^- \ra \pi^+ \pi^-$ &  $e^+e^- \ra \pi^+ \pi^-$ & $e^+e^- \ra \pi^+ \pi^-$ 
& All $e^+e^- \ra \pi^+ \pi^-$\\ 
\hhha ~~~ & (NSK Only) & (NSK Only)  & (NSK+ KLOE) ~&  {\cal +} $e^+e^- \ra (\pi^0/\eta) \gamma$   \\ 
\hhha ~~~ & ~~ & {\cal +} $e^+e^- \ra (\pi^0/\eta) \gamma$& {\cal +} $e^+e^- \ra (\pi^0/\eta) \gamma$
&{\cal +} $e^+e^- \ra  \pi^0 \pi^+ \pi^-$  \\ 
\hline
\hline
Scale New Timelike \hhhb & $0.995 \pm 0.004$ &$0.996 \pm 0.004$ & $0.991 \pm 0.004$
&  $0.991 \pm 0.004$ \\
\hline
Scale Old Timelike  \hhhb & $1.007 \pm 0.009$  &  $ 1.007 \pm 0.009$ & $1.007 \pm 0.009$
 & $1.007 \pm 0.009$ \\
\hline
Scale  KLOE 0 \hhhb &$-$  &$-$ & $1.615 \pm 0.816$ & $1.621 \pm 0.813$\\
\hline
Scale  KLOE 1 \hhhb &$-$  &$-$ &  $-0.041\pm 0.023$ & $-0.041\pm 0.023$\\
\hline
Scale  KLOE 2 \hhhb &$-$  & $-$ & $-0.070 \pm 0.017$ & $-0.070 \pm 0.017$\\
\hline
Scale  KLOE 3 \hhhb &$-$  &$-$ &  $ 0.003 \pm 0.006$ & $ 0.003 \pm 0.006$\\
\hline
Scale  KLOE 4 \hhhb &$-$  &$-$ & $-0.011 \pm 0.014$ & $-0.011 \pm 0.014$\\
\hline
\hline
$a$\hhhb & $2.399 \pm 0.022$ & $2.356 \pm 0.012$ & $2.364 \pm 0.011 $ & $2.365\pm 0.011$ \\
\hline
$g$\hhhb & $5.468 \pm 0.021$& $5.574 \pm 0.019$ & $5.567 \pm 0.013$ & $5.568 \pm 0.011$ \\
\hline
$c_3$\hhhb & $1.018 \pm 0.017$& $0.943 \pm 0.013$ & $0.927 \pm 0.013$ & $0.930 \pm 0.011$ \\
\hline
$x$\hhhb & $0.935 \pm 0.014$ & $ 0.904 \pm 0.014$  & $0.915 \pm 0.014$ & $0.914 \pm 0.014$   \\
\hline
$z_A$\hhhb & $1.577 \pm 0.020$ &  $1.467 \pm 0.034$ &  $1.503 \pm 0.020$&  $1.496 \pm 0.018$ \\
\hline
$z_V$\hhhb &$1.509 \pm 0.020$&  $1.425 \pm 0.045$ &  $1.501 \pm 0.030$&  $1.503 \pm 0.028$  \\
\hline
$z_T$\hhhb & $1.275 \pm 0.053$ & $ 1.301 \pm 0.058$ & $1.340 \pm 0.059$& $1.332 \pm 0.058$  \\
\hline
\hline
\end{tabular}
 
\caption{
\label{T2}
Parameter values in fits performed including  photon VP. The data subsamples
included in the full data sample submitted to fit are indicated on top of the Table.
The number of independent decay widths added to the data sample is 17 (first
data column) or 9 (all other data columns). For the first 3 rescaling coefficients
given in the Table, the corrections are the departure from 1, for all others, 
the rescaling are departures from 0. }
\end{table}

\section{Including The $e^+e^- \ra \pi^+ \pi^-$ Data From KLOE}
\label{pipiKLOE}
\indent \indent The KLOE Collaboration, operating at the $\phi$-factory  DA$\Phi$NE,
has recently published \cite{KLOE_ISR1} the spectrum for the $e^+e^- \ra \pi^+ \pi^-$
form factor. Using the Initial  State Radiation (ISR) mechanism, they produced a spectrum
covering the region $0.60-0.97$ GeV with very small statistical errors ($\simeq 0.5$ \%). 
The systematic errors, also small,  are dominant and have been throroughly 
studied \cite{KLOEnote}.

Several source of systematics are reported as (standard deviation) spectra representing fractions 
 of the measured spectrum for $|F_\pi(s)|^2$. 
Their  \cite{KLOE_ISR1} Table 1 thus gives the uncertainty due to background subtraction 
($\epsilon_1(s)$), their Table 2 displays the  uncertainty due to acceptance corrections ($\epsilon_2(s)$)~; 
Table
3 and 4 respectively give the error due to detector resolution ($\epsilon_3(s)$) and the error
due to the radiator function effects ($\epsilon_4(s)$). From their Table 5, one can derive by adding
in quadrature the various source of systematics (other than the ones just listed)  a global scale
uncertainty ($\epsilon_0(s)$) of 0.76\%  for $|F_\pi(s)|^2$.

All these sources of uncertainty should be considered correlated except -- maybe -- for 
the error due to detector resolution ($\epsilon_3$) which could have to be treated as uncorrelated
\cite{VenanzoniPriv}. This structure of systematic errors is clearly complicated and the
question is how to deal with the error functions $\epsilon_\alpha(s)$
($\alpha=0,\cdots 4$) just defined. 

The most appropriate way seems to follow the method presented in Section \ref{stat},
 {\it i.e.} each function
$\epsilon_\alpha(s)$ is viewed as a gaussian random variable of zero mean and having
$s$--dependent  standard deviations (named $\eta_\alpha(s)$) given by the numbers in the 
Tables 1--5 of \cite{KLOE_ISR1} for each $\Delta s$ bin.  More precisely,
one may assume $<\epsilon_\alpha(s)>=0$, $<[\epsilon_\alpha(s)]^2>=[\eta_\alpha(s)]^2$ and
$<\epsilon_\alpha(s_i)\epsilon_\beta(s_j)>\simeq \delta_{\alpha \beta}  \delta_{ij}$.
This means that these five $\epsilon_\alpha(s)$ functions play as $s$--dependent scale 
uncertainties. Then  a predicted  value  $|F_\pi(s_i)|^2$ should be
associated with a datum $m_i$ modified in the following way~:
\be
m_i \ra m_i^\prime= m_i\left [ 1+ \sum_{\alpha=0,1,2,3,4} \delta \lambda_\alpha(s_i)\right ]
\label{eq39}
\ee
where each $\delta \lambda_\alpha(s_i)$ is  one sampling of the corresponding
random variable $\epsilon_\alpha(s)$ to be fitted. Under these assumptions, 
the error covariance matrix 
writes in the usual way\footnote{
This expression corresponds to having treated $\epsilon_3$ uncorrelated. In this case, the sum
on $\alpha$ in Eq. \ref{eq42} does not extend to $\alpha =3$. We tried both possibilities (uncorrelated
and correlated) without getting significant differences.}~:
\be
V_{ij}=\left [ \sigma_i^2 + [\eta_3(s_i) |F_\pi(s_i)|^2]^2\right ] \delta_{ij}
+\sum_{\alpha=0,1,2,4} 
\left [ \eta_\alpha(s_i) |F_\pi(s_i)|^2 \right ]\left [ \eta_\alpha(s_j) |F_\pi(s_j)|^2 
\right ]
\label{eq40}
\ee
to be inverted numerically for $\chi^2$ estimation. 
$i,~j$  are bin indices,  while $\sigma_i$ is the reported statistical error 
on $m_i$. 

In practice, this turns out to
compare the predicted function $|F_\pi(s)|^2$ following from a model to the
modified data~:
\be
m_i^\prime= m_i \left [1+ \sum_{\alpha=0,1,2,3,4}  q_\alpha  \eta_\alpha (s_i)\right ]
~~~~,~~i=1,\cdots n
\label{eq41}
\ee
where the five constants $q_\alpha$  are to be fit. Then, the partial
$\chi^2$ associated with KLOE data set is ($f_i\equiv |F_\pi(s_i)|^2$)~:
\be
\chi^2_{KLOE}=\sum_{i,j} (m_i^\prime-f_i)(m_j^\prime-f_j) V^{-1}_{ij}+
\sum_{\alpha=0,\cdots~4} q_\alpha^2
\label{eq42}
\ee
In this way, one can check the consistency of 
$q_\alpha$ with respect to expectations as outlined in Section \ref{stat}
and correct, if needed.

Finally, as the data under examination have not been corrected from photon VP
effects, our fitting function is defined by Eqs. (\ref{eq29}) and (\ref{eq30})
given above.

We have submitted to fit the data set consisting of all the previously defined 
data subsets plus the KLOE data. The main results are reported in Table \ref{T1}
(fourth data column) and Table \ref{T2} (third data column).

One first remarks from  Table \ref{T1} that the fit probability is quite favorable (87 \%).
The $\chi^2$ contribution from decays is as already reported in Section \ref{Pgamma} and 
calls for  the same comment (we have $\chi^2(\eta \ra \gamma \gamma)=2.7$ and
$\chi^2(\phi \ra \eta^\prime \gamma)=6.7$). Otherwise, the fit quality of the 
previously introduced (Novosibirsk)  $\pi^+ \pi^-$ timelike data  is marginally
degraded, while the description of 
the $(\pi^0/\eta) \gamma $  data is unchanged. The $\chi^2$ contribution from
the KLOE data set may look  large but is still considered 
acceptable \cite{VenanzoniPriv}~; this may reflect
the unusual property of being highly dominated by systematic errors, always harder
to estimate very precisely than statistical uncertainties. However, even if the $\chi^2_{KLOE}$
is large, the global fit does not degrade significantly the fit quality of the other data sets
and, moreover, the expected physics parameter values are not spoiled.

Some results referring to parameter values are reported in Table \ref{T2}. One observes 
a change in the scale factor of the new timelike data which is shifted by $2\sigma$
from its expected value (0.4 \%), while the scale factor associated with the old
timelike data is unchanged ($\simeq 0.7 \sigma$ from expectation). Out of the 5
scale parameters $q_i$ specific to the KLOE data set, we find a global scale ($q_0$)
correction at $\simeq (1.6 \pm 0.8) \sigma$ from the expected $\sigma=0.76$\%~; 
the parameter $q_2$ associated with the acceptance correction uncertainties
is also significantly non--zero while all 
other corrections are small enough to be neglected.

Parameter values are slightly changed with using KLOE data. One should note the increased precision
on $g$, the universal vector meson coupling. Finally, Fig. \ref{Fig:KLOE} shows
superimposed the KLOE data, the fit function, together with the residuals. One clearly sees that
the description is reasonable. It should be noted however that 
the worst residuals are in the region covered by the $\epsilon_3$ correction,
{\it i.e.} the reported uncertainy on the detector resolution \cite{KLOE_ISR1}. 
As far as we know, this is the first published fit to the available KLOE data.

\section{Including $e^+ e^- \ra \pi^+ \pi^- \pi^0$ Data}
\label{3pionNSK}
\indent \indent
In contrast with the annihilation channels examined so far, the
$e^+ e^- \ra \pi^+ \pi^- \pi^0$ cross section involves 
the box anomaly sector presented in Section \ref{ExtendedModel}.
It also involves the non--anomalous sector \cite{taupaper} and
the triangle anomalies, as the $\pi^0 \gamma$ and $\eta \gamma$
final states. This process allows  to validate all sectors
of the Extended HLS Lagrangian model, as defined at the beginning of this paper. 

Without going into much details, one can list the Lagrangian pieces involved~:
\begin{itemize}
\item The VVP piece (see Eq. (\ref{eq9})) expressed in terms of
ideal vector fields has been already given expanded in \cite{rad}.
This provides diagrams where the photon transforms to physical neutral vector
fields with amplitudes  given in Section \ref{Reminder2}. The transitions
to $\rho \pi$  are given by~:
\be
\left\{
\begin{array}{lll}
{\cal L}_{VVP} = C \epsilon^{\mu \nu \alpha \beta} F_{\mu \nu \alpha \beta}~~~, 
~~C=\displaystyle-\frac{N_c g^2}{8 \pi^2 f_\pi} c_3 \\[0.5cm]
F_{\mu \nu \alpha \beta}=\pa_\mu \omg_\nu^I \left [
\pa_\alpha \rho^I_\beta \pi^0 + \pa_\alpha \rho^+_\beta \pi^- + \pa_\alpha \rho^-_\beta \pi^+ 
\right ] + \cdots
\end{array}
\right.
\label{eq43}
\ee 
where the  physical meson fields $\omg$, $\rho$ and $\phi$ appear when
using Eqs. (\ref{eq23}). The $\rho^\pm$ particles decay with
vertices as given in Eq. (\ref{Lag_tau}), while the neutral physical fields
decay as follows~:
\be
{\cal L}_{VPP}= \displaystyle \frac{ia g}{2} \rho^0_I \cdot \pi^- \parsym \pi^+ = 
\frac{ia g}{2} \left [
\rho^0    - \alpha \omega + \beta \phi \right] \cdot \pi^- \parsym \pi^+
\label{eq44}
\ee
This provides contributions symmetric in all $\rho \pi$ charge combinations.
The argument of $\alpha(s)$,  $\beta(s)$ and $\gamma(s)$ is always the
incoming photon 4--momentum squared.

\item The same VVP piece provides also a diagram
without symmetric partners, where the
$\gamma -\rho^0$ transition connects to $\omg \pi^0$ 
and the $\omg$ meson decays in accordance with Eq. (\ref{eq44}).
In this case the breaking function is $\alpha(s_{+-})$.
The same piece provides also an unusual $\gamma -\rho^0$
term connecting with a $\rho^0 \ra \pi^+ \pi^-$ in the final state, with the same
weight $\alpha(s_{+-})$.

\item The APPP piece (see Eq. (\ref{eq9})) provides a single term 
$\gamma \ra 3 \pi$~:
\be
{\cal L}_{\gamma PPP} = i e D \epsilon^{\mu \nu \alpha \beta} 
A_\mu \pa_\nu \pi^0 \pa_\alpha \pi^+~\pa_\beta \pi^-~~,
~~D=\displaystyle - \frac{N_C}{12 \pi^2 f_\pi^3} \left[
1 -\frac{3}{4} (c_1-c_2+c_3) \right ]
\label{eq45}
\ee
to the amplitude, and depends on the newly introduced parameter $c_1-c_2$.

\item Finally, the VPPP piece contributes through the
following piece of ${\cal L}_{V PPP}$
\be
{\cal L}_{V PPP} = i E \epsilon^{\mu \nu \alpha \beta} 
\omg^I_\mu \partial_\nu \pi^0 \partial_\nu \pi^0 
\partial_\alpha \pi^+ \partial_\beta \pi^-~+ \cdots~~~, 
~~E=\displaystyle-\frac{3 N_c g }{16 \pi^2 f_\pi} (c_1-c_2-c_3)
\label{eq46}
\ee
limiting oneself to the pion sector for the moment. Using Eqs. (\ref{eq23}),
one sees that each of the 3 physical neutral vector mesons may have a direct 3--pion
coupling, however modulated by 1, $\alpha(s)$ and $\gamma(s)$ for resp.
the $\omg$, $\rho^0$ and $\phi$ mesons. $s$ is still the incoming photon  4--momentum.
\end{itemize}

\subsection{Matrix Element  And Cross Section}
\indent \indent  The amplitude for the $\gamma^* \ra \pi^+ \pi^- \pi^0$ 
transition can be written~:
 \be
T (\gamma^* \ra \pi^+ \pi^- \pi^0)= \displaystyle 
[T_{sym} + T_\rho] ~~\epsilon^{\mu \nu \alpha \beta}
\varepsilon_\mu(k) p^0_\nu p^+_\alpha  p^-_\beta
 \label{eq47}
\ee
where $\varepsilon_\mu(k)$ ($k^2=s$) is the (heavy) photon 
polarization vector. $T_{sym}$ is the symmetric part of the amplitude
(in terms of the $\rho \pi$ 'final' states), while $T_\rho$  breaks
this symmetry. 

Using the $\gamma -V$ transition amplitudes given in Section \ref{Reminder2}, let
us define\footnote{If the coupling of the charged $\rho$ to a
pion pair differs from the neutral one, $g$ becoming $g + \delta g$, the last two terms in
$N_2(s)$ should be affected by a factor of the form $1+\delta g/g$.
A test for a non--zero  $\delta g$ was performed in \cite{taupaper}
using $\tau$ data and was not found significant.
}~:
\be
 \left \{
\begin{array}{llll}
N_1(s)=& \displaystyle \left [
\frac{F_{\omg \gamma}(s)}{D_\omg(s)} + \alpha(s)
\frac{F_{\rho \gamma}(s)}{D_{\rho^0}(s)} + \gamma(s)
\frac{F_{\phi \gamma}(s)}{D_\phi(s)}
\right ]\\[0.5cm]
N_2(s)=& \displaystyle \left [
\frac{1}{D_{\rho^0} (s_{+-})}+\frac{1}{D_{\rho^+}(s_{+0})}+\frac{1}{D_{\rho^-}(s_{-0})}
\right ]\\[0.5cm]
N_3(s)=& \displaystyle \alpha(s_{+-}) \left [
\frac{1}{D_{\rho^0}(s_{+-})}  -\frac{1}{D_\omg(s_{+-})}
\right ]
 \end{array}
 \right .
\label{eq48}
\ee
where the meaning of $s_{+-}$, $s_{+0}$ and $s_{-0}$ is obvious. Then, we have~:
 \be
 \left \{
\begin{array}{ll}
 T_{sym}(s) =\displaystyle \frac{ieN_c}{12 \pi^2 f_\pi^3} \left [
 1 -\frac{3}{4} (c_1-c_2+c_3) + \frac{3}{2} m^2 g c_3 N_1(s) N_2(s) 
 - \frac{9}{4} g(c_1-c_2-c_3)N_1(s) 
 
\right]\\[0.5cm]
 T_{\rho}(s) =\displaystyle 
\frac{ieN_c}{12 \pi^2 f_\pi^3}  \left [\frac{3}{2} m^2 g c_3\right] 
\frac{F_{\rho \gamma}(s)}{D_{\rho^0}(s)} N_3(s)
 \end{array}
 \right .
\label{eq49}
\ee

The $\rho$ propagators have already been defined by Eqs. (\ref{eq24}) and  (\ref{eq25}).
In order to study the three pion final state, we assume $\delta m^2$ (see Eq. (\ref{eq24}))
to vanish~; the sensitivity to a non--vanishing $\delta m^2$ is certainly
quite marginal as long as $\tau$ spectra are not considered.

As in Section \ref{Pgamma}, we approximate the $\omg$ and $\phi$ inverse propagators
by~:
 \be
 \left \{
\begin{array}{l}
 D_\omg(q^2)= q^2-m_\omg^2 +im_\omg \Gamma_\omg \\[0.5cm]
 D_\phi(q^2)= q^2-m_\phi^2 +im_\phi \Gamma_\phi
 \end{array}
 \right .
 \label{eq50}
\ee

At the chiral point ($s_{+-}=s_{+0}=s_{-0}=s=0$), $T_{\rho}$ vanishes and one gets~:
\be
  T_{sym}(0) = \displaystyle \frac{ieN_c}{12 \pi^2 f_\pi^3} 
\label{eq51}
\ee
from having required $c_3=c_4$, and having assumed  
$ [D_\omg(s=0))]^{-1} \equiv [D_\rho(s=0))]^{-1}=-m^2$.
Therefore, one recovers the usual WZW term automatically at the chiral limit.

On the other hand, the differential cross section writes~:
\be
\displaystyle
\frac{d^2 \sigma(e^+e^- \ra \pi^+ \pi^- \pi^0)}{dx ~dy}=
\frac{\alpha_{em}}{192 \pi^2} s^2 G(x,y) |T_{sym}+T_\rho|^2
\label{eq52}
\ee
when using the ($x$ and $y$) parametrization proposed by \cite{Kuraev} and summarized
in Appendix \ref{threeBodies}. One should note the abbreviated notation for 
the functions $G(x,y)$, $N_2(s)$ and $N_3(s)$  which are actually all functions of $x$, $y$ and $s$.

\subsection{Properties of the Matrix Element}
\indent \indent The two pieces $ T_{sym}$ and $T_\rho$ of our amplitude are given in
Eqs. (\ref{eq49}). It is easy to check that,
in the limit of Isospin Symmetry conservation, the non--symmetric term $T_\rho$
vanishes and that the symmetric part named $ T_{sym}$ reduces to only one term (with
an intermediate $\omg$).   

As vector meson mixing occurs, 
the symmetric part $T_{sym}$ clearly exhibits the 3  possible $\gamma - V$
transitions allowed by our Extended Lagrangian Model. 
The second term of  $T_{sym}$  is the more
usual one, as it describes the sequence $e^+e^- \ra V^0 $ followed 
by $V^0 \ra \pi^+ \pi^- \pi^0$ through each of the $V^0\rho \pi$ possible couplings. 
One remarks that intermediate $V^0 = \rho^0$ or $V^0=\phi$ are generated by
Isospin Symmetry breaking.

The first term in $ T_{sym}$  is a non--resonant contribution $\gamma \ra\pi^+ \pi^- \pi^0$, 
specific of the HLS Lagrangian. It plays, for the 3--pions decay 
amplitude a role similar to the HLS $a ~(\simeq 2.4)$  parameter 
for the $e^+ e^- \ra \pi \pi$ or $e^+ e^- \ra K \overline{K}$ amplitudes.
The second term is the more usual one and should provide the dominant contribution.

The third term, instead, is the more problematic contact term \cite{Kuraev}~;
one sees that our model predicts that the 3 neutral vector mesons
have a direct decay $V^0 \ra \pi^+ \pi^- \pi^0$ amplitude, with the
$\rho$ and $\phi$ contributions  weighted by the mixing functions $\alpha(s)$ and
$\gamma(s)$. 

If one follows the educated guess expressed in \cite{HLSRef}
and thus fixes $c_3=c_4=c_1-c_2=1$, the resonant contact term
identically vanishes, while the non--resonant one survives with an intensity
of -1/2, as supposed in a previous study on box anomalies \cite{box}.
As we do not expect large departures from the quoted guess, one can
assert that the resonant contact term should indeed be very small. This
is confirmed by the data  and examined below.

Finally, one should note the dependence on $\alpha(s_{+-})$ of
the non--symmetric part of the amplitude  $T_\rho$. Actually, it is
this special dependence which defines the non--symmetric amplitude.
Other approaches generally consider only  an $\omg$ term,
while we also get a $\rho$ term. Taking into account the large width
of the $\rho$ meson, one may guess that this term provides a tiny contribution.

\subsection{The Data And The Fit Procedure}
\indent \indent 
Here also a large number of data sets is available. The most significant are
coming from CMD-2 and SND Collaborations running on the VEPP-2M machine at Novosibirsk.

CMD-2 has published several data sets with quoted uncertainties on 
the cross section merging  statistical errors and uncorrelated systematic errors.
The correlated systematic error, which reflects  the uncertainty on the global scale of
the cross section, is given separately for each CMD--2 data set.
 The CMD-2 data sets to be considered are~:
\begin{itemize}
\item The data set in \cite{CMD2-1995corr} which covers the $\omg$ region
and is claimed being affected by a global scale uncertainty of 1.3\%,
\item The data set given in \cite{CMD2KKb-1} which covers the $\phi$ region,
 with a reported scale error of 4.6\%,
\item The $\phi $ region is also explored in \cite{CMD2-1998}, with
a better reported scale uncertainy~: 1.9\%,
\item  The most recent CMD-2 data set is published in \cite{CMD2-2006}
and still covers the $\phi$ region with a favorable scale uncertainty ( 2.5\%).
\end{itemize}  

On the other hand, the SND Collaboration has published two spectra
covering altogether the region from 0.440 to 1.38 GeV. We have  
considered all data points with  $\sqrt{s} \leq 1.05$ GeV. These are~: 
\begin{itemize}
\item Below 980 MeV, the data set in \cite{SND3pionLow},
\item Above  980 MeV, the data set in \cite{SND3pionHigh} 
\end{itemize}
  
SND preferred providing the statistical and systematic errors separately
in their Tables. They claim that their detection efficiency and their
integrated luminosity measurement merged together are actually a correlated
systematic error. Therefore,
all systematics, but these contributions, were added in quadrature to their
statistical errors and we treated the rest as an uncertainty on the global
scale of the cross section. This amounts to 3.4 \% below 980 MeV \cite{SND3pionLow} 
and we took 5\% above \cite{SND3pionHigh}. 

For both CMD-2 and SND data, we treated the scale uncertainties as explained in Section
\ref{stat}. As in this case, we added penalty terms to the global $\chi^2$ with the
variance just quoted for each data set.

Besides these two groups of data samples, older data sets are worth considering~:
\begin{itemize}
\item The former data sample collected by the ND Collaboration, with a reported
systematic error of 10\%, as given by the corresponding Table in
 the Physics Report by Dolinsky {\it et al.} \cite{DM13pion-1979}.
\item A small CMD data sample \cite{CMD3pion-1989} -- 5 measurements -- 
covering the region in between the $\omg$ and $\phi$ peaks. The reported
systematic error is 15 \%.
\end{itemize}

The former covers the whole energy region between 0.75 and 1.38 GeV, but for our use,
it was truncated at 1.05 GeV. All  data points in this sample lay
outside the $\omg$ and $\phi$ peak regions. Therefore, this allows us to constrain
the most possible the non--resonant region which is well covered by SND, but
more poorly by  CMD-2. The data points being rather unprecise, there was no point in 
splitting  up the systematic error into correlated and uncorrelated pieces~;
we simply added them in quadrature with the reported statistical errors.
We did alike for the CMD measurements \cite{CMD3pion-1989}.

Let us mention a DM1 data sample \cite{DM13pion-1979} which covers
the energy region between 0.75 and 1.098 GeV. It could have been
used, however, we are not sure of the systematics all along the spectrum.
Therefore, we have preferred leaving it aside. 

We still perform a global fit and, therefore, the data set submitted to fit always
includes, together with specified 3 pion data subsets, all reported $\pi^+ \pi^-$ Novosibirsk and 
KLOE data samples and all $e^+e^- \ra (\pi^0/\eta) \gamma$ data sets.

\vspace{0.7cm}
\indent \indent From the point of view of a fit procedure, fitting 3--pion cross sections
using the amplitude given in Eqs. (\ref{eq49}) poses a problem of numerical analysis.
Indeed, as clear from these equations, we have to integrate  on the Kuraev--Siligadze 
\cite{Kuraev} $x$ and $y$ variables at each step 
of the procedure a function which depends also on the parameters subject to minimization.
This makes the computing time\footnote{In order to analyze the behavior of each parameter or 
subset configuration, we need to run the code a large number of times. Without 3--pion
fitting, such a run lasts about 2 minutes. Including the 3--pion samples,
the fastest 'exact' method we found provides a waiting time of about an hour. The simplified
method sketched below reduces the computing time to about 5 minutes per run.} rather prohibitive.

However, having assumed that charged and neutral $\rho$ have the same mass
and the same coupling, the single parameter really influenced by 3--pion data is $c_1-c_2$
(see Eq. (\ref{eq49}))~; all other parameters in functions to be integrated 
have already very precise fit values provided by the other data sets and are marginally
influenced by the  3--pion data.

Therefore, the method we followed is to tabulate, in bins of $\Delta (\sqrt{s}) =0.1$ MeV, 
the integrals over 
$x$ and $y$ of the $G(x,y)$ function alone and combined with the appropriate products of  the $N_i$
(see Eqs. (\ref{eq51})) and $N_j^*$ functions {\it computed first  from a
minimization solution without  3--pion data}. Then one performs the global fit 
-- including 3--pion data -- using these functions. The fit result is then used to 
improve the integral calculations and restart a new minimization step. A priori, this procedure has to
be repeated  until some converge criterium is fulfilled.

 We expected convergence in a very few steps. However, it so happens that there was no need 
 to iterate the procedure, as the $\chi^2$ was never changed by more than 
 $\simeq 0.5$ unit, compared to some  'exact' computing method,
 where the integrals were computed at each minimization step. As the fit parameter
 values were not found to get significant corrections, we found justified
 to use this simplified method for performing our fits.

\begin{table}[!htb]
\vspace{-1.5cm}
\hspace{-1.8cm}
\begin{tabular}{|| c  || c  | | c || c  | c | c ||}
\hline
\hline
\hhhq Data Set  & Basic Fit &   Global Fit  with & \multicolumn{3}{|c|}{ Global Fit with ND+CMD {\cal ++}} \\
\hhhq $\sharp$ (data $+$ conditions) & $\pi \pi$ {\cal +} $3 \pi$  & ND+CMD   &  CMD2+SND &   SND
&  CMD2\\
\hline
\hline
Decays  \hhhq      & $13.93/{\bf 17}$  & $14.55/{\bf 9}$  &  $17.51/{\bf 9}$ 
& $17.93/{\bf 9}$  & $ 14.70/{\bf 9}$ \\
\hline
New Timelike (127+1) \hhhq  & $116.29$   & $127.05$ &  $129.53$ &  $130.07 $ &  $127.76$\\
\hline
Old Timelike (82+1) \hhhq  & $49.89 $    & $49.35 $ &  $49.99$ &  $50.08 $  &  $49.32$ \\
\hline
KLOE (60+5) \hhhq          & ---- & $108.45$  & $105.11$ & $102.69$  &  $108.21$ \\
\hline
$\pi^0 \gamma$ (86)\hhhq & ----  &  $62.06$ & $72.04$  &  $82.61$ &  $65.66$ \\
\hline
$\eta \gamma$ (182)\hhhq & ----&    $129.84$ & $133.64$ &  $131.81 $ &  $135.20$   \\
\hline
\hline
CMD--2 \cite{CMD2-1995corr} (13+1) * \hhhq  & $28.31$ &   ---- & $27.14$ &   ---- &  $25.05$   \\
\hline
CMD--2 \cite{CMD2-2006} (47+1) *\hhhq  & $58.24$ &    ---- & $53.89$ &  ---- &  $52.48$   \\
\hline
CMD--2 \cite{CMD2KKb-1} (16+1) *\hhhq  & $17.84$ &   ---- & $17.78$ & ---- &  $15.76$   \\
\hline
CMD--2 \cite{CMD2-1998} (13+1) *\hhhq  & $21.76$ &  ---- & $20.82$ &  ---- &  $13.86$   \\
\hline
ND+CMD \cite{ND3pion-1991} \cite{CMD3pion-1989}  (37) \hhhq  & $31.47$ &    $25.82$ & $23.78$ &  $23.58$   &  $25.94$   \\
\hline
SND \cite{SND3pionLow} (49+1) * \hhhq  & $47.94$ &   ---- & $67.22$ &  $63.55$ &  ----   \\
\hline
SND \cite{SND3pionHigh} (33+1) * \hhhq  & $54.62$ &   ---- & $77.75$ &  $55.59$ &  ----   \\
\hline
\hline
$g$ \hhhv & $5.641 \pm 0.017$ &    $5.566 \pm 0.010$ & $5.599 \pm 0.011$ &  $ 5.560 \pm 0.010$ &  $5.568 \pm 0.011$   \\
\hline
$c_3$ \hhhv & $0.998 \pm 0.017$ &    $0.927 \pm 0.010$ & $0.898 \pm 0.010$ &  $0.895 \pm 0.007$ &  $0.930 \pm 0.011$   \\
\hline
$c_1-c_2$ \hhhv & $0.766 \pm 0.056$ &    $1.168 \pm 0.069$ & $1.095 \pm 0.039$ &  $1.012 \pm 0.048$ &  $1.210 \pm 0.043$   \\
\hline
corr~($c_3,c_1-c2$ )\hhhv & $ 0.027$ &    $0.252$ & $0.523$ &  $0.152$ &  $0.473$   \\
\hline
\hline
$\chi^2/\rm{dof}$ \hhhq & 454.73/413  & 515.48/564 & 815.40/735 & 667.21/646 & 637.90/653 \\
Probability  & 7.7 \%  & 92.9 \%  &   2.1 \% & 27.4 \% &  65.6 \%\\
\hline
\hline
\end{tabular}
\caption{
\label{T3} 
Fit results using three--pion data sets within global fits.
In the first line the number of decay widths is indicated in boldface.
The so--called 'conditions' are the correlated scale uncertainties which are considered
as data and fit. Comments are given inside the text.
The partial $\chi^2$ in lines flagged by * do not include the additional contribution of the scale factor.
These are, nevertheless, counted inside the final $\chi^2$ together with the number of conditions.  
}
 
\end{table}

\subsection{Exploratory Fits}
\indent \indent
Up to now, we have always used all available data sets for the various annihilation channels
we have examined. The 3--pion data faces us with having to perform a choice between data sets
which has to be motivated.
It is the reason why some exploratory analysis of the 3--pion data sets has been
worth performing.

We start by considering all CMD--2 \cite{CMD2-1995corr,CMD2KKb-1,CMD2-1998,CMD2-2006}
and SND \cite{SND3pionLow,SND3pionHigh} data sets in the less constraining pattern
of our fit procedure~; namely, together with only the so--called old and new Timelike data sets
collecting the standard $e^+ e^- \ra \pi^+ \pi^-$ data sets. In this case, we have to
use \cite{taupaper} also the full set of 17 accepted  decay modes \cite{RPP2008}.
Indeed, doing this way~:
\begin{itemize}
\item
 We lessen at most effects of $e^+ e^- \ra (\pi^0/\eta) \gamma$ on the value
of $c_3$, which plays some role in the 3--pion cross section,
\item We avoid being sensitive to the KLOE data and their particular systematic errors.
\end{itemize}

\indent \indent
While comparing with other models and fits, one should keep in mind that, even
if highly relaxed,  our model is, nevertheless, sharply constrained. It is actually
its lack of flexibility which imposes some good enough understanding of the 
experimental uncertainties and, on the other hand, discloses the correlations between 
the various data sets imposed by the (common) underlying physics.

The fit pattern just described leads to the results reported in the first data column of
Table \ref{T3}. 
The fit qualities for the CMD--2 data sets given in \cite{CMD2KKb-1,CMD2-2006}
are good ($\chi^2/{\rm npoints} \simeq 1$) while the fit qualities of 
the data sets given in 
\cite{CMD2-1995corr,CMD2-1998} are worse ($\chi^2/{\rm npoints} \simeq 1.5 \div 2$).
The fit quality of the old  ND  \cite{ND3pion-1991}  and CMD  \cite{CMD3pion-1989}
data samples is also quite good  ($\chi^2/{\rm npoints} \simeq 1$ altogether). 
The low energy SND data set \cite{SND3pionLow} returns a good fit quality 
($\chi^2/{\rm npoints} \simeq 1$), while the higher  energy SND data set 
\cite{SND3pionHigh} is found worse ($\chi^2/{\rm npoints} \simeq 1.5$).
However, the global fit quality ($\chi^2/{\rm npoints} \simeq  455/413$)
corresponds to a  7.7\% probability.

This relatively poor probability leads us to carefully examine the behavior
of the CMD--2 and SND 3--pion data sets separately. 

\vspace{0.7cm}

\indent \indent
The next step relies on the following remark. All data sets -- except for 3--pion data --
allow to fix all parameters of our model, but $c_1-c_2$. As can be seen from the
third data column in Table \ref{T2}, these are already known with an excellent
accuracy, especially the parameters expected to be influenced by 3--pion
data ($g$, $c_3$ for instance).  Looking at Eqs. (\ref{eq49}), one can
easily see that the missing parameter $c_1-c_2$ occurs in such a way that it is 
highly influenced by the invariant mass region outside both the $\omg$ and $\phi$
peaks. Therefore, one can guess that performing the global fit with the
ND  \cite{ND3pion-1991}  and CMD \cite{CMD3pion-1989} data samples only should allow
a measurement of  $c_1-c_2$ with already a challenging accuracy. On the
other hand, the Breit--Wigner mass and width for the $\omg$ and $\phi$ meson
are accurately fit from the $(\pi^0/\eta) \gamma$ data.

Therefore, one can construct an accurate  3--pion cross section, valid 
for the whole energy range  from threshold to above the $\phi$ mass,
which feeds in the whole correlated physics information. This cross section
can be used as a prediction to be compared to the data sets
provided by the CMD--2 and SND Collaborations separately. 

The global fit following this pattern has provided results shown in the
second data column of Table \ref{T3}. The global fit quality is well illustrated by
the fit probability which reaches 93\%. 
A closer look at the fit properties of the previous data sets shows that they
are still optimally accounted for  (compare this data column with the third data
column in Table \ref{T2}).
The fit quality reached for the ND + CMD data set is also very good~:  
$\chi^2/{\rm npoints} \simeq  26/37$.

The accuracy reached for $g$ and $c_3$ is  unchanged, however this fit provides
a value for  $c_1-c_2 \simeq 1.2$ with an accuracy of $\simeq 4\%$.
Therefore, this allows us to define precisely the 3--pion cross section. This function is
displayed  in Figure \ref{Fig:ND-CMD:omgPhi}
together with the CMD--2 and SND data superimposed in the $\omg$ region
and in the $\phi$ region.

The left plots in Figure \ref{Fig:ND-CMD:omgPhi} show the case for CMD--2 data~; one can see that 
the data are already in good agreement with what can be inferred from ND+CMD data 
together with the physics of the    other annhilation processes. The peak value and lineshape
for the $\omg$ region are in accord with the prediction~; one may note, however, 
minor departures far on the wings.  
Concerning the $\phi$ region, the lineshape is in reasonable agreeement with expectations, 
and one may infer that a fit  will result in a good description. 
 
Rightmost  plots show the case for SND data~; at both the $\omg$ and $\phi$ peaks,
one observes that both tops are $\simeq 10\%$ too large compared with the same expectations as
before. However, a qualitative observation of the behavior in the $\phi$ region, allows one
to infer that one should run into  difficulties to account for both  top  and wings 
with only global (constant) rescalings.

This conclusion is substantiated with performing the global fit with all SND and
CMD--2 data, as reported in the third data column in Table \ref{T3}. In this case, the
global fit probability reached is  $\simeq 2 \%$~; the scale factors returned,
$0.98 \pm 0.03$ for the low energy data sample \cite{SND3pionLow} and
$0.91 \pm 0.04$ for the higher energy one \cite{SND3pionHigh} are in good agreement
with expectations. However, this poor  fit probability, 
associated with poorer SND $\chi^2$'s, indicates that the problem is not
solved by means of  constant rescaling factors.

Of course, this  follows from having requested consistency 
of the 3--pion data sets, not only with each other, but also with all physically
related annihilation processes. Indeed, one clearly understands that a common
fit, using only the CMD-2 and SND 3--pion data, should certainly succeed with a 
solution intermediate between CMD--2 and SND data~; but this success would hide  
the consistency issue  with the rest of related processes.

Therefore, the two global fits performed with merging SND and CMD--2 data tell us that there 
there is some inconsistency between them.
Now, it remains to check separately SND +ND +CMD data, on the one hand and
CMD2 +ND +CMD data, on the other hand, in order to make a motivated choice.

We have performed the global fit with all SND+ND+CMD data~; the results are 
reported in the fourth data column in Table \ref{T3}. The global probability 
becomes more reasonable (27\%)  -- from having removed CMD--2 data -- and the fit rescaling
factors are   $0.96 \pm 0.03$ and $0.89 \pm 0.04$ for resp. the low and high energy
regions.
The fit quality exhibited  for the low energy sample may look reasonable, however,
the corresponding information for the higher energy data sample goes on looking poorer.
This should reflect that the $10 \%$ scale correction, valid on the $\phi$ peak,
is not appropriate on the wings.

The corresponding information from the global
fit performed with all CMD--2+ND+CMD data is reported in the last data column
of the same Table~; one gets a global fit probability of 66 \%. This result,
together with what is displayed in  Figure  \ref{Fig:ND-CMD:omgPhi},
shows that ND+CMD data, on the one hand, and CMD--2 data, on the other hand,
are quite consistent with each other.

Clearly, as the physics contents of the CMD--2 and SND data sets are the same,
the issue just sketched cannot be attributed to our model.
Because  of Figure \ref{Fig:ND-CMD:omgPhi} and 
because of the respective probabilities, 
one has to choose among CMD--2 data and SND data,
and the best motivated choice is
the CMD--2 +ND +CMD solution. This also turns out to remark that
Table \ref{T3} exhibits a poor consistency of SND data compared with 
ND +CMD. SND data are certainly useful, however not 
in the framework of a global fit which requires a good control of
systematics in order to get reasonable consistency
with all other physics process measurements.

\subsection{The Solution With CMD-2, ND and CMD Data}
\indent \indent 
Presently, the choice to remove the SND
samples for 3--pions data considered in the global fit, seems the best motivated one.
The main fit results are shown in the rightmost data column in Table \ref{T3}.
The fit is shown with data superimposed in Fig. \ref{Fig:cmd2-mid}. Top plots
show a zoom on the  peak regions, while the downmost plot focuses on
the region outside the peaks. The fit is clearly satisfactory.

Let us summarize the other physics results of interest~:
\begin{itemize}
\item The rescaling factors for all Novosibirsk and Frascati data
reported in   Table \ref{T2} are recovered 
with    changes affecting the last digit, {\it i.e.}  far inside the
reported  errors,
\item the value for the HLS parameter $a=2.365 \pm 0.105$ is also
unchanged.
\item The breaking parameters $z_A$, $z_V$ and $z_T$ were also left unchanged 
by the fit procedure, with unchanged uncertainties,
compared with the third data  column in Table \ref{T2}.
\item The scale factors found are $0.996 \pm 0.012$ for the data set in \cite{CMD2-1995corr}
(expected departure from 1~: 1.3\%), $0.975 \pm 0.020$ for the data set in \cite{CMD2-2006}
quite in accord with the expected departure from 1 (2.5\%). The other rescaling factors
are resp. $0.953 \pm 0.035$ (data from \cite{CMD2KKb-1} with expected departure from 1 of
4.6\%) and  $0.974 \pm 0.016$ (data from \cite{CMD2-1998} with expected departure from 1 of
1.9\%). Stated otherwise, the first 3 fit scales are 1 $\sigma$ at most from expectations,
while the last scale is found at 1.25 $\sigma$. This also indicates that there is no obvious
signal of missing variance for the 3--pion CMD--2 data samples.
\end{itemize}

The fit quality for the ND data set in \cite{ND3pion-1991} and the old CMD data set
\cite{CMD3pion-1989} is quite satisfactory, The three CMD--2 data sets collected
around the $\phi$ peak benefit from a $\chi^2/{\rm npoints} \simeq 1$~;
instead, the $\chi^2$ for the data set collected at the $\omg$ peak has a much poorer
quality ($\chi^2/{\rm npoints} = 25/13$). A closer look at the results shows that 
more than  half of the $\chi^2$ comes from 3 points on the wings\footnote{
The measurements located at $\simeq 0.76$, $\simeq0.77$ and $0.80$ GeV contribute resp. for
5, 4 and 6 units to the $\chi^2$ .}. 
However, as the global lineshape looks well
reproduced at both peaks and in between them and also because of the good global
fit quality, we consider the fit to the CMD-2, ND and CMD data as satisfactory.

Other fit information, reported in the last data column in Tables \ref{T1} and
\ref{T2}, allows estimating the effects of having included 3--pion data inside our fit data sample. 
Now, with   the full information displayed, Table 1 clearly illustrates
that the only noticeable  changes ($\chi^2$  increased by $\simeq 4$ units)
concern the $\pi^0 \gamma$ and $\eta \gamma$ data samples. Table \ref{T2} instead
shows that there is no modification beyond the 1 $\sigma$ level.

\section{A Few Numerical Results}
\label{numRes}
\indent \indent With the fit presented just above, one ends up 
introducing data affecting the $\omg$ and $\phi$ mesons. Our global fit
already provides interesting physics results  which 
should not be affected by the forthcoming steps of our study.

\subsection{$\omg$ and $\phi$ Masses And Widths}
\indent \indent  We have defined the $\omg$ and $\phi$ propagators
as fixed width Breit--Wigner expressions (see Eqs. (\ref{eq50})). Our fits widely illustrate
that there is no need to go beyond this approximation.  As we account for
both statistical and systematic errors, the uncertainties we quote fold in
both kinds of errors and are thus directly comparable to RPP \cite{RPP2008}
information. Moreover, it should be noted that the RPP data on $\omg$ and $\phi$ masses 
and widths mostly relies on the data we have used. Our final results are~:
\be
\left \{
\begin{array}{llll}
m_\omg=782.42 \pm 0.05 ~~{\rm MeV}~~~&,~~ {\rm RPP}~:~~m_\omg=782.65 \pm 0.12~~{\rm MeV}
\\[0.5cm]
\Gamma_\omg=8.700 \pm 0.084 ~~{\rm MeV}~~~&,~~ {\rm RPP}~:~~\Gamma_\omg=8.49 \pm 0.08~~{\rm MeV}
\\[0.5cm]
m_\phi=1019.173 \pm 0.015~~~{\rm MeV}~~~&,~~ {\rm RPP}~:~~m_\phi=1019.455 \pm 0.020~~{\rm MeV}\\[0.5cm]
\Gamma_\phi=4.259 \pm 0.036~~~{\rm MeV}~~~&,~~ {\rm RPP}~:~~\Gamma_\phi=4.26 \pm 0.04~~{\rm MeV}
\end{array}
\right .
\label{eq54}
\ee
We thus find the $\omg$ mass 2 $\sigma$ below  RPP average value and we get an uncertainty
twice smaller. $\Gamma_\omg$ is found at $2.5 \sigma$ from its recommended value \cite{RPP2008}, but in close
agreement with the fit results of both the CMD--2 and SND Collaborations reported in the Review
of Particle Properties \cite{RPP2008}.

The $\phi$ mass is instead found at 0.282 MeV from its recommended value \cite{RPP2008} --
about 20 $\sigma$~! Interestingly, our extracted $\phi$ mass is also significantly smaller 
compared to the values extracted by CMD--2 and SND from the same data \cite{RPP2008}.
This might partly reflect the contributions of anomalous terms and the effect of vector meson 
mixing. Moreover, as parametrization of the propagator, we used a fixed width Breit--Wigner function,
while CMD--2, for instance, used a varying width Breit--Wigner expression.
Finally,  $\Gamma_\phi$ is found in perfect agreement
with the RPP for both the central value and its uncertainty.

One should note that our global fit is expected to~: 
\begin{itemize}
\item lessen the statistical error. Indeed, each parameter is
constrained  simultaneously by all data sets where it plays a role.
This is, of course, true for the mass and width of the $\omg$ and $\phi$ mesons
which sharply influence the $\pi^0 \gamma$, $\eta \gamma$ and the
$\pi^0 \pi^+ \pi^-$ final state descriptions~; moreover, the sharp drop observed
in $e^+ e^- \ra \pi^+ \pi^-$  is also influencing the $\omg$ meson
parameters.
 
\item perform the folding of systematic and statistical errors in
accord with the whole knowledge of these uncertainties provided
by the various experiments.
\end{itemize}
Therefore, the net expected result is an optimal folding in of 
all reported sources of errors, even when data are coming from different
triggers and/or  groups. 

\subsection{Contact Terms And Anomalous Lagrangians}
\indent \indent A priori, the Extended HLS Model we use has two kinds of
contact terms. One of the form $\gamma P P P$ describes the residual coupling 
of photons to pseudoscalar meson triplets, the other of the form $V PPP$ 
the direct coupling of a vector meson   to a pseudoscalar meson triplet.
The former is given (see Eq. (\ref{eq49})) by~:
\be
C_\gamma= \displaystyle 1 -\frac{3}{4} (c_1-c_2+c_3)= -0.61 \pm 0.10~~~,
\label{eq55}
\ee
and the latter by the coefficient~:
\be
C_V= \displaystyle  -\frac{9}{4} (c_1-c_2-c_3)=-0.63 \pm 0.31~~~;
\label{eq56}
\ee
this second coefficient is additionally modulated by  1, $\alpha(s)$, $\gamma(s)$ for
resp. the $\omg$, $\rho$, $\phi$ mesons when coupling to a pion triplet. The numerical
values just given are derived from the information in Table \ref{T3} and include the
effects of the covariance term $<\delta(c_1-c_2)~\delta c_3>$. These numbers are
very close to the educated guess in \cite{HLSRef} ($ C_\gamma=-1/2$ and $C_V=0$).

One can conclude that the  contact term $C_\gamma$ is certainly significant, while
$C_V$ has only a $2 \sigma$  significance. 

In view of these results, one is interested in 
seeing what happens to the global fit, if one fixes  $c_3=c_1-c_2=1$ in the model. The result is
shown in the second data column of Table \ref{T4}. The subsamples not shown keep, roughly 
speaking,  their usual $\chi^2$ contributions. In this case, as could be expected, the fits
to the $e^+e^- \ra \pi^0 \gamma$ and $e^+e^- \ra \eta \gamma$ cross sections are significantly
degraded and the global fit probability allows to reject this solution.

If one fixes $c_1-c_2=1$ and leaves free $c_3$ (third data column in Table \ref{T4}), 
the good description of $\pi^0 \gamma$ and $\eta \gamma$ data sets is recovered as could 
be expected, but this is done at the expense of a worse decription  of $\pi^0 \pi^+ \pi^-$
data. The situation is similar
if, instead, one fixes $c_3=1$ and let free $c_1-c_2$. 

One may conclude from the results collected in  Table \ref{T4}, that it is
highly meaningful to let $c_1-c_2$ and $c_3$ vary. This means that the
violation of Vector Meson Dominance affects all sectors of the anomalous
HLS Lagrangian \cite{FKTUY,HLSRef}~; this violation is only weaker in the Triangle
sector than in the Box sector.

\begin{table}[!htb]
\vspace{0.cm}
\hspace{0.cm}
\begin{tabular}{|| c  || c  | c | c  | c ||}
\hline
\hline
\hhhu Global fit  &  \multicolumn{4}{|c|}{~~~~~~~~~~~~~~~} \\
\hhhu ($\chi^2$ contributions) &  $c_3$, $c_1-c_2$ free  & $c_3$, $c_1-c_2$ fixed  
&   $c_1-c_2$ fixed &   $c_3$ fixed\\
\hline
\hline
$\pi^0 \gamma$ (86)\hhhu & $65.66 $   &  $106.80$ & $80.72$  &  $69.41$  \\
\hline
$\eta \gamma$ (182)\hhhu  & $135.20 $ &    $156.14$ & $136.57$ &  $162.15$    \\
\hline
$\pi^0 \pi^+ \pi^-$ (130)\hhhu  & $137.06$ &    $205.32$ & $171.24$ &  $159.52$    \\
\hline
\hline
Probability \hhhu  & 65.66 \%  & 0.01 \%  &  12.5\% & 12.7\% \\
\hline
\hline
\end{tabular}
\caption{
\label{T4} 
Influence of the parameters $c_3$ and $c_1-c_2$ on the fit results.
Each entry displays the contribution to the total $\chi^2$ of the quoted
subsample. Last line provides the global fit probability. For the first column
$c_3$ and $c_1-c_2$ values are given in the last column of Table \ref{T3}~;
the second column gives the best solution for $c_3=c_1-c_2=1$. In the third data column one has
set $c_1-c_2=1$ and the fit has returned $c_3=0.830 \pm 0.002$. In the fourth data column
one has fixed $c_3=1$ and fit $c_1-c_2=1.382\pm0.038$.
}
\end{table}

\subsection{The Mixing 'Angles'}
\indent \indent Most of the present data sets examined altogether 
are dominated by the $\omg$ and $\phi$ resonances. Using cross sections
instead of only partial decay widths may produce changes in the mixing functions
$\alpha(s)$, $\beta(s)$ and $\gamma(s)$ which, as noted before, can be interpreted
as resp. the $\omg -\rho$, $\phi -\rho$  and $\omg -\phi$ mixing 'angles'.
These are shown in  resp. Figures  \ref{Fig:alpha},  \ref{Fig:beta} and
\ref{Fig:gamma}. 

Compared to using only partial width decays together with $e^+ e^- \ra \pi^+ \pi^-$
cross section data, one observes that the functions $\alpha(s)$  and $\beta(s)$ 
are unchanged (compare to Figure 7 in \cite{taupaper}). The function $\gamma(s)$,
instead, gets significant changes.  Figure \ref{Fig:gamma} shows that the $\omg -\phi$
mixing angle becomes positive for $s > 0$, while it was negative when using   
partial width decays. However, the variation of $\gamma(s)$ remains significant
between the $\omg$ and $\phi$ regions.
The values which can be read off Figure  \ref{Fig:gamma}
give $\gamma \simeq 2.75^\circ$ at the $\omega$ mass  and $\gamma \simeq 3.84^\circ$ at the $\phi$ mass. 
As, now, one relies on the largest possible data set, this result should be considered
as superseeding the function $\gamma(s)$ in \cite{taupaper}.

\section{Validation Of The Model~: The $\eta/\eta^\prime \ra \pi^+ \pi^- \gamma$ Spectra}
\label{BoxAnomalies}
\indent \indent The amplitudes for $\eta/\eta^\prime \ra \pi^+ \pi^- \gamma$ at the chiral
point can be derived from the WZW Lagrangian \cite{WZ,Witten}. However, as there is no
theoretical knowledge of their momentum dependence, modeling the behavior outside the
chiral point is unavoidable. As shown by Eqs. (\ref{eq19}), the Extended HLS Model allows
to recover the chiral limit and the question is whether the predicted spectra and partial 
widths are in agreement with data. Such a study was already performed a few years ago \cite{box} 
assuming, as inferred by \cite{FKTUY,HLSRef}, that $c_1-c_2=c_3=c_4=1$. 

At the point where we are, all parameters of our Extended HLS Model have definite
values and, thus,  all vector meson and contact term contributions have no longer any
free parameter. The question is now whether the  spectra and the partial widths, which can be
algebraically derived, fit the known physics information.

The most reliable physics information for these 2 decay modes is certainly their 
partial widths \cite{RPP2008}. There exists  two spectra giving the photon momentum
distribution in the $\eta$ rest frame \cite{Gormley,Layter}~; however, the relevant 
information should be read off  plots giving distributions for the  measured spectrum
and for the acceptance/efficiency function. 

Understanding the dipion invariant mass spectrum in the $\eta^\prime \ra \pi^+ \pi^- \gamma$ has been
addressed several times, as most groups were claiming that the $\rho$ mass peak
was observed shifted. Most of these spectra were published only as figures 
\cite{Grigorian,TASSO,ARGUS,TPC2g,LeptonF}~; only Crystal Barrel provided the data points, 
feeding in the information concerning acceptance and efficiency corrections
\cite{CBar}~; some other spectra were only published as preprints or PhD theses. All these
spectra were discussed in \cite{box} and shown to exhibit very different qualities 
due to statistics \cite{Grigorian} or, sometimes, to obvious biases  \cite{TPC2g}, etc...

The issue was whether the $\rho$ peak  in the $\eta^\prime \ra \pi^+ \pi^- \gamma$
spectrum is shifted. Crystal Barrel \cite{CBar} clearly proved that the
$\rho$ peak location was indeed shifted by $\simeq 20$ MeV. The study in \cite{box}
later proved that this shift was actually the way found by fit procedures to
account for a missing constant term which results in a distorsion
of the $\rho$ lineshape.
 This distorsion is produced
by the contact term $\gamma \pi^+ \pi^- \eta^\prime$  which adds up with
$\rho$ meson contribution. Accounting for the contact term,  \cite{box} proved that the
$\rho$ peak in the $\eta^\prime$ spectrum is at the location expected from $e^+e^- \ra \pi^+ \pi^-$ data.
It is this question which is revisited within a framework where the condition
$c_1-c_2=c_3=c_4=1$ is violated and the vector meson mixing at work.

\subsection{Amplitudes For The $\eta/\eta^\prime \ra \pi^+ \pi^- \gamma$  Decays}
\indent \indent Using the Lagrangian pieces given in Section \ref{WZWBrk},
it is now easy to compute the transition amplitudes involved in the
 $\eta/\eta^\prime \ra \pi^+ \pi^- \gamma$ decays. These can be derived from~:
\be
\left\{
\begin{array}{lll}
T(\eta_8 \ra \pi^+ \pi^- \gamma) = & A~ F_8(s) \epsilon^{\mu \nu \alpha \beta} 
\epsilon_\mu(\gamma) q_\nu p^-_\alpha p^+_\beta \\[0.5cm]
T(\eta_0 \ra \pi^+ \pi^- \gamma) = &  x \sqrt{2} ~A~F_0(s) \epsilon^{\mu \nu \alpha \beta} 
\epsilon_\mu(\gamma) q_\nu p^-_\alpha p^+_\beta 
\end{array}
\right .
\label{eq57}
\ee
Having assumed $c_3=c_4$, the octet and singlet functions write~:
\be
\left\{
\begin{array}{lll}
F_8(s) = &
\displaystyle 
\left[
1 - \frac{3 c_3}{2} \left(1+ m^2 \sum_V \frac{c_V^8(s)}{D_V(s)}
\right) \right]\\[0.5cm]
F_0(s) = &
\displaystyle \left[
1 - \frac{3 c_3}{2} \left(1+ m^2 \sum_V \frac{c_V^0(s)}{D_V(s)}
\right) \right]\\[0.5cm]
\displaystyle  A = & \displaystyle  - \frac{i e N_c}{12 \pi^2 f_\pi^3} \frac{1}{\sqrt{3}}
\end{array}
\right .
\label{eq58}
\ee
where $q$ is the photon 4-momentum and $s=(p^+ + p^-)^2$ is the dipion invariant mass.
Neglecting terms of order greater than 1 in the mixing parameters $\alpha$, $\beta$ 
and $\gamma$, one has~:
\be
\left\{
\begin{array}{lll}
\displaystyle \sum_V \frac{c_V^8(s)}{D_V(s)}=
\left[1 +\frac{\alpha(s)}{3} -\frac{2 \sqrt{2}}{3 z_A} \beta(s) \right]
\frac{1}{D_\rho(s)} - \frac{\alpha(s)}{3}  \frac{1}{D_\omg(s)} 
+ \frac{2\sqrt{2}}{3 z_A}  \frac{ \beta(s)}{D_\phi(s)} \\[0.5cm]
\displaystyle \sum_V \frac{c_V^0(s)}{D_V(s)}=
\left[1 +\frac{\alpha(s)}{3} +\frac{\sqrt{2}}{3 z_A} \beta(s) \right]
\frac{1}{D_\rho(s)} - \frac{\alpha(s)}{3}  \frac{1}{D_\omg(s)} 
-\frac{\sqrt{2}}{3 z_A} \frac{ \beta(s)}{D_\phi(s)}
 \end{array}
\right .
\label{eq59}
\ee
which fulfill~:
\be
\displaystyle 
 \sum_V \frac{c_V^0(s)}{D_V(s)} -\sum_V \frac{c_V^8(s)}{D_V(s)} =
 \frac{\sqrt{2}}{ z_A} \beta 
 \left [ \frac{1}{D_\rho(s)} -\frac{1}{D_\phi(s)} \right]
\label{eq60}
\ee

The $\phi$ contributions in these expressions can be dropped out, because of the phase
space cuts at  $s=m_{\eta^\prime}^2$ or $s=m_{\eta}^2$, much below the $\phi$ mass.
Defining~:
\be
\left ( 
\begin{array}{lll}
F_\eta(s) \\[0.5cm]
F_{\eta^\prime}(s)
\end{array}
\right )
=
\left ( 
\begin{array}{lll}
\cos{\theta_P}  & -\sin{\theta_P}  \\[0.5cm]
\sin{\theta_P}& \cos{\theta_P} 
\end{array}
\right )
~~~
\left ( 
\begin{array}{lll}
F_8(s) \\[0.5cm]
x \sqrt{2}F_0(s)
\end{array}
\right )
\label{eq61}
\ee
the decay amplitudes for $\eta$ and $\eta^\prime$ can be written~:
\be
T(\eta/\eta^\prime \ra \pi^+ \pi^- \gamma)=A F_{\eta/\eta^\prime}(s)
~~\epsilon^{\mu \nu \alpha \beta} 
\epsilon_\mu(\gamma) q_\nu p^-_\alpha p^+_\beta
\label{eq62}
\ee
and the decay partial width can be obtained by integrating~:
 \be
\displaystyle \frac{d\Gamma(X \ra \pi^+ \pi^- \gamma)}{d \sqrt{s}} =
\frac{1}{9} \frac{\alpha_{em}}{[2 \pi f_\pi]^6}
\left| F_X(s) \right|^2 q_\gamma^3 p_\pi^3~~~~,~~~ X=\eta,~\eta^\prime
\label{eq63}
\ee
from the two--pion threshold to the $\eta$ or $\eta^\prime$ mass. One
has also defined $q_\gamma = (m_X^2-s)/2 m_X$ and $p_\pi=\sqrt{s-4 m_\pi^2}/2$. 

\subsection{Properties of the Amplitudes}
\indent \indent The amplitudes just given have several interesting properties.
As in the model developped in  \cite{box}, one finds in Eqs (\ref{eq58})
a vector meson amplitude which adds up with a non--resonant amplitude.

One might be surprised that these expressions do not depend on $c_1-c_2$
as the decay amplitude for $e^+ e^- \ra \pi^0 \pi^+ \pi^-$. This is due to
an unexpected conspiracy between the $VPPP$ and $APPP$ couplings 
(see Eqs. (\ref{eq16}) and  (\ref{eq17}))  which both play a
role in the transitions  $\eta/\eta^\prime \ra \pi^+ \pi^- \gamma$.

As for the amplitude for $e^+ e^- \ra \pi^+ \pi^-$, the dipion invariant mass
distribution exhibits an interference pattern between the $\rho$ and $\omg$ meson 
contributions.  However, all experimental data sets listed above indicate that
this usual interference pattern should not be met, as there is no observed drop of the
distributions around $\simeq 782$ MeV. The single difference between
the $e^+ e^- \ra \pi^+ \pi^-$ and $\eta^\prime \ra \pi^+ \pi^- \gamma$
processes is that the term of order $\alpha$ 
is weighted by $1/3$ in the $\eta^\prime$ decay amplitude.
 It is thus interesting to see whether this different weighting
alone allows to cancel out the (expected) drop in the dipion spectrum.

\subsection{Comparison With Data}
\indent \indent
In Figure \ref{Fig:box}, one displays the dipion mass spectra in the  
$\eta^\prime \ra \pi^+ \pi^- \gamma$ decay with, superimposed, the predictions
coming out of the model fitting $e^+e^- \ra \pi^+ \pi^-,~\pi^0 \gamma,~ \eta \gamma,
~\pi^+ \pi^- \pi^0$. We thus show the case for ARGUS data \cite{ARGUS}, for 
an experiment run at Serpukhov on the Lepton F facility \cite{LeptonF} 
with relatively large statistics, and for
Crystal Barrel \cite{CBar}, which is certainly the most precise 
$\eta^\prime$ spectrum. The last plot in Figure  \ref{Fig:box} is the spectrum
of the photon momentum in the $\eta$ rest frame for the 
$\eta \ra \pi^+ \pi^- \gamma$  decay \cite{Layter}, together with our prediction.

The agreement is satisfactory. One should note that some effect 
of the $\omg-\rho$  interference on the predicted lineshape 
can be detected but is not contradicted by the data.
Actually, the data are binned and the curves shown are not averaged over the 
bin widths. One clearly sees that the peak location, as well as the global 
lineshapes are well predicted by our model. One may remark that the 20 MeV
shift of the $\rho$ peak confirmed by \cite{CBar} is correctly reproduced.
Therefore, having $c_3 \ne 1$ and vector meson mixing 
does not degrade the agreement already obtained in \cite{box} with a much
simpler model.

However, the quality of these data is not good enough to allow including the 
corresponding spectra
into the  global fit\footnote{Actually, the Crystal Barrel spectrum could be safely fit,
except for a single point -- at 812.5 MeV -- which degrades severely the $\chi^2$
without changing the parameter values at minimum.
The Crystal Barrel  spectrum is actually the merging of 4 spectra collected in
$p \overline{p}$ annihilations at rest. Two of these
(from $p \overline{p} \ra \omega (\ra \pi^0 \gamma /\pi^0 \pi^+ \pi^-)  \eta^\prime$)
exhibit  this faulty point, for two others ($p \overline{p} \ra \pi^0 \pi^0 \eta^\prime$ and 
$p \overline{p} \ra \pi^+ \pi^- \eta^\prime$) the  corresponding 
measurement is located on the predicted curve.
 We have preferred avoiding to include a truncated spectrum inside our data set.}.

Other pieces of information are the partial widths for $\eta$ and $\eta^\prime$ decays
into this final state. The predicted values for these are~:
\be
\left \{
\begin{array}{ll}
\Gamma(\eta^\prime \ra \pi^+ \pi^- \gamma)= 53.11 \pm 1.47 ~~{\rm keV}\\[0.5cm]
\Gamma(\eta \ra \pi^+ \pi^- \gamma)= 55.82\pm  0.83  ~~{\rm eV}
\end{array}
\right .
\label{eq64}
\ee
quite close to the corresponding recommended values \cite{RPP2008}~:
$60 \pm 5$ keV and $60 \pm 4$ eV. 
The predictions are clearly quite accurate and 
provide support to the parameter values we derived from our global fit. 
This also indicates that we have exhausted our parameter
freedom with the data set we already considered. Indeed, including these two 
pieces of information inside our global fit does not return results 
substantially different from the predicted values given just above.
Therefore, the  $\eta/\eta^\prime \ra \pi^+ \pi^- \gamma$ decay modes
represent a good test  and better data on these decays
would certainly be valuable in order to further constrain
our model.

\section{Decay Partial Widths Of Vector Mesons}  
\label{partWidths}
\indent \indent Having allowed $c_3,c_1-c_2 \neq 1$ in the HLS model \cite{HLSRef}, 
one allows other anomalous contributions than $VVP$ to partial widths. Then, beside
the diagrams represented by Fig. (\ref{graphes}a), radiative decays of vector mesons
get also contributions from diagrams as shown in  Fig. (\ref{graphes}b). 
These contributions are proportional to
$1-c_3$  and are generated by the $AAP$ Lagrangian piece of our
Extended Model.  Correspondingly, the decay widths to 3 pions
get  their dominant contributions from diagrams generically represented by 
Fig. (\ref{graphes}c), but
also subleading  $VPPP$ contributions  proportional to $c_1-c_2-c_3$
from diagrams shown in Fig. (\ref{graphes}d)
 and $APPP$ contributions (see  Fig. (\ref{graphes}e)) proportional
to $1-3(c_1-c_2+c_3)/4$. 

\subsection{Radiative Decays Of Vector Mesons}
\indent \indent
As far as radiative decays are concerned, let us denote $g_{VP\gamma}$ the couplings
generated by the $VVP$ couplings and given in Appendix E of \cite{taupaper} where
$c_3=1$ was assumed. In our Extended HLS Model, they  become~:
\be
\begin{array}{ll}
G_{VP\gamma}=g_{VP\gamma}~c_3 ~~,&~~~V=K^{*0},K^{*\pm},\rho^\pm~~~.
\end{array}
\label{eqD1}
\ee
However, the couplings to ideal fields ($V_I = \rho_I,~\omg_I,~\phi_I$)
yield additional corrections~:
\be
\left \{
\begin{array}{ll}
\displaystyle
G_{V_I\pi^0\gamma}=g_{V_I \pi^0\gamma}~c_3  -\frac{eD}{3}
\frac{F_{V\gamma}(m_V^2)}{m_V^2}~~~,
~~D=-\frac{N_c e^2}{4 \pi^2 f_\pi} (1-c_3)\\[0.5cm]
\displaystyle
G_{V_I\eta^8\gamma}=g_{V_I \eta^8\gamma}~c_3  
-\frac{e~D}{9\sqrt{3}}\left[ \frac{5 z_A-2}{z_A}\right ]
\frac{F_{V\gamma}(m_V^2)}{m_V^2} \\[0.5cm]
\displaystyle
G_{V_I\eta^0\gamma}=g_{V_I \eta^0\gamma}~c_3  
-\frac{e~x~D}{9} \sqrt{\frac{2}{3}}\left[ \frac{5 z_A+1}{z_A}\right ]
\frac{F_{V\gamma}(m_V^2)} {m_V^2} 
\end{array}
\right .
\label{eqD2}
\ee

The dominant $VVP$ contributions are first order  in $g$ and are weighted by $c_3 \simeq 1$. 
The additional $AAP$ contributions are of order $e^2$ and are additionally suppressed 
by $(1-c_3)$. In the annihilation processes $e^+e^- \ra \gamma P$, the $AAP$ terms are
subleading and are actually  absorbed by the intermediate photon vacuum polarization 
in the $e^+e^-$ annihilation amplitude.
From a numerical point of view, these additional terms, all proportional to the $\gamma-V$
transition amplitudes $F_{V\gamma}(m_V^2)$, give a negligible contribution. 
These transitions amplitudes are given in Section \ref{Reminder2}. In order to
compute the radiative decays of vector mesons, ideal field combinations have to be constructed
\cite{taupaper}, as reminded in Section \ref{Reminder1}. Also, pseudoscalar 
singlet and octet couplings have to be combined in order to derive the $\eta$
and $\eta^\prime$ couplings \cite{taupaper}. The dependence upon the mixing 'angles' of vector 
mesons is  hidden inside the $g_{VP\gamma}$ and $F_{V\gamma}$ functions.

\subsection{Three Pion Decays of Vector Mesons}
\indent \indent The coupling constants for three pion decays of neutral vector 
mesons can be derived in close correspondence with the $\gamma^* \ra \pi^+ \pi^- \pi^0$
amplitude constructed in Section \ref{3pionNSK}. Let us define~:
\be
\displaystyle
\left \{
\begin{array}{ll}
\displaystyle
M(x,y,s) =\frac{3}{2} m^2 g c_3 N_2(x,y,s) -\frac{9}{4} g (c_1-c_2-c_3)\\[0.5cm]
\displaystyle
C=1 -\frac{3}{4} (c_1-c_2+c_3)\\[0.5cm]
\displaystyle
E=-\frac{N_c}{12 \pi^2 f_\pi^3}
\end{array}
\right .
\label{eqD3}
\ee
fully displaying the dependence upon the Kuraev--Siligadze \cite{Kuraev} variables.
The function $N_2(x,y,s)$ has been defined in Section \ref{3pionNSK} and $m^2=a g^2 f_\pi^2$.
Then, the amplitudes for $V \ra \pi^+ \pi^- \pi^0$ are, at leading order in the 
symmetry breaking parameters~:
\be
\displaystyle
\left \{
\begin{array}{ll}
\displaystyle
A_\omg(x,y) = E \left [
M(x,y,m_\omg^2) + e^2 C \frac{F_{\omg \gamma}(m_\omg^2)}{m_\omg^2}
\right ]\\[0.5cm]
\displaystyle
A_\phi(x,y) = E \left [\gamma(m_\phi^2) 
M(x,y,m_\phi^2) + e^2 C \frac{F_{\phi  \gamma}(m_\phi^2 )}{m_\phi^2}
\right ]\\[0.5cm]
\displaystyle
A_\rho(x,y) = E \left [\alpha(m_\rho^2) 
M(x,y,m_\rho^2) + \frac{3}{2} m^2 g c_3 N_3(x,y,m_\rho^2) +
e^2 C \frac{F_{\rho  \gamma}(m_\rho^2 )}{m_\rho^2}
\right ]
 \end{array}
\right .
\label{eqD4}
\ee
where  $N_3(x,y,s)$ has also been defined in  Section \ref{3pionNSK} and hides a 
dependence upon $\alpha(s_{+-})$ ($s_{+-}$ being the $\pi^+ \pi^-$ squared
invariant mass of the decay products).  One may observe that the dominant terms
for $\rho$ and $\phi$ are generated by vector meson mixing and that
there is a specific term in the $\rho$ amplitude compared to those
for $\omg$ and $\phi$. One should also note
that the $APPP$ coupling generates a subleading term (of order $e^2$)
which exists  even if there were no vector meson mixing. 
The  present model is more complicated than the one developped by \cite{Kuraev}
because of these $APPP$ terms and of the vector meson mixing.

Finally, in terms of the amplitudes given in Eqs. (\ref{eqD4}), the 3--pion partial 
widths are given by \cite{Kuraev}~:

\be
\displaystyle
\Gamma(V\ra \pi \pi \pi)= \frac{m_V^7}{768 \pi^3} \int \int dx~ dy~ G(x,y) |A_V(x,y)|^2
\label{eqD5}
\ee
where the function $G(x,y)$  \cite{Kuraev} has been reminded in Appendix \ref{threeBodies}.

\subsection{Effects Of A ${\cal O}(p^4)$ Lagrangian Piece}
\indent \indent In  \cite{Harada2000}, Harada and Yamawaki have studied
the Wilsonian matching of the HLS model with QCD. For this purpose they
have identified 35  ${\cal O}(p^4)$  Lagrangian pieces
of the HLS Lagrangian which are provided explicitly  in  \cite{HLSRef}. Among these, one is of
special interest for our purpose, the so--called $z_3$ term which is of concern for the
$\gamma-V$ and $W-V$ transitions. Discarding effects of SU(3) symmetry breaking, the 
additional Lagrangian piece writes~:
\be
\left \{
\begin{array}{ll}
\displaystyle {\cal L}_4 =  {\cal  L}_4^\gamma  + {\cal L} _4^{W^+} + {\cal L} _4^{W^-} \\[0.5cm]
\displaystyle {\cal  L}_4^\gamma=2 z_3 g e ~\partial_\mu A_\nu \left[
(\partial_\mu \rho^I_\nu -\partial_\nu \rho^I_\mu)
+\frac{1}{3}  (\partial_\mu \omg^I_\nu -\partial_\nu \omg^I_\mu)
-\frac{\sqrt{2}}{3}(\partial_\mu \phi^I_\nu -\partial_\nu \phi^I_\mu)
\right] +\cdots \\[0.5cm]
\displaystyle {\cal  L}_4^{W^+}  = z_3 g g_2 V_{ud} ~\partial_\mu W^+_\nu 
(\partial_\mu \rho^-_\nu -\partial_\nu \rho^-_\mu)  +\cdots \\[0.5cm]
\displaystyle {\cal  L}_4^{W^-}  = z_3 g g_2 \overline{V}_{ud} ~\partial_\mu W^-_\nu 
(\partial_\mu \rho^+_\nu -\partial_\nu \rho^+_\mu)  +\cdots  
 \end{array}
\right .
\label{eqD6}
\ee
using notations already defined in \cite{taupaper} and reminded above.
This turns out to modify in exactly the same way the transition amplitudes for
$\gamma-V$ and $W-V$ by adding a $s$--dependent term to the constant part of the couplings.
More precisely, one gets (see Eqs. (30) in \cite{taupaper} for $\tau$ decays 
and Section \ref{Reminder2} above for $e^+ e^-$ annihilations)~:
\be
\left \{
\begin{array}{ll}
\displaystyle 
f_\rho^\tau \Longrightarrow f_\rho^\tau \left ( 1-\frac{z_3}{a f_\pi^2} s \right )\\[0.5cm]
\displaystyle 
f_{V\gamma} \Longrightarrow f_{V\gamma}  \left ( 1-\frac{z_3}{a f_\pi^2} s \right)
 \end{array}
\right .
\label{eqD7}
\ee
where $z_3$ can be related with the Low Energy Constant (LEC) $L_9$ of the Chiral Perturbation
Theory. 
When  on mass shell, the additional factor is simply $(1-g^2 z_3)$ and can be guessed \cite{HLSRef} 
of the order 1.10. Indeed, even if $z_3$ is expected small, the $g^2$ term produces
an important enhancement factor.

 However, in processes like $e^+e^-$ annihilations or $\tau$ decay, $f_\rho^\tau$ or
$f_{V\gamma}$ always come in combination with the loop dressing functions \cite{taupaper} $\Pi_W(s)$ or  
$\Pi_{V\gamma}(s)$ which contain subtraction polynomials to be fitted. Therefore, the $z_3$
contribution comes in entangled with the first degree term of the subtraction polynomial
and cannot be singled out. As a consequence, the fit value
for $g$  may absorb effects of the $z_3$ correction and may produce too low values
by as much as  10\% for the amplitudes \cite{HLSRef}. 

\vspace{0.7cm}
\indent \indent
In the $e^+ e^-$ annihilation channels considered in the present work,
our fit procedure 
is sensitive to the product  $F_{V\gamma} \times g$  for each resonance rather  
than to the $F_{V\gamma}$'s and $g$ separately \footnote{In our full data set, only the
radiative decays of the $K^{*0},~K^{*\pm} ,~\rho^\pm$ do not belong to
this category, but are not sufficient to modify the picture.}.
 The fit quality we have reached  allows us to consider that these products are well
understood. Therefore, all pieces of information relying on these products can be considered
reliable. This covers  all products of widths like $\Gamma(V \ra P \gamma) \Gamma(V\ra e^+ e^-)$ or 
$\Gamma(V \ra \pi^+ \pi^- \pi^0) \Gamma(V\ra e^+ e^-)$. Ratios of widths performed 
from the  $\Gamma(V \ra P \gamma)$'s and  $\Gamma(V \ra \pi^+ \pi^- \pi^0)$'s are also
free from the disease mentioned above and can be considered as secure. It is the reason why  we 
will not go beyond these products and ratios until a satisfactory solution 
to the issue just mentioned is found.

Actually, some possibilities exist to solve this ambiguity, at least in principle. Indeed, processes like
$e^+ e^- \ra \mu^+ \mu^-$ are sensitive to the $[F_{V\gamma}]^2$'s and could be used.
Some scarce (accurate) data (3 measurement points) at the $\phi$ mass are reported 
\cite{KLOE_lept} but, unfortunately,  nothing with the required accuracy is
reported  around the $\omg$ mass. Another possibility could be to examine channels like
$e^+ e^- \ra \omg(\ra \pi^0 \gamma) \pi^0$  which are rather sensitive to some of the 
$F_{V\gamma} \times g^2$'s. Finally, one could also fix the leptonic widths as
coming from the fits of the Particle Data Group \cite{RPP2008}, which basically corresponds
to our results published in \cite{taupaper}. This last solution, even if legitimate, is not
completely satisfactory. 

It should be stressed that the effect mentioned is actually common to any parametrization 
of $e^+e^-$ and $\tau$ data and is by no means specific to the HLS Model (see, for instance,
the discussion in \cite{Kuraev}). On the other hand, reporting on products and ratios is the
standard way experimental groups proceed with the kind of data we are dealing with and we may compare
to these.

\begin{table}[!htb]
\hspace{0.cm}
\begin{tabular}{|| c  || c  |  c  ||}
\hline
Data\hhhu  &  Our  Fit &  PDG 2008 \\
\hline
\hline
\hhhv
$\Gamma^\prime(\rho^0 \ra \pi \pi)$ $\times 10^5$ & $4.72 \pm 0.02$ & $4.876 \pm 0.023 \pm 0.064$  \\
\hline 
\hhhv
$\Gamma^\prime(\omg \ra \pi \pi)$ $ \times 10^6$  & $1.146 \pm 0.057$ & $1.225 \pm 0.058\pm 0.041$ \\
\hline
\hhhv
$\Gamma^\prime(\rho^0 \ra \pi^0 \gamma)$ $ \times 10^8$   & $1.875 \pm 0.026$   & $2.8 \pm 0.4$ \\
\hline
\hhhv
$\Gamma^\prime(\omg \ra \pi^0 \gamma)$ $ \times 10^6$   & $6.80 \pm 0.13$   & $6.39 \pm 0.15$~~ **\\
\hline
\hhhv
$\Gamma^\prime(\phi \ra \pi^0 \gamma)$ $ \times 10^7$   & $4.29 \pm 0.11$   & $3.75 \pm 0.18$ \\
\hline
\hhhv
$\Gamma^\prime(\rho^0 \ra \eta \gamma)$ $ \times 10^8$   & $1.05 \pm 0.02$   & $1.42 \pm 0.10$ \\
\hline
\hhhv
$\Gamma^\prime(\omg \ra \eta \gamma)$ $ \times 10^8$   & $4.50 \pm 0.10$   & $3.31 \pm 0.28$~~ **\\
\hline
\hhhv
$\Gamma^\prime(\phi \ra \eta \gamma)$ $ \times 10^6$   & $4.19 \pm 0.06$   & $3.87 \pm 0.07$~~ **\\
\hline
\hhhv
$\Gamma^\prime(\rho^0 \ra \pi \pi \pi)$ $ \times 10^{10}$ & $9.03 \pm 0.76$ & $45.8^{+24.6}_{-16.4} \pm 15.6 $ ~~ **\\
\hline 
\hhhv
$\Gamma^\prime(\omg \ra \pi \pi \pi)$ $ \times 10^5$  & $6.20 \pm 0.13$ & $6.39  \pm 0.10$ \\
\hline
\hhhv
$\Gamma^\prime(\phi \ra \pi \pi \pi)$  $ \times 10^5$ & $4.38 \pm 0.12$& $ 4.53 \pm 0.10$\\
\hline
 \end{tabular}
\caption{
\label{T7} Results for $\Gamma^\prime(V \ra f) \equiv \Gamma(V \ra e^+ e^-) \Gamma(V \ra f)/ \Gamma^2_{tot} $ for each 
each $V$ vector meson and each final state $f$. The most significant results are flagged by 
  $**$.}
\end{table}

\subsection{Results For Partial Widths}
\indent \indent
In Table \ref{T7}, we give our fit results for~~:
$$\Gamma^\prime(V \ra f) \equiv \Gamma(V \ra e^+ e^-) \Gamma(V \ra f)/ \Gamma^2_{tot} $$
and provide the measurements collected in the RPP \cite{RPP2008}. The results on which we report 
have been derived from a fit to all  $e^+e^-$ annihilation data, except for KLOE data \cite{KLOE},
by fixing the scale corrections to zero in accordance with Section \ref{stat}.

One thus notes that, for most of
the reported products, one gets results in accordance with what is extracted from the same data.
One should remember that the errors we quote fold in systematic and statistical errors 
and can be compared to the averages performed by the Particle Data Group.

Concerning $\rho^0 \ra \pi \pi \pi$, our result, mostly determined by CMD--2 data, is in agreement
with the SND result, but with a much lower central value and a much better accuracy.
Significant differences  mostly concerning  $\omg/\phi \ra \eta \gamma$ can be read off
Table \ref{T7}. They correspond to performing a simultaneous fit of all the involved data
which happens  to be of very high quality as shown in Section \ref{3pionNSK} above.
Finally, as a general statement, our uncertainties reported in Table \ref{T7} are either
of magnitude comparable to the existing information or  of much better accuracy. 
 
\begin{table}[!htb]
\hspace{0.cm}
\begin{tabular}{|| c  || c  |  c  ||}
\hline
Data\hhhu  &  Our  Fit &  PDG 2008 \\
\hline
\hline
\hhhv
$\Gamma(\rho^0 \ra \pi \pi)/\Gamma(\rho^0 \ra \pi \pi \pi)$  & $(5.26 \pm 0.45)~10^4$ & $> 100$  \\
\hline
\hhhv
 $\Gamma(\omg \ra \pi \pi)/\Gamma(\omg \ra \pi \pi \pi)$ & $ (18.5 \pm 1.0)~10^{-3}$ & $(17.2 \pm 1.4)~10^{-3}$  \\
\hline
\hline
\hhhv
 $\Gamma(\rho^0 \ra \pi \gamma)/\Gamma(\rho^0 \ra \pi \pi \pi)$ & $ 20.91 \pm 1.60$
  & $[{\bf 5.94 \pm 3.36}]$~** \\
\hline
\hhhv
 $\Gamma(\omg \ra \pi \gamma)/\Gamma(\omg \ra \pi \pi \pi)$ & $ 0.110 \pm 0.003 $ & $0.0999 \pm 0.0026$ ~** \\
\hline
\hhhv
 $\Gamma(\phi \ra \pi \gamma)/\Gamma(\phi \ra \pi \pi \pi)$ & $ (9.80 \pm 0.29)~10^{-3} $ 
 & $({\bf 8.3 \pm 0.6})~10^{-3}$  \\
\hline
\hline
\hhhv
 $\Gamma(\rho^0 \ra \pi \pi)/\Gamma(\rho^0 \ra \pi^0 \gamma)$ & $ (2.52 \pm 0.03)~10^3$ 
 & $({\bf 1.74 \pm 0.27 )~10^3}$\\
\hline
\hhhv
$\Gamma(\omg \ra \pi \pi)/\Gamma(\omg \ra \pi^0 \gamma)$  & $ 0.169 \pm 0.009 $ & $0.20 \pm 0.04$  \\
\hline
\hline
\hhhv
 $\Gamma(\rho^0 \ra \eta \gamma)/\Gamma(\rho^0 \ra \pi^0 \gamma)$ & $ 0.561 \pm 0.008$ & 
  $({\bf 0.50 \pm 0.08 )}$ \\
\hline
\hhhv
$\Gamma(\omg \ra \eta \gamma)/\Gamma(\omg \ra \pi^0 \gamma)$  & $ (6.62 \pm 0.17)~10^{-3} $ & $(9.8 \pm 2.4) ~10^{-3}$   ~**\\
\hline
\hhhv
$\Gamma(\phi \ra \eta \gamma)/\Gamma(\phi \ra \pi^0 \gamma)$  & $ 9.77 \pm 0.20$ & $10.9 \pm 0.3\pm 0.7$  \\
\hline
\hline
 \end{tabular}
\caption{
\label{T8} Ratios of Widths. In the second data column, number written in plain style 
are directly extracted from the Review of Particle Properties, the boldface ones
are derived from making the ratios of the accepted branching ratios \cite{RPP2008}.
The most significant results are flagged by   $**$.}
\end{table}

 As our fit procedure is global, it also allows to relate different decay modes of the same vector meson.
 These are provided in Table \ref{T8} with information extracted from the most recent RPP \cite{RPP2008}.
 Taking into account the fit quality of the corresponding cross sections, the difference
 with RPP expectations should be considered significant.
  
  One should first remark the ratio $\Gamma(\rho^0 \ra \pi \gamma)/\Gamma(\rho^0 \ra \pi \pi \pi)$
  which is found $\simeq 3.5$ larger than expected from the ratio of branching ratios as
  given in \cite{RPP2008}, with a significance of $\simeq 4 \sigma$.  Less impressive but,
nevertheless, significant is the corresponding ratio for the $\omg$ meson. 

More interesting, however, is the ratio  
 $\Gamma(\omg \ra \eta \gamma)/\Gamma(\omg \ra \pi^0 \gamma)$ which is found much more
 precise than  (old) existing measurements and benefits from being derived by means of
  a simultaneous fit to all recent relevant data sets.

\section{Conclusion}
\label{conclude}
\indent \indent
The model developped in \cite{taupaper} succeeded in accounting for 17 decay modes
of vector mesons and in providing a good simultaneous description of the 
$e^+ e^- \ra \pi^+\pi^-$ cross section and of the dipion spectrum in $\tau$ decay.
The present paper has shown that it can be easily extended in order to provide
also a successful description of the $e^+ e^- \ra \pi^0\gamma$,
$e^+ e^- \ra \eta\gamma$ and $e^+ e^- \ra \pi^0 \pi^+\pi^-$ cross sections
and account precisely for the properties of the $\eta/\eta^\prime \ra \pi^+ \pi^- \gamma$
decays. Its application to $e^+ e^- \ra K \overline{K}$ decay mode is postponed
to another study, because of the problem stated by \cite{BGPter}. 

This extension
has been performed at the expense of only 2 more free parameters $c_1-c_2$ and
$c_3$ which describe the amount by which the pure VMD assumption is violated.
The precise numerical value of these are found close to the guess of Harada and 
Yamawaki \cite{HLSRef} and amount to a small violation of the full VMD assumption
in the Triangle Anomaly Sector and to a large violation for the Box Anomalies.
On the other hand,
we did not find any clear and unambiguous evidence for a high mass vector 
meson influence in $e^+ e^-$  annihilations up to $\simeq 1.05$ GeV.

We have also shown that all experimental data, except for a data set on  
$e^+ e^- \ra \pi^0 \pi^+\pi^-$ annihilation, are successfully described
by a global fit within a unified (HLS) framework. This is clearly more
constraining than only accounting for partial decay widths.
Systematic uncertainties, especially correlated errors, have been 
carefully studied because of their importance in numerical estimation
of some physics parameters. It has thus been shown that all data samples 
have already the expected absolute scale, except for the high statistics
two-pion data sample from KLOE which has to be rescaled  according to expectations.
Then, all $e^+ e^-$ data samples exhibit a good consistency with  each other,
including the KLOE  sample which only marginally degrades the global fit quality.

In summary the Extended HLS Model, supplied with the vector meson mixing
mechanism defined in \cite{taupaper} provides a successful simultaneous 
description of all the data falling inside its scope. One should note
that this model allows for the first simultaneous fit to all
the low energy data not affected by scalar mesons or higher 
mass vector mesons.

The last part of this study was devoted to results on partial widths of
vector mesons as derived from our global fit. We have argued on the difficulty
to firmly assess partial width values of vector mesons to some final state $f$
relying only on $e^+e^-$ annihilation to this final state.  We have 
preferred limiting ourselves to providing our estimations for products
like $\Gamma(V \ra r^+e^-) \times \Gamma(V \ra f)$ or ratios of the
form $\Gamma(V \ra f_1)/\Gamma(V \ra f_2)$ where the uncertainties
in estimating $\Gamma(V \ra e^+e^-)$ cancel out. Taking into account
the quality of our fits and the additional information provided by
the physics underlying various processes, we  believe our results are
reliable and could supersede several older reported results.

\section*{Acknowledgements}
\indent \indent
We gratefully acknowledge, G. Venanzoni, Laboratori Nazionali di Frascati,
Italy, for having provided us with the KLOE data and errors, for
helpful correspondance on their handling and for useful comments on the 
manuscript. We also acknowledge S. Eidelman, Budker Institute, Novosibirsk,
for important information concerning CMD2 and CMD data.

\section*{\Large{Appendices}}
\appendix
\section{The Non--Anomalous HLS Lagrangian}
\label{HLSLagrangian}

\indent \indent
We remind \cite{taupaper} in this Appendix the HLS Lagrangian piece describing 
the photon sector (traditional VMD)~:

\be
\begin{array}{ll}
{\cal L}_{VMD} &=  \displaystyle ie (1-\frac{a}{2})  A \cdot \pi^- \parsym \pi^+
+ \displaystyle i\frac{e}{z_A} (z_A-\frac{a}{2} -b) A \cdot K^- \parsym K^+
+ \displaystyle i\frac{e}{z_A} b A \cdot K^0 \parsym  \overline{K}^0
 \\[0.5cm]
\hspace{-3.cm} ~&\displaystyle 
+\frac{ia g}{2} \rho^0_I \cdot \pi^- \parsym \pi^+ + \displaystyle \frac{ia g}{4 z_A}
(\rho^0_I + \omg_I -\sqrt{2}  z_V \phi_I ) K^- \parsym K^+ 
+\displaystyle \frac{ia g}{4z_A}
(\rho^0_I - \omg_I +\sqrt{2}  z_V \phi_I ) K^0 \parsym \overline{K}^0 \\[0.5cm]
\hspace{-3.cm} ~& \displaystyle -e a g f^2_\pi \left[
\displaystyle \rho^0_I + \frac{1}{3} \omg_I - \frac{\sqrt{2}}{3} z_V \phi_I
\right] \cdot A + 
\displaystyle \frac{1}{9} a f^2_\pi e^2 (5+z_V) A^2 
+  \frac{a f^2_\pi g^2 }{2}
\left[\displaystyle (\rho^0_I)^2 +\omg^2_I + z_V \phi^2_I \right]
\end{array}
\label{Lag_ee}
\ee

The parameter   $g$ is the traditional universal vector meson coupling constant~;
the parameter  $a$ specific of the  HLS model, is expected equal to 2 in standard VMD approaches
but   rather fitted  to $a \simeq 2.3\div 2.5$ \cite{ffVeryOld,ffOld,taupaper}.
The parameter $b$ in Eq. (\ref{Lag_ee}) is $b=a(z_V-1)/6$ where $z_V$ is  the SU(3) 
breaking parameter of the ${\cal L}_V$ part of the HLS Lagrangian, while 
$z_A=[f_K/f_\pi]^2=1.495 \pm 0.031$  \cite{RPP2006} is the SU(3) breaking parameter of 
its ${\cal L}_A$ part \cite{HLSOrigin,HLSRef}. 

A subscript $I$   on the fields, standing for ``ideal'',
affects the neutral vector meson fields. It indicates that the corresponding
fields occuring in the Lagrangian are not the physical fields.

On the other hand, the part of the HLS model involved in $\tau$ decays writes \cite{taupaper}~:
\be
\begin{array}{ll}
{\cal L}_{\tau}& =- \displaystyle \frac{i g_2}{2}  V_{ud} W^+\cdot
\left[ (1-\frac{a}{2}) \pi^- \parsym \pi^0 + 
\displaystyle (z_A-\frac{a}{2}) \frac{1}{z_A \sqrt{2}} K^0 \parsym K^-
\right]\\[0.5cm]
\hspace{-3.cm} ~&  -  \displaystyle  \frac{af_\pi^2 g g_2}{2} V_{ud} W^+ \cdot \rho^- 
-\displaystyle \frac{iag}{2} \rho^- 
\left[ \pi^0 \parsym \pi^+ -\displaystyle \frac{1}{z_A \sqrt{2}} \overline{K}^0 \parsym K^+
\right]\\[0.5cm]
\hspace{-1.cm} ~& \displaystyle + f^2_\pi g_2^2 \left \{ \frac{1+a}{4} 
\left[\displaystyle z_A|V_{us}|^2 + |V_{ud}|^2\right] +
\frac{a}{4}[\sqrt{z_V}-z_A] |V_{us}|^2  \right \} W^+ \cdot W^-
\displaystyle  + a f^2_\pi g^2 \rho^+\rho^- 
\end{array}
\label{Lag_tau}
\ee
\noindent plus the conjugate of the interaction term (the $W^-$ term, not displayed). 
 This Lagrangian piece depends  on the CKM matrix element
$V_{ud}= 0.97377 \pm 0.00027$ \cite{RPP2006}, on $g_2$ (fixed by
its relation with the Fermi constant)~:
\be
g_2 = 2 m_W ~ \sqrt{G_F \sqrt{2}}~~~~~,
\label{eqA3}
\ee
on the universal coupling $g$ and on the breaking
parameters $z_A$  and $z_V$ already defined. One should note, balancing the photon mass
term in ${\cal L}_{VMD}$, a small mass term complementing the $W$ mass of the Standard Model
which could be removed by appropriate field redefinitions.
At the $\tau$ lepton mass scale one has \cite{RPP2006}~:
 $$ g_2=0.629 ~~~~ ({\rm and}~~~~ e=0.30286)~~.$$

\section{Effective Lagrangian Piece for $e^+e^- \ra P \gamma$}
\label{HLSPgamma}

\indent \indent 
The VVP Lagrangian piece relevant for the  $P \gamma$ final state 
can be derived from the first Lagrangian in Eqs (\ref{eq4})~;
it is given in terms of ideal fields by~:
\be
\begin{array}{llll}
{\cal L}_{VVP} = & \displaystyle
B \epsilon^{\mu \nu \alpha \beta} H_{\mu \nu \alpha \beta}
\hspace{3cm} \left ( B=-\frac{N_c g^2~c_3}{8 \pi^2 f_\pi} \right )\\[0.5cm]
H_{\mu \nu \alpha \beta} = & \displaystyle
(\partial_\mu \rho_{I\nu} \partial_\alpha \rho_{I\beta}
+\partial_\mu \omg_{I\nu}\partial_\alpha \omg_{I\beta}) \left[
\frac{\eta_8}{2 \sqrt{3}} +x\frac{\eta_0}{\sqrt{6}}
\right] + \frac{z_W z_T^2}{z_A}
\partial_\mu \phi_{I\nu}\partial_\alpha \phi_{I\beta}
\left[\frac{-\eta_8}{\sqrt{3}} +x\frac{\eta_0}{\sqrt{6}}\right]\\[0.5cm]
~~& + \partial_\mu \rho_{I\nu} \partial_\alpha \omg_{I\beta} \pi^0
\end{array}
\label{eqB1}
\ee
where the pseudoscalar  fields are still the bare fields.

As stated in the main text, the condition  $z_W z_T^2=1$ is requested. Using this 
piece and Eq. (\ref{Lag_ee}), one can define  an equivalent effective term which allows
for a simpler derivation of the $VP\gamma$ couplings~: 
\be
{\cal L}_{AVP}^\prime = \displaystyle B^\prime \epsilon^{\mu \nu \alpha \beta} 
{\rm Tr} \left[
\partial_\mu(eQ A_\nu+g c_3 V_\nu)\partial_\alpha(eQ A_\beta+g c_3 V_\beta)
X_A^{-1/2} (P_8+xP_0)X_A^{-1/2}
\right]
\label{eqB2}
\ee
where now the pseudoscalar fields  are the renormalized ones, $Q$ is the 
quark charge matrix, $A$ is the electromagnetic field and $B^\prime$ is given
below. The effective piece of interest can further be written~:
\be
\left \{
\begin{array}{llll}
{\cal L}_{AVP}^\prime = &\displaystyle B^\prime \epsilon^{\mu \nu \alpha \beta} 
\partial_\mu A_\nu H_{\alpha \beta} 
\hspace{1.5cm} \left ( {\rm with}~~~
B^\prime =-\frac{N_c e g c_3}{4 \pi^2 f_\pi} \right )\\[0.5cm]
 H_{\alpha \beta }= & \displaystyle 
 \partial_\alpha \rho^I_\beta  \left[
 \frac{\pi^0}{6} + \frac{\eta_8}{2\sqrt{3}} + \frac{x \eta_0}{ \sqrt{6}}
 \right] +
  \partial_\alpha \omg^I_\beta  ~\left[
 \frac{\pi^0}{2} + \frac{\eta_8}{6\sqrt{3}} + \frac{x \eta_0}{ 3\sqrt{6}}
 \right] \\[0.5cm]
  ~~~& +\displaystyle  \partial_\alpha \phi^I_\beta \frac{1}{3 z_A} 
 \left[   
\sqrt{\frac{2}{3}} \eta_8 - \frac{x  }{\sqrt{3}}\eta_0 
 \right] 
\end{array}
\right .
\label{eqB3}
\ee
in terms of renormalized ideal and pseudoscalar fields, renormalized
after applying the SU(3)/U(3) breaking mechanism.
One can now replace the ideal vector fields by their physical partners 
using Eqs. (\ref{eq22}) and gets~:
\be
\displaystyle  {\cal L}_{AVP}^\prime = B^\prime
\sum_{i,j}  ~~ H_{V_i}^{P_j} ~~ P_j~~  \epsilon^{\mu \nu \alpha \beta} 
 \partial_\mu A_\nu  \partial_\alpha V_{i ~\beta }
\label{eqB4}
\ee
where the sum extends over the physical (neutral) vector fields and 
the neutral pseudoscalar mesons. We have defined ($\Lambda=\sqrt{2}/z_A$)~:
\be
\left\{
\begin{array}{llll}
H_\rho^{\pi^0}= \displaystyle \frac{1+ 3 \alpha}{6}~~,~~&
H_\omg^{\pi^0}= \displaystyle \frac{3 - \alpha}{6}~~,~~&
H_\phi^{\pi^0}= \displaystyle \frac{\beta+ 3\gamma}{6}\\[0.5cm]
H_\rho^{\eta_8}= \displaystyle \frac{3 + \alpha
-2 \Lambda \beta}{6 \sqrt{3}}~~,~~&
H_\omg^{\eta_8}= \displaystyle \frac{1-3 \alpha
-2 \Lambda \gamma}{6 \sqrt{3}}~~,~~&
H_\phi^{\eta_8}= \displaystyle \frac{\gamma + 3 \beta
+2 \Lambda}{6 \sqrt{3}}\\[0.5cm]
H_\rho^{\eta_0}= x\displaystyle \frac{3 + \alpha
+\Lambda \beta}{3 \sqrt{6}}~~,~~&
H_\omg^{\eta_0}= x\displaystyle \frac{1 - 3\alpha
+\Lambda \gamma}{3 \sqrt{6}}~~,~~&
H_\phi^{\eta_0}= x\displaystyle \frac{3 \beta + \gamma
-\Lambda}{3 \sqrt{6}}~
\end{array}
\right.
\label{eqB5}
\ee
where the dependence of the "angles" upon $s$ has been omitted. 

The other Lagrangian piece contributing to $e^+ e^- \ra P \gamma$ 
annihilations is (see Eqs. (\ref{eq4})~:

\be
\left\{
\begin{array}{llll}
{\cal L}_{AAP}= \displaystyle B^{\prime \prime} \epsilon^{\mu \nu \alpha \beta}
\partial_\mu A_\nu \partial_\alpha A_\beta~~K ~~~, 
\hspace{1cm}  \displaystyle   \left(B^{\prime \prime}
=-\frac{e^2N_c}{4 \pi^2 f_\pi} (1-c_3) \right )\\[0.5cm]
K= \displaystyle
\left[ 
\frac{\pi^0}{6} + \frac{5 z_A -2}{3 z_A} \frac{\eta_8}{6\sqrt{3}}
+\frac{5 z_A +1}{3 z_A} \frac{x}{3} \frac{\eta_0}{\sqrt{6}}
\right]
\end{array}
\right.
\label{eqB6}
\ee
which provides the constant contributions to the amplitudes given by Eqs. (\ref{eq34}).

\section{Integrated Cross Section For $e^+ e^- \ra \pi^+ \pi^- \pi^0$}
\label{threeBodies}
\indent \indent
Using the parametrization of Kuraev and Siligadze \cite{Kuraev}, 
the differential cross section is written~:
\be
\displaystyle
\frac{d^2 \sigma(e^+e^- \ra \pi^+ \pi^- \pi^0)}{dx ~dy}=
\frac{\alpha_{em}}{192 \pi^2} s^2 G(x,y) |T_{sym}+T_\rho|^2
\label{eqC1}
\ee
where~:
 \be
 \displaystyle
G(x,y)=  4 
(x^2-\frac{m_\pi^2}{s})(y^2-\frac{m_\pi^2}{s})
-\left (1 -2x -2y +2 x y + \frac{2m_{\pi}^2 - m_0^2}{s}
\right)^2
 \label{eqC2}
\ee

The variables $x$ and $y$ are defined by ($m_0=m_{\pi^0}$,
$m_\pi=m_{\pi^\pm}$) by~:
 \be
 \left \{
\begin{array}{lll}
\displaystyle
s_{+-}=& s(2x+2y-1)+m_0^2\\[0.5cm]
\displaystyle
s_{+0}=& s(1-2y )+m_{\pi}^2\\[0.5cm]
\displaystyle
s_{-0}=& s(1-2x)+m_{\pi}^2
 \end{array}
 \right .
 \label{eqC3}
\ee
The integrated cross section is~:
\be
\displaystyle
\sigma(e^+e^- \ra \pi^+ \pi^- \pi^0)=
\int_{x_{min}}^{x_{max}} dx \int_{y_{min}}^{y_{max}} dy 
\frac{d^2 \sigma(e^+e^- \ra \pi^+ \pi^- \pi^0)}{dx ~dy}
\label{eqC4}
\ee
with~:
\be
 \left \{
\begin{array}{lll}
\displaystyle
x_{min}=\frac{m_\pi}{\sqrt{s}}~~~~,
~~ x_{max}=\frac{1}{2} \left(
1 -\frac{m_0(2 m_\pi+m_0)}{s} \right )
\\[0.5cm]
\displaystyle
y_{min/max}= \frac{1}{2(1-2x +x_{min}^2)}
\left\{
(1-x) (1 -2x +\frac{(2 m_\pi^2-m_0^2)}{s}) \right . \\[0.5cm]
\displaystyle
\left.
{\cal{+/-}} \left[
(x^2-\frac{m_\pi^2}{s}) (1 -2x+\frac{m_0(2 m_\pi-m_0)}{s})
\left(
1 -2x - \frac{m_0(2m_\pi+m_0)}{s}
\right)
\right]^{1/2}
\right\}
 \end{array}
 \right .
 \label{eqC5}
\ee

                     \bibliographystyle{h-physrev}
                     \bibliography{vmd}

\newpage
\begin{figure}[!ht]
\begin{minipage}{\textwidth}
\begin{center}
\resizebox{\textwidth}{!}
{\includegraphics*{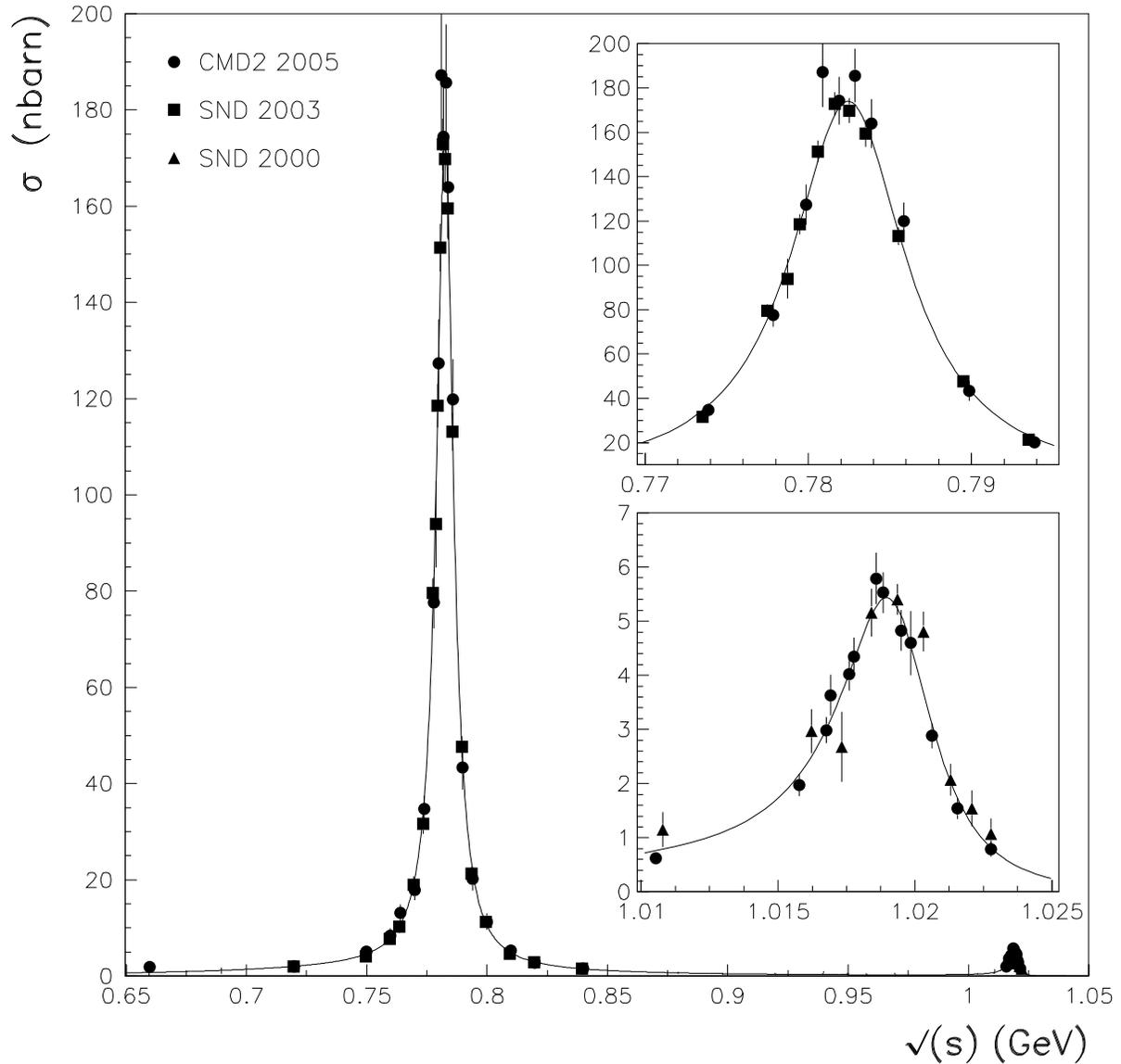}}
\end{center}
\end{minipage}
\begin{center}
\vspace{-0.3cm}
\caption{\label{Fig:pi0g}
Annihilation process $e^+e^- \ra \pi^0 \gamma$. 
The errors plotted combine the reported systematic and
statistical errors in quadrature. The insets magnify
the top part of the $\omg$ region on the one hand and the $\phi$ region 
on the other hand. "CMD--2 2005" refers to \cite{CMD2Pg2005},
"SND 2003" to \cite{sndPg2003} and "SND 2000" to  \cite{sndPg2000}.}
\end{center}
\end{figure}

\begin{figure}[!ht]
\begin{minipage}{\textwidth}
\begin{center}
\resizebox{\textwidth}{!}
{\includegraphics*{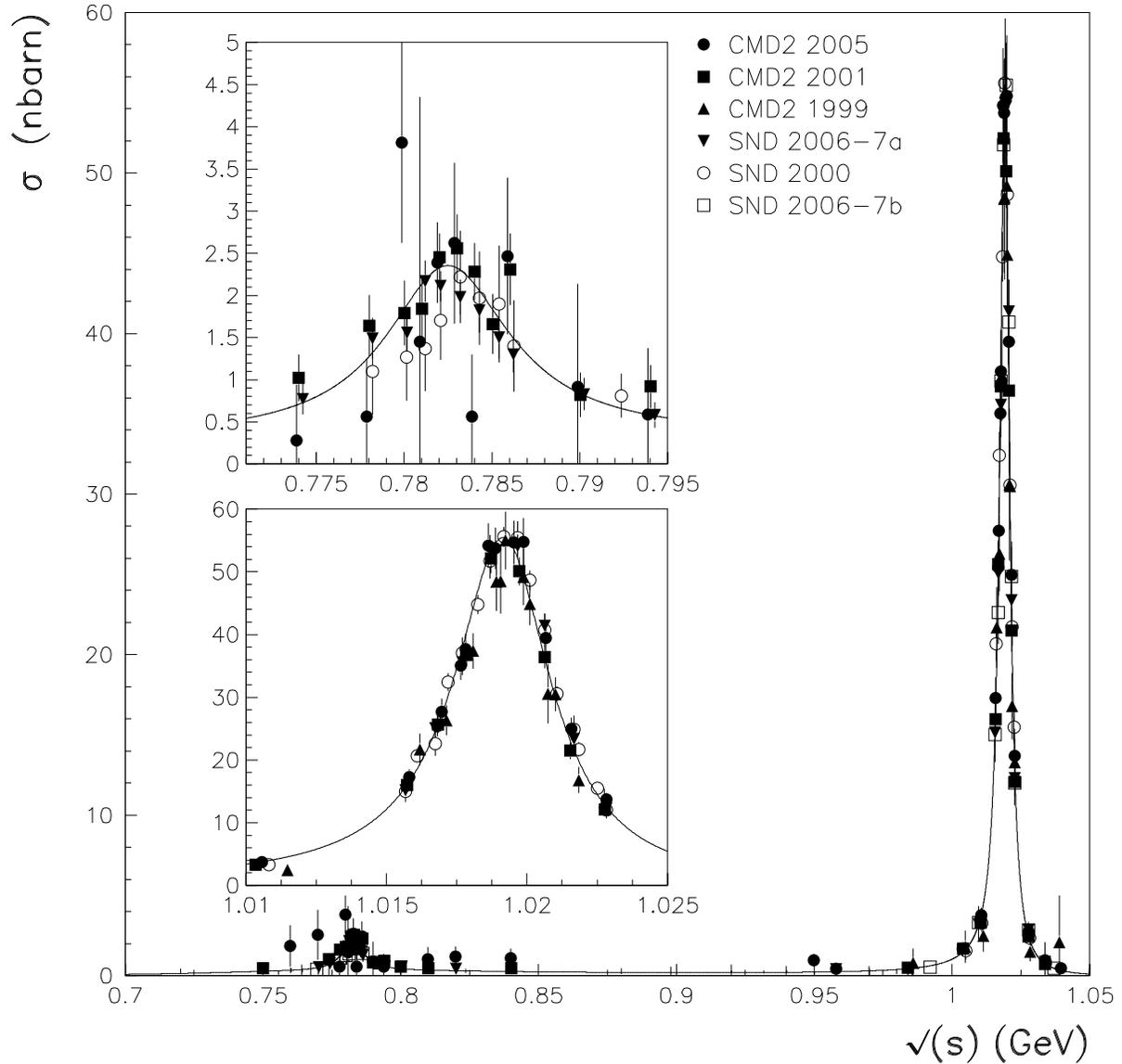}}
\end{center}
\end{minipage}
\begin{center}
\vspace{-0.3cm}
\caption{\label{Fig:etag}
Annihilation process $e^+e^- \ra \eta \gamma$. 
The errors plotted combine the reported systematic and
statistical errors in quadrature. The insets magnify
the  $\omg$ region and the $\phi$ region. 
 "CMD--2 2005" refers to \cite{CMD2Pg2005}
(with $\eta \ra 2 \gamma$), .
"CMD--2 2001"   to \cite{CMD2Pg2001} (with $\eta \ra 3 \pi^0$),
"CMD--2 1999"   to \cite{CMD2Pg1999} (with $\eta \ra \pi^+ \pi^- \pi^0$),
"SND 2006--7a/b" refer to the two data sets published in \cite{sndPg2007}
(with $\eta \ra 3 \pi^0$ and $\eta \ra \pi^+ \pi^- \pi^0$)
and "SND 2000" to the data set published in \cite{sndPg2000}
(with $\eta \ra 2 \gamma$).
 }
\end{center}
\end{figure}

\begin{figure}[!ht]
\begin{minipage}{\textwidth}
\begin{center}
\resizebox{\textwidth}{!}
{\includegraphics*{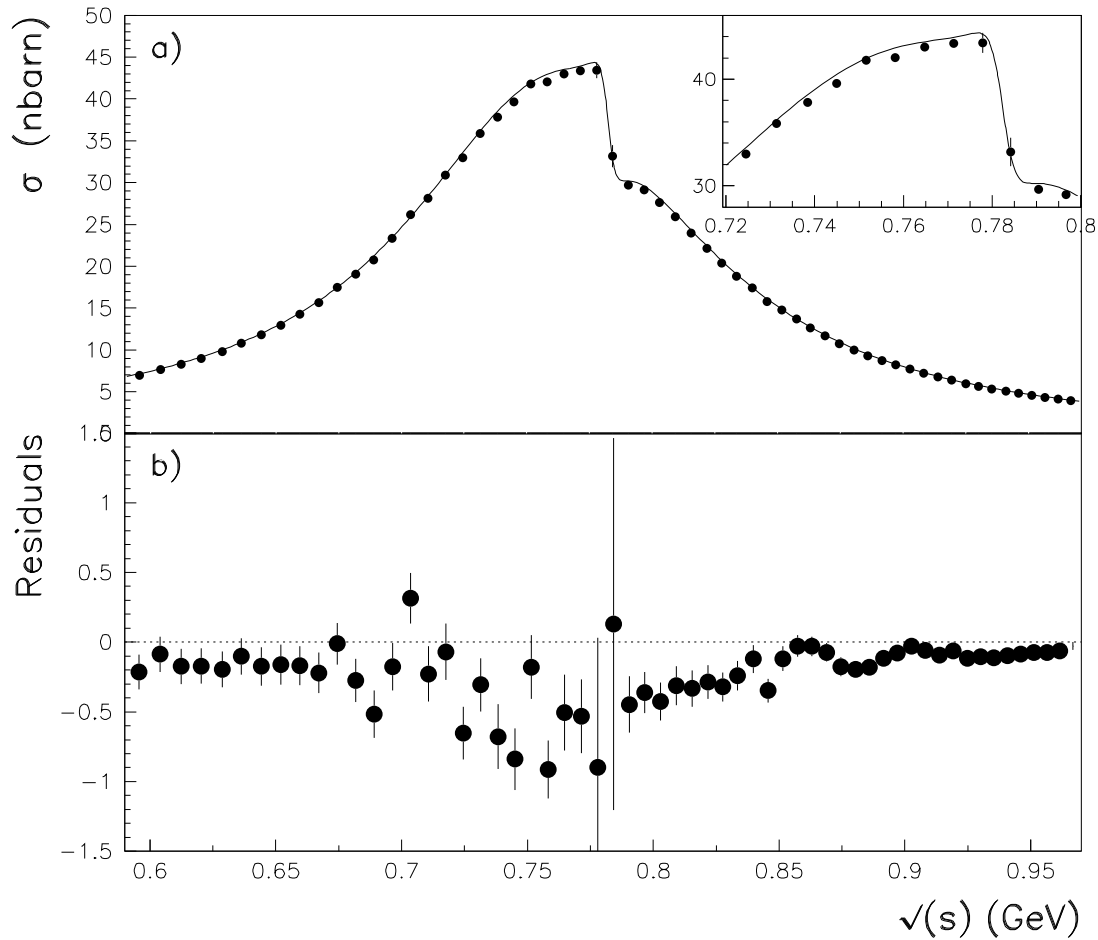}}
\end{center}
\end{minipage}
\begin{center}
\vspace{-0.3cm}
\caption{\label{Fig:KLOE}
ISR data  \cite{KLOE_ISR1} on the  $e^+e^- \ra \pi^+ \pi^-$ cross section collected
by the KLOE  Collaboration at DA$\Phi$NE with the fit superimposed. The 
plotted errors are the square roots of the diagonal terms in the full error covariance 
matrix (see text). The inset  magnifies the  $\rho-\omg$ region. Downmost plot shows
the residual spectrum (with same units) coming out from the global fit performed with all Novosibirsk
$e^+e^- \ra \pi^+ \pi^-$ and $e^+e^- \ra (\pi^0/\eta) \gamma$ data.
 }
\end{center}
\end{figure}

\begin{figure}[!ht]
\begin{minipage}{\textwidth}
\begin{center}
\resizebox{\textwidth}{!}
{\includegraphics*{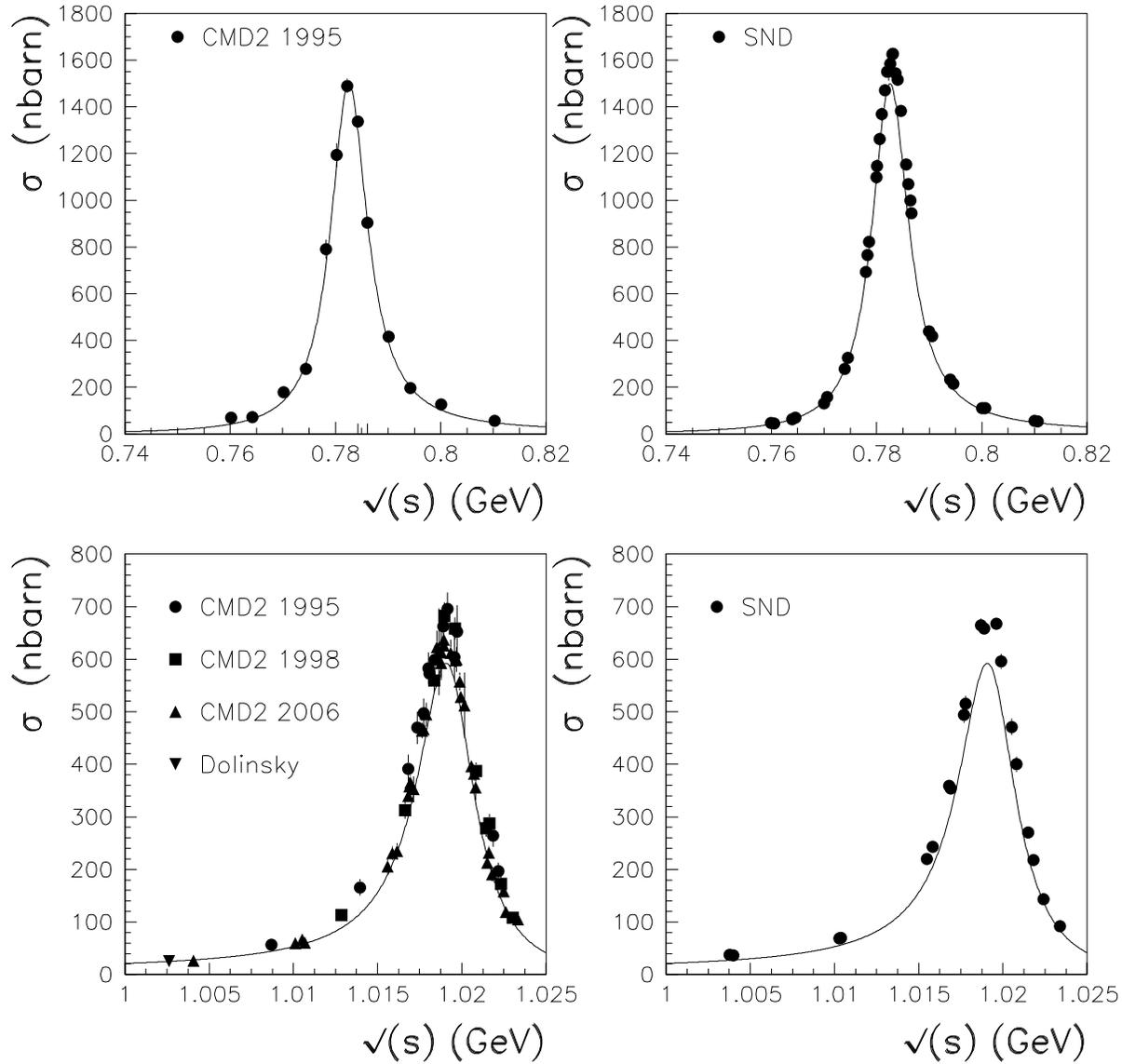}} 
\end{center}
\end{minipage}
\begin{center}
\vspace{-0.3cm}
\caption{\label{Fig:ND-CMD:omgPhi}
Plots of CMD--2 data (left figures) and SND data (right figures)
superimposed with the cross section predicted from running the
global fit with the ND and CMD data as single 3--pion data sets.
Top figures show the case in the $\omg$ peak region, downmost
figures show the $\phi$ peak region.
See text for reference and comments.
}
\end{center}
\end{figure}

\begin{figure}[!ht]
\begin{minipage}{\textwidth}
\begin{center}
\resizebox{\textwidth}{!}
{\includegraphics*{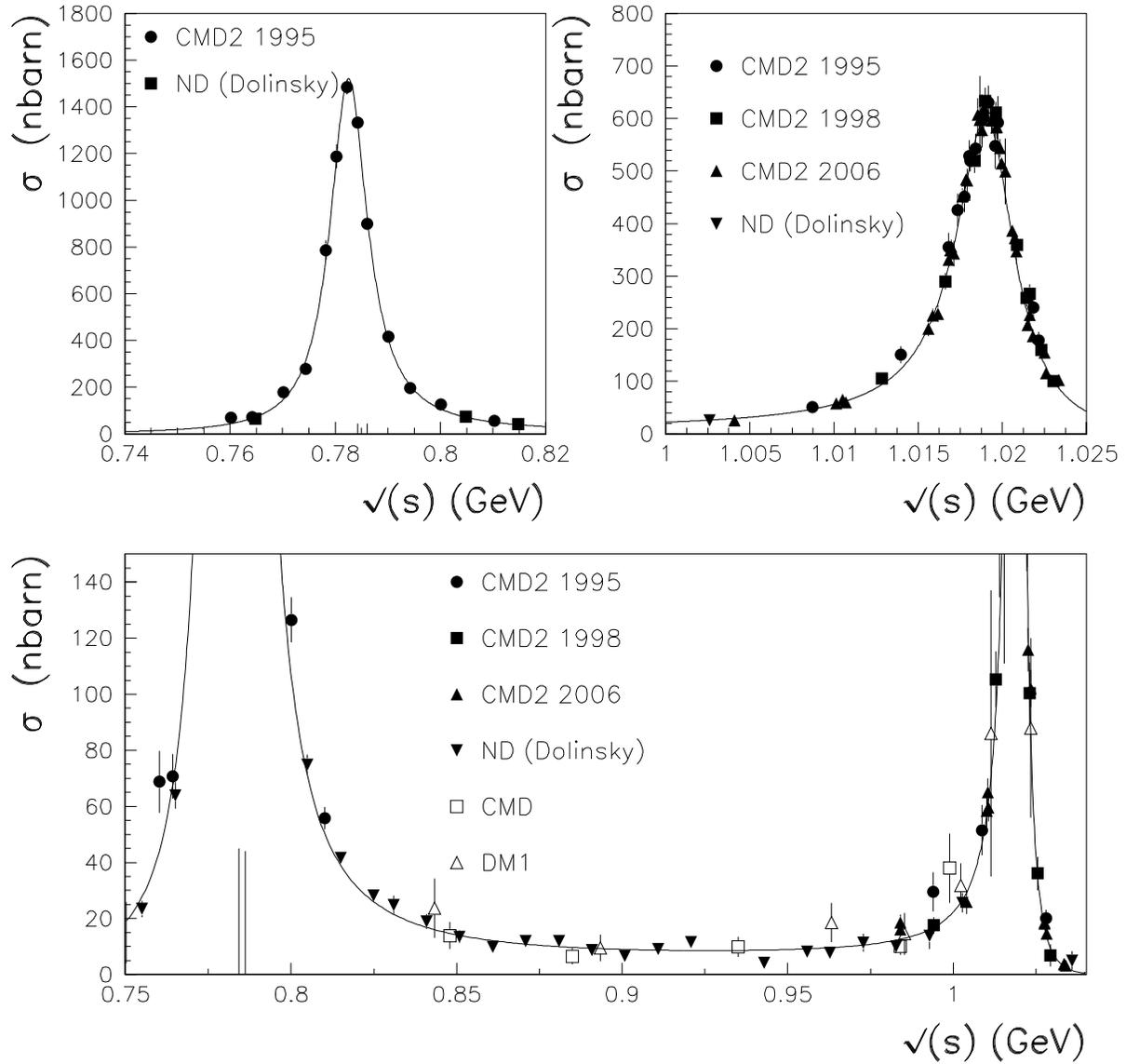}}
\end{center}
\end{minipage}
\begin{center}
\vspace{-0.3cm}
\caption{\label{Fig:cmd2-mid}
Fit performed using CMD--2, ND and the old CMD data (see text).
Top figures show the $\omg$ and $\phi$ peak regions, the downmost
plot shows, magnified, the region outside the peaks. The measurements
of the former DM1 data set have not been used in the fit, but are 
nevertheless displayed in this plot. See text for
references and comments.
}
\end{center}
\end{figure}

\begin{figure}[!ht]
\begin{minipage}{\textwidth}
\begin{center}
\resizebox{\textwidth}{!}
{\includegraphics*{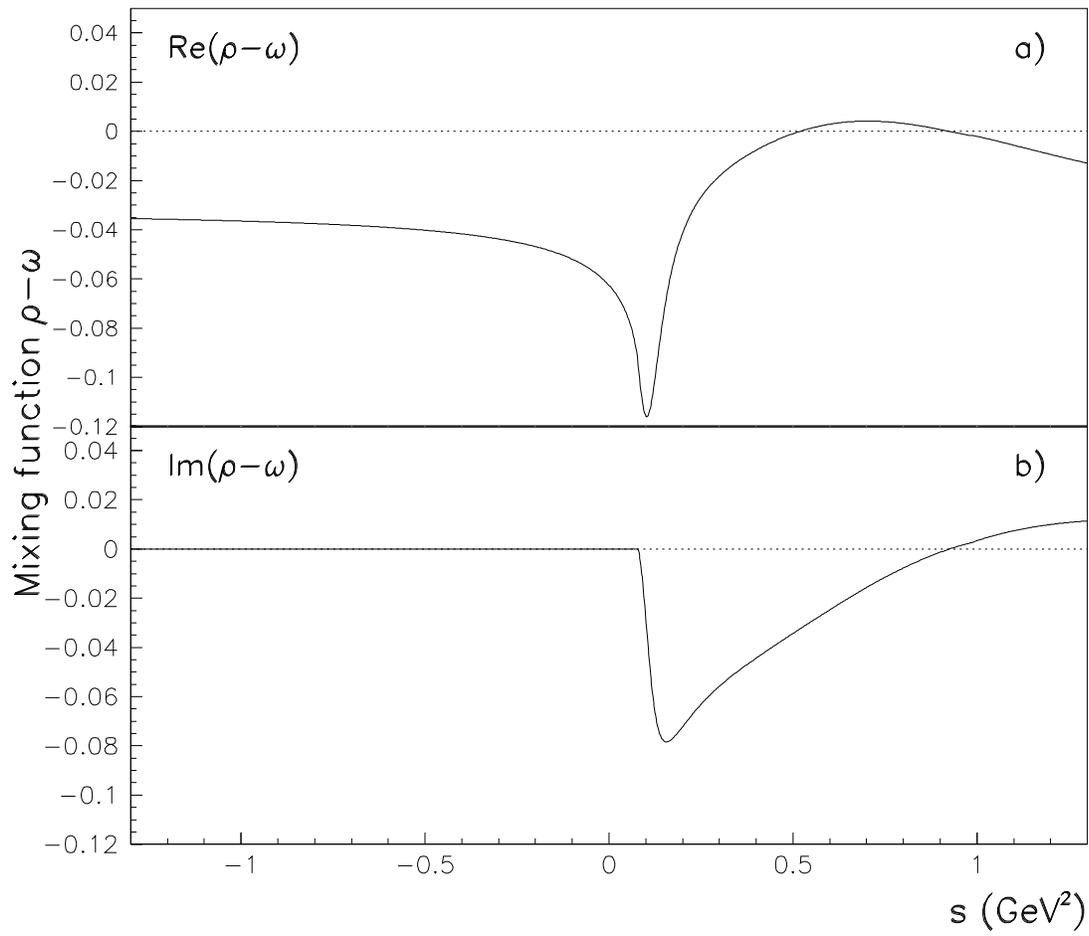}}
\end{center}
\end{minipage}
\begin{center}
\vspace{-0.3cm}
\caption{\label{Fig:alpha}
The isospin breaking parameter $\alpha(s)$. 
}
\end{center}
\end{figure}

\begin{figure}[!ht]
\begin{minipage}{\textwidth}
\begin{center}
\resizebox{\textwidth}{!}
{\includegraphics*{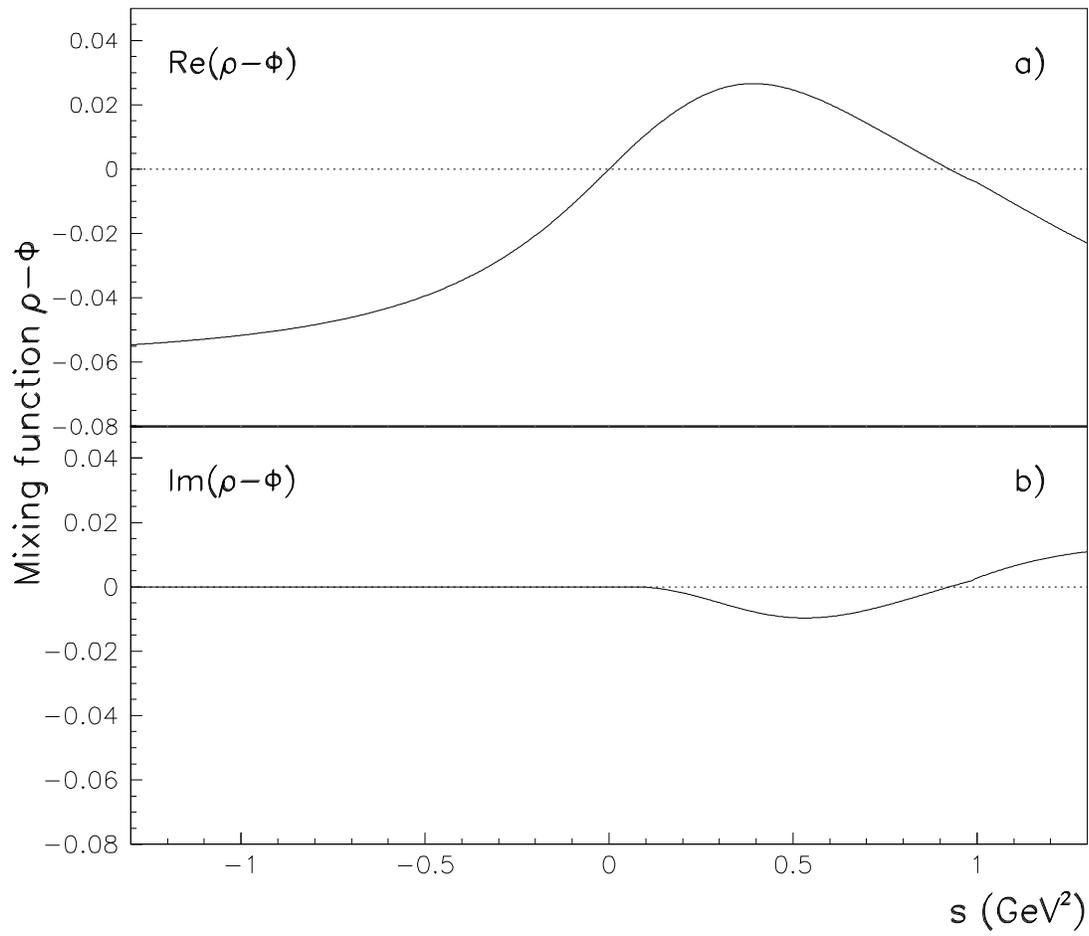}}
\end{center}
\end{minipage}
\begin{center}
\vspace{-0.3cm}
\caption{\label{Fig:beta}
The isospin breaking parameter $\beta(s)$. 
}
\end{center}
\end{figure}

\begin{figure}[!ht]
\begin{minipage}{\textwidth}
\begin{center}
\resizebox{\textwidth}{!}
{\includegraphics*{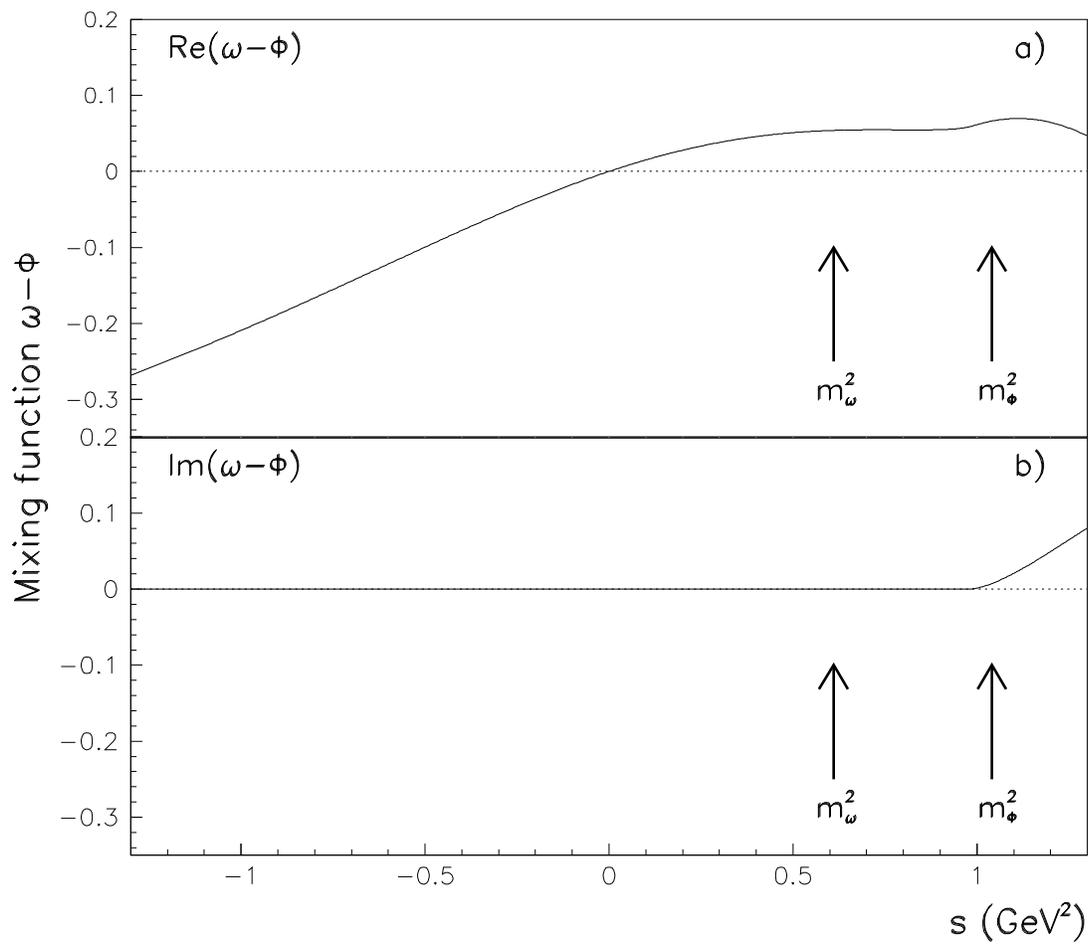}}
\end{center}
\end{minipage}
\begin{center}
\vspace{-0.3cm}
\caption{\label{Fig:gamma}
The isospin breaking parameter $\gamma(s)$. 
}
\end{center}
\end{figure}

\begin{figure}[!ht]
\begin{minipage}{\textwidth}
\begin{center}
\resizebox{\textwidth}{!}
{\includegraphics*{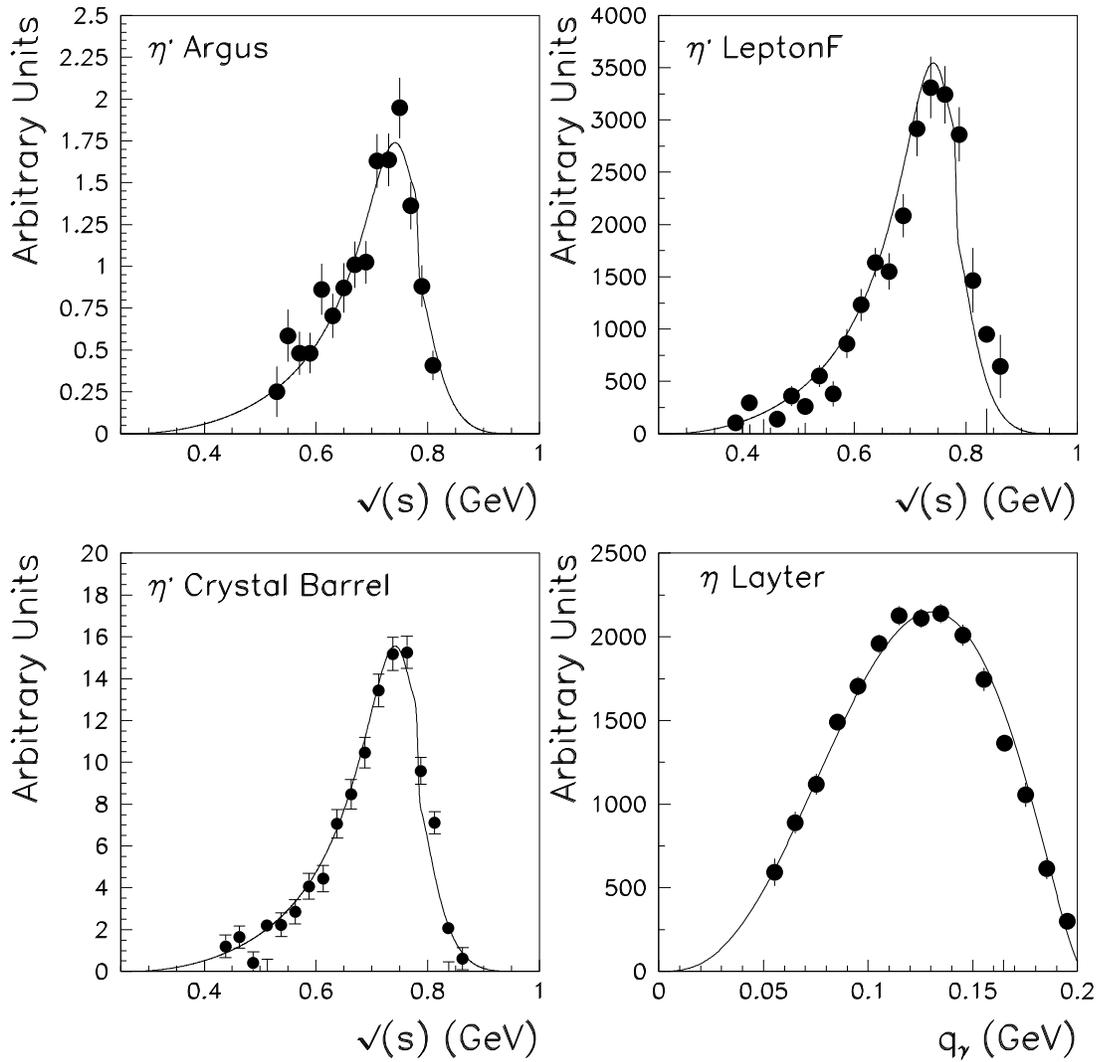}}
\end{center}
\end{minipage}
\begin{center}
\vspace{-0.3cm}
\caption{\label{Fig:box}
Dipion invariant mass spectra in the $\eta^\prime \ra \pi^+ \pi^- \gamma$
decays from Argus \cite{ARGUS}, Lepton F \cite{LeptonF}
and Crystal Barrel \cite{CBar} experiments.The last plot shows the
photon momentum spectrum in the decay 
$\eta \ra \pi^+ \pi^- \gamma$ from \cite{Layter}.
The curves superimposed are predictions, not fits.
}
\end{center}
\end{figure}
\begin{figure}[!ht]
\vspace{0.5cm}
\begin{center}
\includegraphics*[angle=0,width=1.\columnwidth]{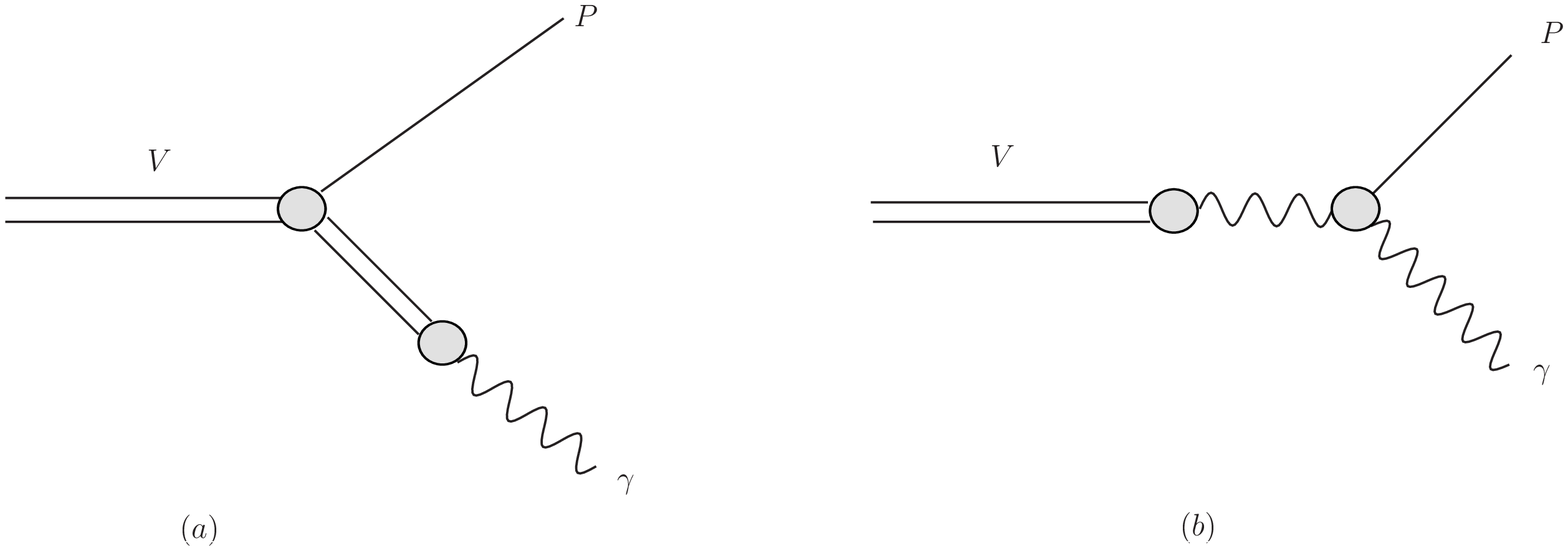}\\
\includegraphics*[angle=0,width=1.\columnwidth]{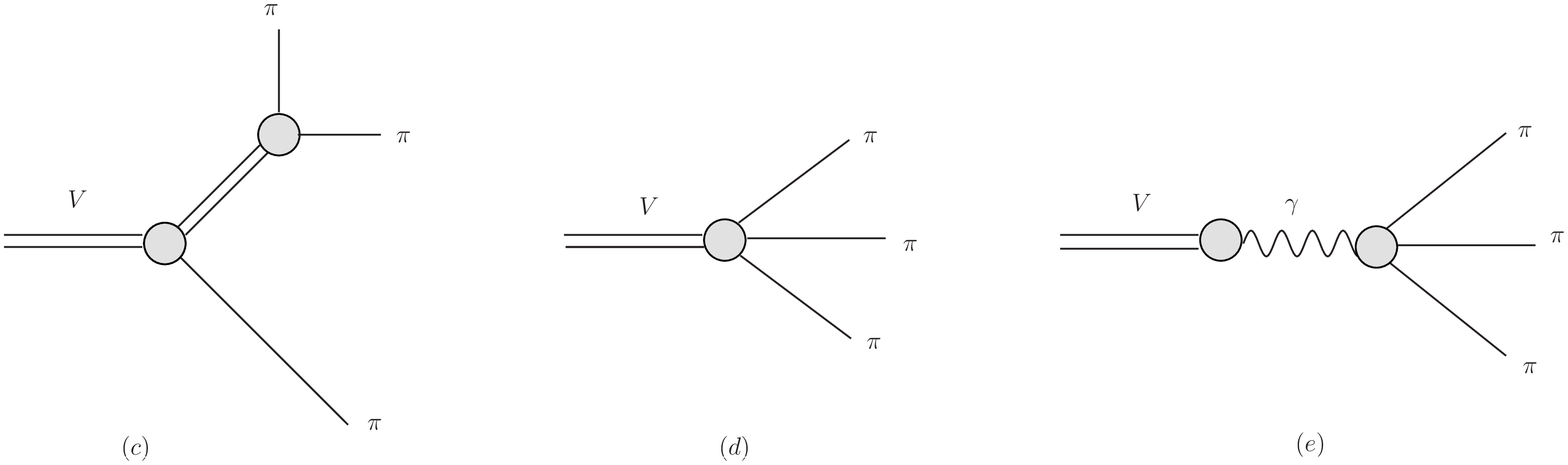}
\caption{\label{graphes} Diagrams sketching the contributions to radiative decays
 -- (a) and (b)-- and to three pion decays -- (c), (d), (e)--
of vector mesons. }
\end{center}
\end{figure}

\end{document}